\documentclass[a4paper,11pt]{article}
\pdfoutput=1 

\usepackage{jheppub} 

\usepackage[T1]{fontenc} 
\usepackage{amssymb, bm, graphicx, graphics, color, mathrsfs, multirow}
\usepackage{subfigure} \subfiguretopcaptrue

\newcommand{\be}{\begin{equation}}
\newcommand{\ee}{\end{equation}}
\newcommand{\ben}{\begin{eqnarray}}
\newcommand{\een}{\end{eqnarray}}

\newcommand{\la}{{\lambda}}

\newcommand{\cL}{{\cal L}}

\newcommand{\na}{\nabla}

\newcommand{\tpe}{{\tilde p}}

\newcommand{\hg}{\hat g}
\newcommand{\hR}{\hat R}
\newcommand{\hD}{\hat D}
\newcommand{\htD}{\hat{\widetilde{D}}}
\newcommand{\hna}{\hat \nabla}

\newcommand{\zpsi}{\psi^{\ast}}
\newcommand{\zchi}{\chi^{\ast}}

\newcommand{\tD}{{\widetilde D}}

\newcommand{\hS}{\hat S}

\newcommand{\ak}{\tpe_k}
\newcommand{\as}{\tpe_s}
\newcommand{\ah}{\tpe_h}
\newcommand{\POv}{_{,v}}
\newcommand{\POu}{_{,u}}
\newcommand{\POuv}{_{,uv}}

\makeatletter
\newcommand{\raisemath}[1]{\mathpalette{\raisem@th{#1}}}
\newcommand{\raisem@th}[3]{\raisebox{#1}{$#2#3$}}
\makeatother

\title{\boldmath Dark sector impact on~gravitational collapse of~an~electrically charged scalar field}

\author[a,b]{Anna Nakonieczna,}
\author[a]{Marek Rogatko,}
\author[c]{and {\L}ukasz Nakonieczny}

\affiliation[a]{Institute of~Physics, Maria Curie-Sk{\l}odowska University, \\
Plac Marii Curie-Sk{\l}odowskiej 1, 20-031 Lublin, Poland}
\affiliation[b]{Institute of~Agrophysics, Polish Academy of~Sciences,\\
Do{\'s}wiadczalna 4, 20-290 Lublin, Poland}
\affiliation[c]{Institute of~Theoretical Physics, Faculty of~Physics, University of~Warsaw, \\
Pasteura 5, 02-093 Warszawa, Poland}

\emailAdd{aborkow@kft.umcs.lublin.pl}
\emailAdd{rogat@kft.umcs.lublin.pl}
\emailAdd{Lukasz.Nakonieczny@fuw.edu.pl}

\abstract{
Dark matter and~dark energy are dominating components of~the~Universe. Their presence affects the~course and~results of~processes, which are driven by~the~gravitational interaction. The~objective of~the~paper was to~examine the~influence of~the~dark sector on~the~gravitational collapse of~an~electrically charged scalar field. A~phantom scalar field was used as~a~model of~dark energy in~the~system. Dark matter was modeled by~a~complex scalar field with a~quartic potential, charged under a~$U(1)$-gauge field. The~dark components were coupled to~the~electrically charged scalar field via the~exponential coupling and~the~gauge field-Maxwell field kinetic mixing, respectively. Complete non-linear simulations of~the~investigated process were performed. They were conducted from regular initial data to~the~end state, which was the~matter dispersal or~a~singularity formation in~a~spacetime. During the~collapse in~the~presence of~dark energy dynamical wormholes and~naked singularities were formed in~emerging spacetimes. The~wormhole throats were stabilized by~the~violation of~the~null energy condition, which occurred due to~a~significant increase of~a~value of~the~phantom scalar field function in~its vicinity. The~square of~mass parameter of~the~dark matter scalar field potential controlled the~formation of~a~Cauchy horizon or~wormhole throats in~the~spacetime. The~joint impact of~dark energy and~dark matter on~the~examined process indicated that the~former decides what type of~an~object forms, while the~latter controls the~amount of~time needed for~the~object to~form. Additionally, the~dark sector suppresses the~natural tendency of~an~electrically charged scalar field to~form a~dynamical Reissner-Nordstr\"{o}m spacetime during the~gravitational collapse.
}

\begin{document} 
\maketitle
\flushbottom

\section{Introduction}
\label{sec:intro}

Black holes which exist in~the~Universe are rotating and~electrically neutral, due to~the~presence of~an interstellar medium. They are formed during gravitational collapse of~a~star, whose mass exceeds the~Tolman-Oppenheimer-Volkoff limit~\cite{Tolman1939-364,OppenheimerVolkoff1939-374}. One of~theoretical toy~models of~this process, which allows to~describe the~resulting spacetime structures and~properties of~the~emerging objects, is~a~dynamical evolution of~an electrically charged scalar field~\cite{SorkinPiran2001-084006,SorkinPiran2001-124024}. A~considerable advantage of~its investigations is~that they allow to~describe physics of~both outer and~inner regions of~the~arising objects.

Dynamical gravitational collapse attracts scientific interest also because the~structures of~dynamically formed spacetimes differ significantly from their non-dynamical counterparts. This means that the~knowledge of~static solutions, which describe a~particular system, is~insufficient to~characterize the~results of~dynamical processes, which take place within~it. For~this reason, the~evolution of~a~massless electrically charged scalar field under the~influence of~gravity was examined~\cite{HodPiran1998-1554,OrenPiran2003-044013}. It~turned out that the~dynamical Reissner-Nordstr\"{o}m spacetime is~substantially different from the~static one due to~the~existence of~a~central spacelike singularity and~a~null singularity along the~Cauchy horizon, which results from the~mass inflation phenomenon~\cite{PoissonIsrael1990-1796}. The~effect of~pair creation during the~collapse and~the~subsequent evaporation of~a~black hole were also considered~\cite{SorkinPiran2001-084006,SorkinPiran2001-124024,HongHwangStewartYeom2010-045014,HwangYeom2011-064020}. The~collapse of~a~Brans-Dicke field was elaborated in~\cite{HwangYeom2010-205002,HwangLeeYeom2011-006,HansenYeom2014-040,HansenYeom2015-arxiv}, while the~results of~the~gravitational evolution in~Einstein-Maxwell-dilaton theory with standard and~phantom couplings of~the~involved fields were described in~\cite{BorkowskaRogatkoModerski2011-084007,NakoniecznaRogatko2012-3175,NakoniecznaRogatkoModerski2012-044043}.

According to~current cosmic microwave background~(CMB) measurements performed by~Planck, the~amount of~dark matter and~dark energy in~the~Universe is~about 27\% and~69\%, respectively~\cite{Ade2014-A16}. Such high abundance of~the~dark components, both of~which interact gravitationally, undoubtedly affects the~course and~results of~the~gravitational collapse, which was one of~the~motivations for~the~analyses presented in~the~paper. The~conducted research concerning the~dark sector influence on~the~relatively small-scale process supplement vast investigations on~its significance during the~cosmological evolution.

Until now, the~collapse in~the~presence of~the~dark sector was studied mainly in~the~context of~the~large scale structure formation and~the~course of~a~few astrophysical processes. The~evolution of~dark matter concentration in~the~form of~a~spherically symmetric dust cloud of~a~finite radius was examined on~a~background spacetime containing dark energy modeled by~isotropic and~anisotropic fluids, as~well as~the~Chaplygin gas~\cite{CaiWang2006-063005,ChakrabortyBandyopadhyay2010-151,RudraDebnath2014-2668}. The~structure formation during the~evolution of~quintessence dark energy induced by~the~collapse of~dark matter halos was described in~\cite{WangFan2009-123012}. The~spherical collapse model of~the~structure formation on~cosmological scales in~the~presence of~the~dark sector was also considered~\cite{MotaBruck2004-71,DelliouBarreiro2013-037}. The~evolution was followed on~a~background of~a~flat, homogeneous and~isotropic universe, while the~matter content was modeled by~a~multicomponent fluid. The~clustering process was investigated during the~non-linear spherical 'top hat' collapse of~the~generalized Chaplygin gas, which unifies dark energy and~dark matter into a~single matter component~\cite{CaramesFabrisVelten2014-083533}.

Recently, the~role of~the~dark sector in~gravitational collapse has been studied from the~astrophysical perspective. The~issue of~a~neutron star collapse due to~capture and~sedimentation of~dark matter within its core was considered as~a~solution to~two interesting observational puzzles~\cite{FullerOtt2015-L71}. One~of~them are fast radio bursts, which are probably extragalactic, transient bright radio pulses of~an unexplained origin~\cite{LorimerEtAl2007-777}. The~second is a~missing pulsar problem, which is related to~the~non-detection of~pulsars within the~inner core of~the~Galactic Center of~the~radius close to~$10$~parsecs~\cite{JohnstonEtAl2006-L6}. The~early stage and~the~long-term phase of~the~formation of~supermassive black holes progenitors via a~direct gas collapse in~dark matter halos at~high redshifts was studied in~\cite{ChoiShlosmanBegelman2015-4411,ShlosmanChoiBegelmanNagamine2015-arxiv}.

The~above-mentioned analyses were mostly focused on~the~behavior of~matter during its gravitational collapse. The~research was performed within~the~framework of~fluid dynamics on~either fixed or~evolving background spacetimes. The~current studies, by~contrast, involve examining dynamical matter-geometry systems with the~matter content of~spacetime described in~terms of~classical field theory. To~our knowledge, such a~methodology has not yet been used in~the~analyses of~the~role of~the~dark sector in~gravitational collapse. It~allowed us not~only to~investigate the~matter behavior, but~also to~follow the~geometry dynamics observed during the~evolution. The~latter was reflected in~the~obtained spacetime structures. Due~to~the~fact that they resulted from the~fully non-linear computations conducted within the~whole spacetime region subjected to~the~collapse dynamics, they were not~limited by~the~capabilities of~the~previously employed local techniques~\cite{CaiWang2006-063005,ChakrabortyBandyopadhyay2010-151} and~allowed presenting global characteristics of~the~forming spacetimes. Their interpretation enabled us to~describe causal relations between distinct spacetime areas and~to~test the~geometrical properties of~both outer and~inner regions of~the~emerging objects.

The~first indirect evidence suggesting the~existence of~an~electromagnetically non-in\-te\-rac\-ting unknown matter component of~the~Universe came from observations of~the~galactic rotation curves~\cite{Zwicky1933-110,RubinFord1970-379}. Since then, the~concept of~dark matter has put down its roots in~modern cosmology. Apart from the~aforementioned experiments, the~gravitational interaction of~dark matter with the~remaining content of~the~Universe may be inferred from the~observations of~CMB, baryon acoustic oscillations~(BAO), the~rate of~the~large scale structures formation~\cite{Ade2014-A16,Ade2014-A17} or~galactic collisions, like the~Bullet Cluster~\cite{RandallMarkevitchCloweGonzalezBradac2008-1173}. These observations also provide information of~the~dark matter abundance and~distribution in~the~Universe, as~well as~its non-gravitational nature, i.e.,~possible interaction types and~the~lifetime of~the~dark matter particle. Unfortunately, they are insufficient to~fully identify the~particle nature of~dark matter, which is~a~dominant matter component of~the~Universe. However, the~non-gravitational interactions of~dark matter disfavoring some of~the~proposed extensions of~the~Standard Model were revealed recently~\cite{massey2015-science,massey2015-mnras}.

There is~yet another branch of~modern physics that is~immensely interested in~the~topic, namely particle physics. The~Standard Model of~particle physics, despite being very successful in~describing the~visible matter, lacks any dark matter candidates. From this perspective the~quest of~gathering and~analyzing information about dark matter is~the~quest of~understanding the~beyond-Standard Model physics. This situation and~an~enormous technological advancement triggered in~the~last few decades made the~subject of~dark matter one of~the~most interesting in~modern physics. Astrophysical observations (other than these mentioned above) provide hints that dark matter may be responsible for~the~$\gamma$-ray bursts from galactic centers~\cite{HooperGoodenough2011-412,AbazajianKaplinghat2012-083511,GordonMacias2013-083521} and~keV photons from galactic clusters~\cite{BulbulEtAl2014-13,BoyarskyRuchayskiyIakubovskyiFranse2014-251301}. The~terrestrial experiments in~supercolliders such as~the~Large Hadron Collider~(LHC) also strive to~identify dark matter or~at~least constrain its models~\cite{Mitsou2013-1330052}, for~a~current list of~the~dark matter models proposed to~search for~in~the~LHC run-2 data see~\cite{Abercrombie2015-arxiv}. Apart from the~above two sources of~the~information on~dark matter, a~new one appeared recently. It~is~concerned with the~theoretical investigation of~the~role of~dark~matter in~the~missing pulsar problem~\cite{BramanteLinden2014-191301,FullerOtt2015-L71} and~its impact on~the~composition and~lifetime of~the~early generation of~stars~\cite{LopesSilk2014-25}.

As~a~dark matter model we used one of~the~viable, from the~particle physics perspective, models of~the~dark matter sector. It~is~composed of~a~complex scalar field with a~quartic self-interaction, which is~charged under an~additional Abelian gauge field. Its~coupling to~the~standard matter sector (represented by~an~electrically charged scalar field) was implemented via a~kinetic mixing between the~introduced gauge field and~the~electromagnetic field~\cite{ChangMaYuan2014-054}. A~new experiment constraining such a~coupling is~currently under preparation~\cite{HeavyPhotonSearchExperiment}, while some previous results can be found, e.g.,~in~\cite{LeesEtAl2014-201801}. The~considered model describes one or~two dark matter candidates, depending on~the~vacuum expectation value~(vev) of~the~complex scalar field. If~the~vev is~non-zero, the~dark matter candidate is~the~massive gauge boson mentioned earlier (in~the~particle physics literature often called $Z^\prime$ or~dark photon). When the~scalar does not~possess the~vev the~candidate can be either the~scalar or~the~gauge boson. A~slightly modified version of~this model was analyzed in~the~context of~dark matter phenomenology~\cite{BaekKoPark2013-013,BaekKoPark2014-73} and~the~GeV-scale $\gamma$-ray excess in~the~Galactic Center~\cite{BaekKoPark2015-255}. On~the~other hand, a~new theoretical tool for~detecting the~influence of~the~dark matter sector on~characteristic quantities of~holographic superconductors was proposed~\cite{NakoniecznyRogatko2014-106004,NakoniecznyRogatkoWysokinski2015-046007,NakoniecznyRogatkoWysokinski2015-066008}. It~happens that the~coupling constant of~the~dark matter sector to~the~ordinary Maxwell field imprints its existence during holographic phase transitions and~these phenomena may constitute a~possible way of~understanding the~nature of~the~dark matter sector.

Dark energy is~a~notion assigned within the~$\Lambda$CDM cosmological model to~the~content of~the~Universe, which remains after determining the~amount of~matter detectable either directly (photons, neutrinos and~baryonic matter) or~indirectly (dark matter). It~is~responsible for~the~present-day accelerated expansion of~the~Universe~\cite{PerlmutterEtAl1999-565,RiessEtAl1998-1009,SchmidtEtA1998l-46,TonryEtAl2003-1}. Experimental restrictions on~dark energy models are imposed by~observations of~supernovae~Ia, combined with the~CMB and~BAO measurements~\cite{CaldwellKamionkowski2009-397}. The~obtained constraints allow of~the~existence of~dark energy in~the~form of~phantom matter~\cite{Caldwell2002-23}. Its~pressure~$P$ and~density~$\rho$ satisfy the~relation $P<-\rho$, which means that the~value of~its barotropic index $w=P\rho^{-1}$ is~less than~$-1$. Matter of~this type is~coupled repulsively to~gravity and~thus exerts negative pressure and~causes cosmic acceleration. Such~a~phantom scalar field coupled exponentially to~the~electrically charged scalar field was chosen as~a~dark energy model in~our research.

Investigating a~dynamical evolution in~the~presence of~phantom fields is~also justified due to~a~wide variety of~non-dynamical solutions to~Einstein equations involving such fields. New classes of~spherically symmetric solutions of~Einstein-Maxwell-dilaton theory with a~phantom coupling were considered in~\cite{GibbonsRasheed1996-515,ClementFabrisRodrigues2009-064021}, while static multicentered solutions in~this theory were discussed in~\cite{AzregAinouClementFabrisRodrigues2011-124001}. A~vast array of~diversified and~complex solutions opens new possibilities of~finding their interesting dynamical counterparts.

The~concept of~a~static wormhole was introduced by~Wheeler~\cite{WheelerGeometrodynamics} and~extended by~Morris and~Thorne~\cite{MorrisThorne1988-395}, who invented a~class of~potentially traversable wormholes. The~wormhole throat is~kept open by~the~presence of~exotic matter, whose energy-momentum tensor violates the~energy conditions. A~scalar field coupled to~gravity in~a~phantom manner is~a~good candidate for~such a~type of~matter, which supports wormhole traversability~\cite{LoboParsaelRiazi2013-084030}. The~quoted works initiated studies of~wormholes as~topological bridges connecting two distinct regions of~spacetime. There were proposed various models such as~thin shells in~cosmological background~\cite{LemosLoboOliveira2003-064004}, in~Brans-Dicke gravity \cite{LoboOliveiro2010-067501}, in~the~low-energy limit of~the~string theory~\cite{EiroaSimeone2005-127501}, as~well as~wormholes supported by~scalar phantom fields~\cite{GonzalesGuzmanMontelongoGarciaZannias2009-064027,Sushkov2005-043520}, tachyonic matter~\cite{DasKar2005-3045} or~non-linear electrodynamics~\cite{BalakinLemosZayats2010-084015}. Recently, to~circumvent the~problem of~exotic matter, the~modified-gravity theories were considered as~suitable candidates to~obtain stable traversable wormholes~\cite{KantiKleihausKunz2011-271101,KantiKleihausKunz2012-044007,HarkoLoboMakSushkov2013-067504,MehdizadehZangenehLobo2015-084004}. It~is~worth emphasizing that these researches deal with constructing eternal, i.e.,~non-dynamical wormholes. There also exists a~research branch, which is concerned with physics of~evolving wormholes~\cite{CataldoMeza2013-064012,ZangenehLoboRiazi2014-024072,PanChakraborty2015-1}. The~aim of~our studies was to~test the~dynamical formation of~such objects during fully non-linear investigations of~the~gravitational collapse.

Because of~the~ambiguous nature of~wormholes in~the~view of~astrophysical observations, a~considerable resurgence of~interests in~their physics was observed recently. Experimental studies of~phenomena and~processes which result from the~presence of~a~black hole in~spacetime are based on~the~effects related to~its external gravitational field. Wormholes possess the~same gravitational properties as~black holes, which are characteristic for~them from the~observational viewpoint~\cite{DamourSolodukhin2007-024016}. For~this reason, a~wormhole is~an~alternative to~a~black hole and~hence various possibilities of~the~experimental distinction between these two types of~objects are being analyzed~\cite{Bambi2013-107501,LiBambi2014-024071}.

Apart from the~efforts to~observationally distinguish wormholes from black holes, the~issue of~experimental distinction between naked singularities and~black holes has also been pursued. These two types of~objects could be~differentiated through their gravitational lensing features~\cite{VirbhadraEllis2000-084003}, such as~a~number of~generated images of~a~particular light source~\cite{VirbhadraEllis2002-103004}, their orientation in~space, total magnifications and~time delays~\cite{VirbhadraKeeton2008-124014,Virbhadra2009-083004}. Another possibility is investigating the~properties of~accretion disks, which form around naked singularities and~black holes, as~their luminosities and~angular velocities of~the~particles within them depend on~the~type of~the~central object~\cite{KovacsHarko2010-124047,JoshiMalafarinaNarayan2014-015002}. The~above findings indicated that the~observational consequences of~the~existence of~naked singularities and~black holes differ. It~opens a~possibility of~testing the~cosmic censorship hypothesis experimentally through astronomical measurements~\cite{SahuPatilNarasimhaJoshi2012-063010}.

Astrophysical objects present in~the~Universe are dynamical, as~they constantly undergo changes due to~interactions with the~surrounding environment. Thus one of~the~elements necessary to~achieve the~aforementioned goal, that is observational determination of~a~true nature of~the~astrophysical objects with strong gravitational fields, is~to~investigate the~formation and~behavior of~black holes, wormholes and~naked singularities during dynamical evolutions.

The~leading research objective of~the~current studies was to~test the~influence of~dark energy and~dark matter on~the~collapse of~an electrically charged scalar field. The~role of~the~dark sector during the~analyzed process was investigated through the~prism of~the~forming spacetime structures, the~behavior of~fields within them and~properties of~the~emerging dynamical objects. The~paper is~organized as~follows. In~section~\ref{sec:model} we present the~theoretical model, which enables us to~study the~dynamical gravitational collapse of~an electrically charged scalar field accompanied by~the~dark sector. Section~\ref{sec:particulars} contains basic information on~solving the~derived equations of~motion and~particulars of~the~results presentation. In~sections~\ref{sec:de} and~\ref{sec:dm} we discuss the~course and~outcomes of~the~investigated process running in~the~presence of~dark energy and~dark matter, respectively. Section~\ref{sec:dmde} is~devoted to~studies of~the~joint influence of~these two components on~the~evolution. The~summary of~the~obtained results and~prospects of~further research are~placed in~section~\ref{sec:conclusions}. Comments on~numerical methods used for~solving the~equations of~motion and~the~code accuracy checks can be found in~appendix~\ref{sec:appendix}.

\section{Theoretical model of~the~evolution}
\label{sec:model}

The~model used to~investigate the~dynamical collapse of~interest consists of~three parts, whose construction is~based on~self-interacting scalar fields. The~first one refers to~a~complex scalar field~$\psi$ coupled with the~Maxwell field~$A_\mu$. The~second one is~a~phantom scalar field~$\phi$, which represents dark energy in~the~system. Its~exponential coupling to~the~previous field is~consistent with the~low-energy string theory regime. The~third component is~a~massive complex scalar field~$\chi$ with a~quartic potential coupled with a~$U(1)$-gauge field~$P_\mu$, which is~coupled to~the~Maxwell field. It~serves as~dark matter in~the~system. Since the~model incorporates notions taken from string theory, it~is convenient to~write the~most general form of~the~action in~the~string frame
\ben
\hS = \int d^{4} x \sqrt{-\hg} \left\{ e^{- 2 \phi}
\left[ \hR - 2 \xi \left( \hna \phi \right)^{\raisemath{-1.15pt}{\hspace{-0.05cm}2}} + e^{2 \alpha \phi} \hat{\cL}_{SF} \right] 
+ \hat{\cL}_{DM} \right\},
\label{a1}
\een
where the~constant $\xi$ determines the~nature of~the~coupling between~$\phi$ and~gravity. It~is~set as~equal to~$-1$, which means that the~field is~phantom~\cite{ClementFabrisRodrigues2009-064021,AzregAinouClementFabrisRodrigues2011-124001}. The~constant $\alpha$ characterizes the~coupling between the~scalar field~$\phi$ and~the~electrically charged one. Since the~low-energy limit of~the~string theory is~considered, we set $\alpha$ as~equal to~$-1$~\cite{Rakhmanov1994-5155}. This value was incorporated in~the~equations below. The~dark energy and~dark matter sectors are chosen to~be independent, so~dark matter represented by~the~Lagrangian $\hat{\cL}_{DM}$ is~not coupled to~the~phantom scalar field via the~exponential coupling~$e^{-2\phi}$. The~presented model enables us to~study the~impact of~dark components on~the~evolution both separately and~jointly. Throughout the~computations, we~used the~geometrized units system, in~which \mbox{$8\pi G=c=1$}.

The~Lagrangian of~the~electrically charged scalar field $\psi$ is~given by~the~expression
\ben
\hat{\cL}_{SF} = - \frac{1}{2} \hD_{\beta} \psi \left(\hD^\beta \psi\right)^{\raisemath{-1pt}{\hspace{-0.05cm}\ast}} 
- F_{\beta \sigma} F^{\beta \sigma},
\een
where $F_{\beta \sigma}$ is~the~strength tensor of~the~Maxwell field. The~covariant derivative has the~form $\hD_\beta = \hna_{\beta} + ieA_{\beta}$, where $e$ is~the~electric coupling constant, $A_\beta$ is~the~four-potential and~$i$~denotes the~imaginary unit.

The~dark matter Lagrangian is~the~following:
\ben
\hat{\cL}_{DM} = - \htD_\beta \chi \left(\htD^{\raisemath{-5pt}{\beta}} \chi\right)^{\raisemath{-1pt}{\hspace{-0.05cm}\ast}}
- \frac{1}{4} B_{\beta \sigma} B^{\beta \sigma}
- \frac{\alpha_{DM}}{4} B_{\beta \sigma} F^{\beta \sigma} - V\left(|\chi|^2\right).
\een
The~covariant derivative is~$\htD_\beta = \hna_\beta + i \tilde{e} P_\beta$, while $P_\beta$ is~the~four-potential of~a~$U(1)$-gauge field, $\tilde{e}$ is~its coupling constant with $\chi$ and~$B_{\beta \sigma}\equiv\partial_\beta P_\sigma-\partial_\sigma P_\beta$. The~scalar field potential is~given by~$V\left(|\chi|^2\right) = \frac{m^2}{2}|\chi|^2 + \frac{\lambda_{DM}}{4}|\chi|^4$. The~constants $m^2$ and~$\lambda_{DM}$ are a~square of~the~mass parameter of~the~scalar field and~its quartic self-interaction coupling constant, respectively. The~square of~the~mass parameter can be either positive or~negative, when it~is negative the~scalar field possesses a~non-zero vacuum expectation value. On~a~side note, due~to~the~system of~units, which was used in~our calculations, it~is impossible to~compare the~parameter $m^2$ with the~particles masses directly. The~perturbativity of~the~quantized theory requires that $\lambda_{DM}$ is~in~general smaller than~$4\pi$. In~the~considered theory it~is restricted to~values not~exceeding $0.2$~\cite{BaekKoPark2013-013}. The~parameter $\alpha_{DM}$ stands for~the~kinetic mixing coupling constant, which controls the~strength of~the~mixing between the~photon and~the~dark photon. Taking into account various experimental scenarios, its upper value is~constrained to~be not~larger than $10^{-3}$~\cite{GuptaPrimulandoSaraswat2015-079}.

The~searches for~the~dark photon have been conducted with the~use of~diverse experimental techniques. The~current constraints on~its mass and~its coupling constant with photon come from a~collection of~experiments, which are adapted to~inspecting their various value ranges. The~respective upper limits obtained during the~dish antenna experiments are $3.1$~eV and~$6\cdot 10^{-12}$~\cite{SuzukiHorieInoueMinowa2015-042}. Studying the~supernovae excess cooling covers the~$\alpha_{DM}$ values from $10^{-10}$ to~$10^{-6}$ and~masses within the~range $10^{-3}-1$~GeV~\cite{DreinerFortinHanhartUbaldi2014-105015}. The~beam dump experiments~\cite{Blumlein2011-155,Blumlein2014-320}, searching for~lepton jets in~$pp$ collisions~\cite{AadEtAl2014-088}, investigating heavy neutrino decays~\cite{Gninenko2012-244} and~analyses of~the~distortion of~the~CMB~\cite{MirizziRedondoSigl2009-026} allowed examining the~kinetic mixing parameter from within the~range $10^{-7}-10^{-4}$ for~the~dark photon mass ranges $0.03-0.63$~GeV, $0.4-1.1$~GeV, $1-500$~MeV and~$10^{-14}-10^{-7}$~eV, respectively. Optical experiments dealt with the~dark photon masses between $10^{-5}$ and~$10^{-2}$~eV and~the~values of~the~mixing parameter greater than $10^{-7}$~\cite{Afanasev2009-317}. The~upper limit of~$\alpha_{DM}$ was established as~equal to~$1.7\cdot 10^{-5}$ for~the~mass range $5-470$~MeV in~experiments focused on~analyzing meson decays~\cite{Archilli2012-251,Babusci2013-111,Adlarson2013-187}. Investigating $e^+e^-$ collisions allowed searching the~mass range $0.02-10.2$~GeV~\cite{LeesEtAl2014-201801}, while analyses of~the~$e^-$ and~$p$ scattering on~nuclei focused on~$175-550$~MeV range~\cite{AbrahamyanEtAl2011-191804,Merkel2011-251802,Agakishiev2014-265}, both with the~values of~the~coupling constant with photon of~the~order of~$10^{-4}-10^{-3}$.

The~variation of~the~action \eqref{a1} with respect to~adequate fields, namely phantom scalar~$\phi$, Maxwell~$A_\mu$, complex scalar fields $\psi$ and~$\chi$, as~well as~the~$P_\mu$ gauge field leads to~the~following set of~equations of~motion:
\ben
\na^{2} \phi + \frac{1}{2\xi} e^{ -2 \phi} F_{\beta \sigma} F^{\beta \sigma} &= 0,
\label{aaa} \\
 \na_{\mu} \left( e^{ -2\phi} F^{\mu \nu} \right) 
+ \frac{1}{4} \Big[
i e \zpsi D^\nu \psi
- i e \psi \left( D^\nu \psi \right)^\ast \Big] 
+\frac{1}{2} \alpha_{DM} \na_\mu B^{\mu\nu} &= 0,
\label{bbb} \\
\na^{2} \psi + i e A^{\beta} \left(
2 \na_{\beta} + i e A_{\beta} \right) \psi
+ i e \na_{\beta} A^{\beta} \psi &= 0,
\label{ccc} \\
\na^{2} \zpsi - i e A^{\beta} \left(
2 \na_{\beta} - i e A_{\beta} \right) \zpsi
- i e \na_{\beta} A^{\beta} \zpsi &= 0,
\label{c1c1c1} \\
\na^{2} \chi - \tilde{e}^2 P_{\beta} P^{\beta} \chi + \frac{1}{2} i \tilde{e} 
\Big[ \na^\beta \left(\chi P_\beta\right) + P_\beta \left(\na^\beta \chi \right) \Big]
-\frac{m^2}{2} \chi - \frac{\lambda_{DM}}{2} |\chi|^2 \chi &= 0,
\label{ddd} \\
\na^{2} \zchi - \tilde{e}^2 P_{\beta} P^{\beta} \zchi - \frac{1}{2} i \tilde{e} 
\Big[ \na^\beta \left(\zchi P_\beta\right) + P_\beta \left(\na^\beta \zchi \right) \Big]
-\frac{m^2}{2} \zchi - \frac{\lambda_{DM}}{2} |\chi|^2 \zchi &= 0,
\label{d1d1d1} \\
\na_{\mu} B^{\mu \nu} - 2\tilde{e}^2 P^\nu |\chi|^2
-i \tilde{e} \left( \chi \na^{\nu} \zchi - \zchi \na^{\nu} \chi \right) 
+\frac{1}{2} \alpha_{DM} \na_\mu F^{\mu\nu} &= 0.
\label{eee}
\een
During the~derivation of~the~above evolution equations the~string frame was converted into the~Einstein frame in~order to~get manageable equations of~motion. The~metrics in~these two frames are related via the~conformal transformation
\ben
g_{\mu \nu} = e^{- 2 \phi} \hg_{\mu \nu},
\label{eqn:conf-tr}
\een
where $g_{\mu \nu}$ and~$\hg_{\mu \nu}$ denote metric tensors in~the~Einstein and~string frames, respectively~\cite{Ortin}. The~transformation between frames~\eqref{eqn:conf-tr} preserves causality. Since the~interpretation of~the~obtained results is~based mainly on~investigating notions related to~causal structures of~emerging spacetimes, we do~not~expect the~transformation to~influence the~ultimate conclusions, which can thus be~regarded as~physically relevant. From now~on, all quantities are written in~the~Einstein frame.

The~issue of~choosing an~adequate conformal frame to~describe physics of~a~particular system has been so~far addressed on~numerous occasions in~research related primarily to~cosmology and~particle physics (see,~e.g.,~\cite{AlvarezConde2002-413,Veneziano2002-581,Flanagan2004-071101,Vollick2004-3813}). A~general conclusion on~the~equivalence of~conformal frames at~the~classical level is that although the~frames are physically equivalent, the~interpretations of~results obtained within them may differ~\cite{DomenechSasaki2015-022}. However, as~was stated above, this drawback does not refer to~our case, as~the~results analysis is~based mainly on~describing causal properties of~spacetimes.

The~most convincing arguments on~physical relevance of~a~particular frame could be~obtained by~comparing theoretical and~experimental data. Some attempts in~this respect have been made recently in~relation to~cosmological observations~\cite{WhiteMinamitsujiSasaki2013-015,ChibaYamaguchi2013-040,DomenechSasaki2015-022}. However, they are not yet complete enough to~resolve the~issue of~the~conformal frames equivalence satisfactorily and~hence the~problem of~its experimental confirmation still remains open.

The~Einstein equations derived by~varying the~action \eqref{a1} with respect to~gravitational field complement the~above set of~equations, which describes the~examined dynamical system. For~studying the~evolutions of~interest, the~double null spherically symmetric line element~\cite{MisnerThorneWheeler} is~selected
\ben
ds^2 = - a(u, v)^2 du dv + r^2(u, v) d \Omega^2,
\label{m}
\een
where~$u$ and~$v$ are retarded and~advanced time null coordinates, respectively, and~$d \Omega^2 = d\Theta^2 + \sin^2\Theta d\Phi^2 $ is~the~line element of~the~unit sphere, where $\Theta$ and~$\Phi$ are angular coordinates. Such a~coordinate choice determines the~spacetime foliation for~conducting computations, which is~2+2~\cite{InvernoSmallwood1980-1223}. The~double null formalism was successfully employed in~dynamical gravitational collapse investigations, e.g.,~\cite{SorkinPiran2001-084006,SorkinPiran2001-124024,HodPiran1998-1554,OrenPiran2003-044013,HongHwangStewartYeom2010-045014,HwangYeom2011-064020}, because it~enables to~follow the~evolution from approximately past null infinity, through the~formation of~horizons up to~the~final central singularity in~the~case of~singular spacetimes.

Regarding spherical symmetry, the~only non-vanishing components of~the~field tensors are $F_{uv}$, $F_{vu}$, $B_{uv}$ and~$B_{vu}$. Due~to~the~gauge freedom $A_{u} \to A_{u} + \na_{u} \theta^\prime$ and~$P_{u} \to P_{u} + \na_{u} \theta^{\prime\prime}$, where $\theta^\prime = \int A_{v}dv$ and~$\theta^{\prime\prime} = \int P_{v}dv$, the~only non-zero four-vector components are~$A_{u}$ and~$P_{u}$. They are functions of~retarded and~advanced time.

The~equation of~motion for~the~phantom scalar field \eqref{aaa} in~the~chosen coordinate system is~provided~by
\ben
r\POu \phi\POv + r\POv \phi\POu + r \phi\POuv 
+ {1 \over \xi} e^{ -2 \phi} {Q^2 a^2 \over 4 r^3} = 0,
\label{d}
\een
where we set
\be
Q = 2 {A_{u, v}~r^2 \over a^2}.
\label{charge}
\ee
$Q$ is~a~function of~retarded and~advanced time, which corresponds to~electric charge within a~sphere of~a~radius $r(u,v)$, on~a~spacelike hypersurface
containing the~point~$(u,v)$. Partial derivatives with respect to~the~null coordinates are marked as~$\POu$ and~$\POv$. Concerning the~assumed line element~\eqref{m} and~the~definition of~electric charge~\eqref{charge}, the~$v$-component of~Maxwell equations~\eqref{bbb} can be separated into two first-order differential equations. The~first one governs the~evolution of~the~only non-zero component of~the~four-vector of~the~Maxwell field
\ben
A_{u, v} - {Q a^{2} \over 2 r^{2}} = 0,
\label{p}
\een
while the~second one describes the~dynamical behavior of~$Q$. Namely, one has
\ben
Q\POv - 2 \phi\POv Q + {i e r^2 \over 4} e^{2 \phi}
\left( \zpsi \psi\POv - \psi \zpsi\POv \right) 
+ \frac{1}{8} \alpha_{DM} e^{2 \phi} T\POv = 0,
\label{l}
\een
where we denoted by~$T$ the~expression
\be
T = {2 r^2 \over a^2} P_{u, v},
\label{charge-dm}
\ee
being related to the~charge associated with the~$P_\mu$ field within a~sphere of~a~radius $r(u,v)$ on~a~specific spacelike hypersurface containing~$(u,v)$. The~relations for~the~complex scalar field \eqref{ccc}--\eqref{c1c1c1} imply
\ben
r\POu \psi\POv + r\POv \psi\POu + r \psi\POuv
+ i e r A_{u}\psi\POv + ier\POv A_{u}\psi 
+ {ieQa^2 \over 4 r} \psi &= 0,
\label{s1} \\
r\POu \zpsi\POv + r\POv \zpsi\POu + r \zpsi\POuv 
- i e r A_{u} \zpsi\POv - i e r\POv A_{u} \zpsi 
- {i e Q a^2 \over 4 r} \zpsi &= 0.
\label{s2}
\een
The~equations of~the~scalar field $\chi$, its complex conjugate $\zchi$ and~the~related $U(1)$-gauge field are given by
\ben
r\chi\POuv + r\POu\chi\POv + r\POv\chi\POu 
+ \frac{1}{4} i \tilde{e} \chi r P_{u,v} + \frac{1}{2} i \tilde{e} r P_u \chi\POv + \frac{1}{2} i \tilde{e} r\POv P_u \chi + \nonumber\\
+ \frac{1}{8} m^2 a^2 r\chi + \frac{\lambda_{DM}}{8} a^2 r |\chi|^2 \chi &= 0,
\label{h1} \\
r\zchi\POuv + r\POu\zchi\POv + r\POv\zchi\POu 
- \frac{1}{4} i \tilde{e} \zchi r P_{u,v} - \frac{1}{2} i \tilde{e} r P_u \zchi\POv - \frac{1}{2} i \tilde{e} r\POv P_u \zchi + \nonumber\\
+ \frac{1}{8} m^2 a^2 r\zchi + \frac{\lambda_{DM}}{8} a^2 r |\chi|^2 \zchi &= 0,
\label{h2} \\
T\POv - 2i \tilde{e} r^2 \left(\chi \zchi\POv - \zchi \chi\POv\right) 
+ \frac{1}{2} \alpha_{DM} Q\POv &= 0, \\
P_{u, v} - {T a^{2} \over 2 r^{2}} &= 0.
\label{p1}
\een

The~stress-energy tensor for~the~considered theory is~the~following:
\ben
T_{\mu \nu} &=& 2 \xi \phi_{,\mu} \phi_{,\nu} - g_{\mu \nu} \xi \phi_{,\beta}\phi^{,\beta} + e^{-2 \phi} \bigg(
2 F_{\mu \beta} F_{\nu}{}{}^{\beta} - {1 \over 2} g_{\mu \nu} F_{\beta \sigma} F^{\beta \sigma} \bigg) + \nonumber \\
&& - \frac{1}{4} g_{\mu \nu} D_\beta \psi \left(D^\beta \psi\right)^{\raisemath{-1pt}{\hspace{-0.05cm}\ast}}
+ \frac{1}{4} \Big[ D_\mu \psi \left(D_\nu \psi\right)^\ast
+ \left(D_\mu \psi\right)^\ast D_\nu \psi \Big] + \nonumber \\
&& + \frac{1}{2} \left[ \tD_\mu \chi \left(\tD_\nu \chi\right)^{\raisemath{-1pt}{\hspace{-0.05cm}\ast}}
+ \left(\tD_\mu \chi\right)^{\raisemath{-1pt}{\hspace{-0.05cm}\ast}} \tD_\nu \chi \right] + B_{\mu\beta} B_\nu^{\ \beta} 
+ \alpha_{DM} B_{\mu\beta} F_\nu^{\ \beta} + \nonumber \\
&& - g_{\mu\nu} \Big[ \tD_\beta \chi \left(\tD^\beta \chi\right)^{\raisemath{-1pt}{\hspace{-0.05cm}\ast}}
+ \frac{1}{4} B_{\beta \sigma} B^{\beta \sigma}
+ \frac{\alpha_{DM}}{4} B_{\beta \sigma} F^{\beta \sigma} + V\left(|\chi|^2\right) \Big].
\label{ten}
\een
Its non-vanishing components calculated in~double null coordinates are of~the~forms
\ben
\label{eqn:Tuu}
T_{uu} &=& 2\xi \phi\POu^2 + \frac{1}{2}
\Big[\psi\POu\zpsi\POu+ieA_u\left(\psi\zpsi\POu-\zpsi\psi\POu\right)+e^2A_u^2\psi\zpsi\Big]+ \nonumber \\
&&+\chi\POu\zchi\POu+i\tilde{e}P_u\left(\chi\zchi\POu-\zchi\chi\POu\right)+\tilde{e}^2P_u^2\chi\zchi, \\
T_{vv} &=& 2\xi \phi\POv^2 + \frac{1}{2} \psi\POv\zpsi\POv + \chi\POv\zchi\POv, \\
T_{uv} &=& e^{-2 \phi} \frac{Q^2a^2}{2r^4} + \frac{T^2a^2}{4r^4} + \alpha_{DM}\frac{TQa^2}{4r^4}
+ \frac{a^2}{4} \left( m^2|\chi|^2 + \frac{\lambda_{DM}}{2}|\chi|^4 \right), \\
T_{\theta\theta} &=& 4\xi\frac{r^2}{a^2}\phi\POu\phi\POv + e^{-2 \phi} \frac{Q^2}{r^2}
+\frac{1}{2}\frac{r^2}{a^2} \Big[\psi\POu\zpsi\POv+\psi\POv\zpsi\POu +ieA_u\left(\psi\zpsi\POv-\zpsi\psi\POv\right)\Big] + \nonumber \\
&&+ \frac{T^2}{2r^2} + \alpha_{DM}\frac{TQ}{2r^2} + \frac{2r^2}{a^2} \Big[\chi\POu\zchi\POv+\chi\POv\zchi\POu +i\tilde{e}P_u\left(\chi\zchi\POv-\zchi\chi\POv\right)\Big].
\label{eqn:Ttt}
\een
Combining the~adequate components of~the~Einstein tensor resulting from the~metric \eqref{m} and~the~above stress-energy tensor components, the~Einstein equations of~the~gravitational field are obtained
{\allowdisplaybreaks
\ben
{2 a\POu r\POu \over a} - r_{,u u}
&=& {r \over 4} \Big[
\psi\POu \zpsi\POu + i e A_{u} \left(
\psi\zpsi\POu - \zpsi\psi\POu \right) + e^2 A_{u}^2 \psi \zpsi \Big] + \nonumber \\
&& + \xi r \phi\POu^2 + {r \over 2} \Big[ \chi\POu \zchi\POu + i \tilde{e} P_{u} \left(
\chi \zchi\POu - \zchi \chi\POu \right) + \tilde{e}^2 P_{u}^2 \chi \zchi 
\Big],
\label{e1} \\
{2 a\POv r\POv \over a} - r_{,vv} &=&
\xi r {\phi_{v}}^2 + {r \over 4} \psi\POv \zpsi\POv
+ {r \over 2} \chi\POv \zchi\POv, \\
{a^2 \over 4r} + {r\POu r_{v} \over r} + r\POuv &=& e^{-2 \phi}{Q^2 a^2 \over 4 r^3} + \nonumber \\
&& + {T^2 a^2 \over 8 r^3} + \alpha_{DM} {TQ a^2 \over 8 r^3} 
+ \frac{r a^2}{4} \left( \frac{m^2}{2}|\chi|^2 + \frac{\lambda_{DM}}{4}|\chi|^4 \right), \\
{a\POu a\POv \over a^2}
- {a\POuv \over a} - {r\POuv \over r} &=&
e^{-2 \phi} {Q^2 a^2 \over 4 r^4} + \xi \phi\POu \phi\POv + {1 \over 8} \Big[
\psi\POu \zpsi\POv + \zpsi\POu \psi\POv
+ i e A_{u} \left(\psi \zpsi_{v} - \zpsi\psi\POv \right) \Big] + \nonumber \\
&& + {T^2 a^2 \over 8 r^4} + \alpha_{DM} {TQ a^2 \over 8 r^4} + \nonumber \\
&& +{1 \over 2} \Big[ \chi\POu \zchi\POv + \zchi\POu \chi\POv
+ i \tilde{e} P_{u} \left(\chi \zchi_{v} - \zchi \chi\POv \right) \Big].
\label{e4}
\een}
They complement the~preceding relations in~order to~obtain the~complete set of~equations of~motion for~the~examined system, which are \eqref{d}--\eqref{e4}, excluding the~relations \eqref{charge}, \eqref{charge-dm} and~\eqref{ten}, which determine the~physical quantities $Q$, $T$ and~$T_{\mu\nu}$.

In~order to~make the~solution of~the~obtained equations of~motion attainable, one~introduces a~set of~auxiliary variables
\ben\label{eqn:substitution}
\begin{split}
c &= \frac{a\POu}{a}, \qquad &d&= \frac{a\POv}{a},
\qquad &f&= r\POu, \qquad &g&= r\POv, \\
k &= \phi, \qquad &x&= \phi\POu, \qquad &y&= \phi\POv, \\
s &= \psi, \qquad &p&= \psi\POu, \qquad &q&= \psi\POv, \qquad &\beta&= A_u, \\
h &= \chi, \qquad &w&= \chi\POu, \qquad &z&= \chi\POv, \qquad &\gamma &= P_u,
\end{split}
\een
and the~quantities denoted by
\ben
\la \equiv \frac{a^2}{4} + f g, \qquad \mu \equiv f q + g p, \qquad \kappa \equiv g w + f z,
\een
which make it~possible to~rewrite the~second-order differential equations \eqref{d}, \eqref{s1}--\eqref{h2} and~\eqref{e1}--\eqref{e4} as~first-order ones. Moreover, the~real fields $\psi_1$, $\psi_2$, $\chi_1$ and~$\chi_2$ are introduced instead of~conjugate fields $\psi$, $\zpsi$, $\chi$ and~$\zchi$ according to~$\psi = \psi_{1} + i \psi_{2}$, $\zpsi = \psi_{1} - i \psi_{2}$, $\chi = \chi_{1} + i \chi_{2}$ and~$\zchi = \chi_{1} - i \chi_{2}$. These relations result in
\ben
\begin{split}
s &= s_1 + i s_2, \qquad &p& = p_1 + i p_2, \qquad &q& = q_1 + i q_2, \\
h &= h_1 + i h_2, \qquad &w& = w_1 + i w_2, \qquad &z& = z_1 + i z_2, \\
\mu &= \mu_1 + i \mu_2, \qquad &\mu_1& = f q_1 + g p_1, \qquad &\mu_2& = fq_2 + gp_2,\\
\kappa &= \kappa_1 + i \kappa_2, \qquad &\kappa_1& = f z_1 + g w_1, \qquad &\kappa_2& = fz_2 + gw_2.
\label{defdef}
\end{split}
\een

The~final system of~equations of~motion, which governs the~investigated evolution yields
{\allowdisplaybreaks
\ben \label{eqn:P1-2}
P1: && a\POu = ac,\\
P2: && a\POv = ad,\\
P3: && r\POu = f,\\
P4: && r\POv = g,\\
P5: && s_{1(2),u} = p_{1(2)},\\
P6: && s_{1(2),v} = q_{1(2)},\\
P7: && h_{1(2),u} = w_{1(2)},\\
P8: && h_{1(2),v} = z_{1(2)},\\
E1: && f\POu = 2 c f - r x^2 - \frac{r}{4}
\Big[p_1^{\: 2} + p_2^{\: 2} + 2e\beta \left(s_1 p_2 - s_2 p_1 \right)
+ e^2 \beta^2 \left(s_1^{\: 2} + s_2^{\: 2} \right) \Big] + \nonumber \\
 && - \frac{1}{2} r \Big[w_1^{\: 2} + w_2^{\: 2} + 2 \tilde{e}\gamma \left(h_1 w_2 - h_2 w_1 \right)
+ \tilde{e}^2 \gamma^2 \left(h_1^{\: 2} + h_2^{\: 2} \right) \Big],
\label{eqn:E1} \\
E2: && g\POv = 2 d g - r y^2 - \frac{r}{4} \left(q_1^{\: 2} + q_2^{\: 2}\right)
- \frac{r}{2} \left(z_1^{\: 2} + z_2^{\: 2}\right),\\
E3: && g\POu = f\POv = - \frac{\la}{r} + e^{-2 k} \frac{Q^2 a^2}{4 r^3} + \frac{T^2 a^2}{8 r^3} 
+ \alpha_{DM}\frac{TQ a^2}{8 r^3} + \nonumber\\
 && + \frac{ra^2}{4} \bigg[\frac{m^2}{2} \left(h_1^{\: 2} + h_2^{\: 2} \right) 
+ \frac{\lambda_{DM}}{4} \left(h_1^{\: 2} + h_2^{\: 2} \right)^2 \bigg], \\
E4: && d\POu = c\POv = \frac{\la}{r^2} - xy -\frac{1}{4}
\Big[p_1 q_1 + p_2 q_2 + e\beta \left(s_1 q_2 - s_2 q_1 \right) \Big] - e^{-2 k} \frac{Q^2 a^2}{2 r^4} + \nonumber \\
 && - \frac{T^2 a^2}{8 r^4} - \alpha_{DM}\frac{TQ a^2}{8 r^4} - w_1 z_1 - w_2 z_2 
+ \tilde{e}\gamma\left(h_2z_1 - h_1z_2\right),
\label{eqn:E4} \\
S_{_{\left(Re\right)}}: && q_{1,u} = p_{1,v} = - \frac{\mu_1}{r} 
+ e \beta q_2 + e s_2 \beta \frac{g}{r} + e s_2 \frac{Qa^2}{4r^2}, \\
S_{_{\left(Im\right)}}: && q_{2,u} = p_{2,v} = - \frac{\mu_2}{r} 
- e \beta q_1 - e s_1 \beta \frac{g}{r} - e s_1 \frac{Qa^2}{4r^2}, \\
M1: && \beta\POv = \frac{Qa^2}{2 r^2},\\
M2: && Q\POv = \bigg[ 2 y Q + {1 \over 2} e r^2 e^{2k} \left( s_1 q_2 - s_2 q_1 \right) + \nonumber\\
 && + \frac{\alpha_{DM}}{4} \tilde{e} r^2 e^{2 k} \left( z_1 h_2 - z_2 h_1 \right) \bigg] 
\left( 1 - e^{2 k} \frac{\alpha_{DM}^2}{16} \right)^{-1},
\label{eqn:M2} \\
H_{_{\left(Re\right)}}: && z_{1,u} = w_{1,v} = - \frac{\kappa_1}{r} 
+ \tilde{e} h_2 \frac{Ta^2}{8r^2} + {1 \over 2} \tilde{e} z_2 \gamma + {1 \over 2} \tilde{e} h_2 \gamma {g \over r} + \nonumber\\
 && - \frac{1}{8} a^2 h_1 \Big[ m^2 + \lambda_{DM} \left(h_1^2 + h_2^2\right) \Big],
\label{eqn:HR} \\
H_{_{\left(Im\right)}}: && z_{2,u} = w_{2,v} = - \frac{\kappa_2}{r} 
- \tilde{e} h_1 \frac{Ta^2}{8r^2} - {1 \over 2} \tilde{e} z_1 \gamma - {1 \over 2} \tilde{e} h_1 \gamma {g \over r} + \nonumber\\
 && - \frac{1}{8} a^2 h_2 \Big[ m^2 + \lambda_{DM} \left(h_1^2 + h_2^2\right) \Big], \\
C1: && \gamma\POv = \frac{Ta^2}{2 r^2}, \\
C2: && T\POv = 2 \tilde{e} r^2 \left( h_1 z_2 - h_2 z_1 \right) 
- \frac{\alpha_{DM}}{2} \bigg[ 2 y Q + {1 \over 2} e r^2 e^{2k} \left( s_1 q_2 - s_2 q_1 \right) + \nonumber\\
 && + \frac{\alpha_{DM}}{4} \tilde{e} r^2 e^{2 k} \left( z_1 h_2 - z_2 h_1 \right) \bigg]
\left( 1 - e^{2 k} \frac{\alpha_{DM}^2}{16} \right)^{-1}.
\label{eqn:C2}
\een}

\section{Details of~computer simulations and~results analysis}
\label{sec:particulars}

Because of~its complexity, the~system of~the~obtained differential equations \eqref{eqn:P1-2}--\eqref{eqn:C2} needs to~be solved numerically. The~details of~the~numerical code and~performed tests are presented in~appendix~\ref{sec:appendix}.

\begin{figure}[tbp]
\begin{minipage}{0.35\textwidth}
\centering
\includegraphics[width=0.75\textwidth]{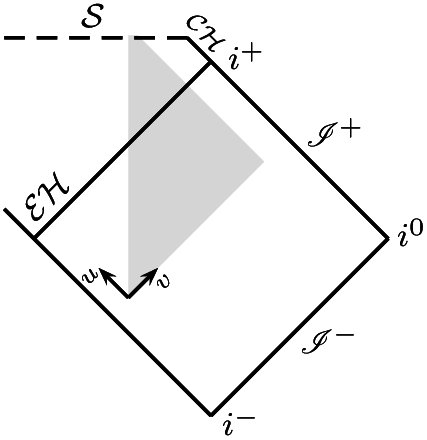}
\end{minipage}
\hfill
\begin{minipage}{0.625\textwidth}
\caption{The~computational domain (marked gray) on~the~background of~the~Carter-Penrose diagram of~the~dynamical Reissner-Nordstr\"{o}m spacetime. The~central singularity along $r=0$, the~event and~Cauchy horizons are denoted as~$\mathcal{S}$, $\mathcal{EH}$ and~$\mathcal{CH}$, respectively. $\mathscr{I}^\pm$ and~$i^\pm$ are null and~timelike infinities, while $i^0$ is~a~spacelike infinity.}
\label{fig:domain}
\end{minipage}
\end{figure}

The~evolution equations were solved in~the~region of~the~$\left(vu\right)$-plane, which is~shown on~the~background of~a~dynamical Reissner-Nordstr\"{o}m spacetime~\cite{OrenPiran2003-044013} in~figure~\ref{fig:domain}. In~all conducted simulations it~was confined to~$0\leqslant v\leqslant 7.5$ and~$0\leqslant u\leqslant 7.5$. The~only arbitrary input data of~the~computations were profiles of~the~evolving fields, posed on~the~initial null hypersurface denoted as~$u=0$. The~initial profile of~the~phantom scalar field was Gaussian
\ben
\phi = \ak\cdot v^2\cdot e^{-\left(\frac{v-c_1}{c_2}\right)^2},
\label{phi-prof}
\een
while the~complex fields were modeled by~the~trigonometric profile of~the~following form:
\ben
\psi(\textrm{or }\chi) = \as(\textrm{or }\ah)\cdot \sin^2\left(\pi\frac{v}{v_f}\right)
\cdot\Bigg[\cos\left(\pi\frac{2v}{v_f}\right)+i\cos\left(\pi\frac{2v}{v_f}+\delta\right)\Bigg].
\label{psichi-prof}
\een
The~above profiles were selected so that they described the~behavior of~real and~complex scalar fields properly~\cite{HamadeStewart1996-497,AyalPiran1997-4768,HodPiran1998-1554,OrenPiran2003-044013}. 
The~profiles were treated as~one-parameter families with amplitudes~$\ak$,~$\as$ and~$\ah$ as~free family parameters. The~amplitudes are indicators of~the~strength of~the~gravitational self-interaction of~the~particular field~\cite{Choptuik1993-9}. The~existence of~a~specific matter type in~the~examined system was guaranteed by~the~non-zero value of~the~respective amplitude. Three systems were investigated, i.e.,
\begin{itemize}
 \item electrically charged scalar field -- dark energy ($SF$--$DE$) with $\as\neq 0$,~$\ak\neq 0$\linebreak and~$\ah=0$,
 \item electrically charged scalar field -- dark matter ($SF$--$DM$) with $\as\neq 0$,~$\ak=0$\linebreak and~$\ah\neq 0$,
 \item electrically charged scalar field -- dark energy -- dark matter ($SF$--$DE$--$DM$) with $\as\neq 0$,~$\ak\neq 0$ and~$\ah\neq0$.
\end{itemize}

The~remaining constants were arbitrarily chosen and~invariable during computations. To be precise, we set $c_1=1.3$, $c_2=0.21$ and~the~parameter determining the~amount of~initial charge~$\delta=\frac{\pi}{2}$. The~maximum value of~advanced time was~$v_f=7.5$. The~choice of~initial conditions is~representative for~the~conducted evolutions, because their outcomes are independent of~the~types of~profiles provided that they are regular, i.e.,~they result in~a~regular spacetime slice at~the~initial null hypersurface. This condition is~fulfilled by~the~selected profiles \eqref{phi-prof} and~\eqref{psichi-prof}.

Provided that the~value of~the~electric coupling constant is~non-zero, it~does not~affect the~results of~the~collapse~\cite{BorkowskaRogatkoModerski2011-084007}, which was confirmed for~the~investigated cases. For~this reason, it~was set as~equal to~$e=0.5$ in~all evolutions. A~similar property was observed for~the~coupling constant $\tilde{e}$, so it~was also kept constant and~equal to~$0.5$ during simulations. The~cases of~vanishing $e$ and~$\tilde{e}$ were not~examined, because they refer to~uncharged complex scalar fields, whose behavior during the~collapse is~beyond the~scope of~the~current research.

The~dynamical spacetime structures resulting from the~considered evolutions will be presented on~Penrose diagrams. These diagrams contain contours 
of $r=const.$ lines plotted in~the~$\left(vu\right)$-plane. The~outermost thick line refers to~$r=0$, which is~non-singular when coinciding with the~$u=v$ line and~singular in~the~remaining part. The~lines indicating the~vanishing expansion 
\ben
\theta_i\equiv\frac{2}{r}r_{,i},
\label{eqn:expansion}
\een
in~the~respective null directions $i=u,v$ will be also presented on~the~diagrams and~their role in~a~spacetime will be commented in~each case separately. They will be denoted as~$r\POv=0$ and~$r\POu=0$ and~marked on~the~diagrams as~red and~blue solid lines, respectively. The~presentation of~Penrose diagrams will be limited to~spacetime regions, which are crucial for~the~conducted results analysis.

\begin{figure}[tbp]
\subfigure[][]{\includegraphics[width=0.445\textwidth]{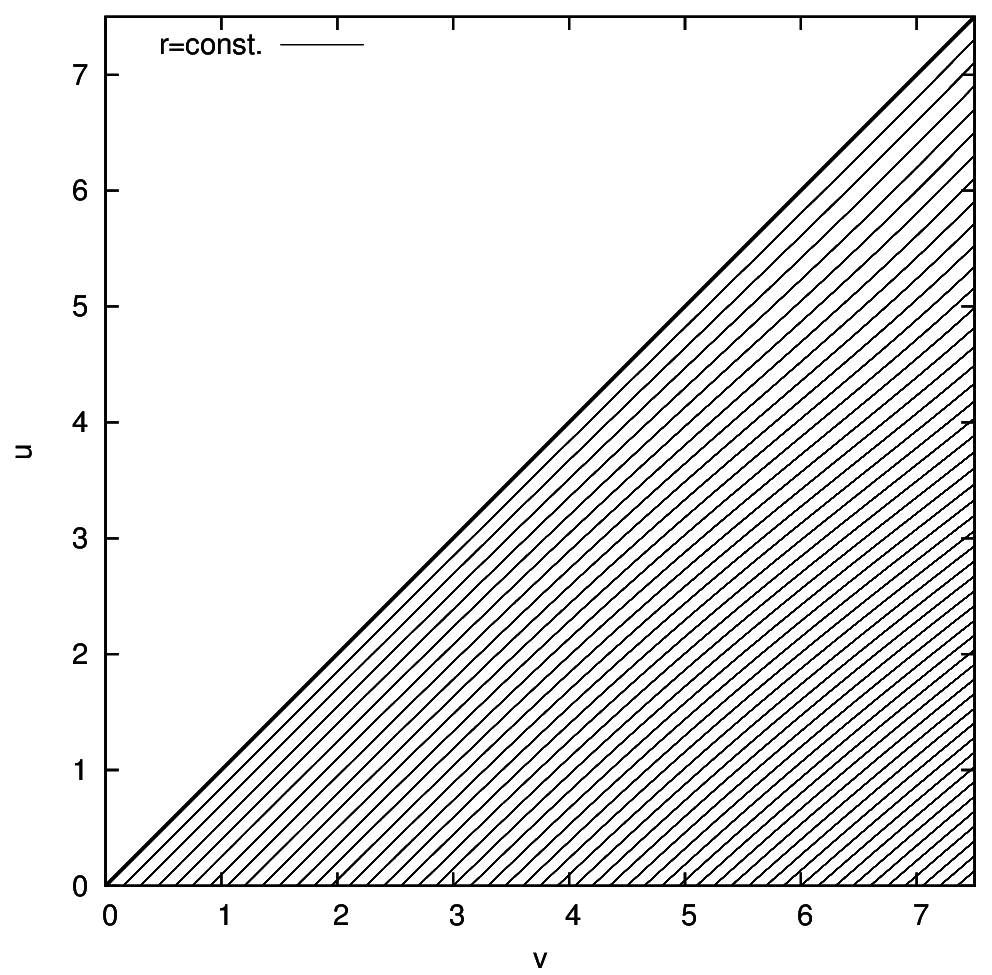}\label{fig:sf-02}}
\hfill
\subfigure[][]{\includegraphics[width=0.46\textwidth]{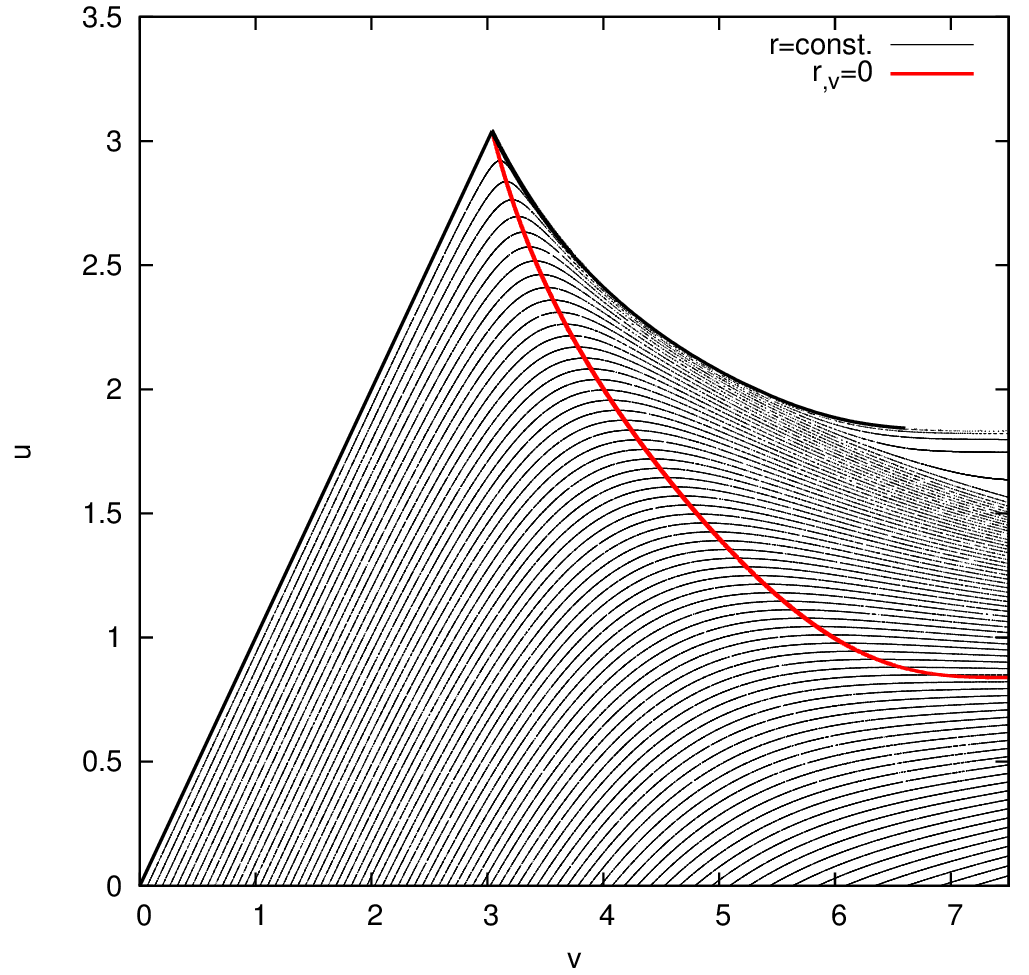}\label{fig:sf-06}}
\caption{(color online) Spacetime structures emerging from the~collapse of~an electrically charged scalar field of~the~amplitude~(a)~$\as=0.2$ and~(b)~$\as=0.6$.}
\label{fig:sf}
\end{figure}

The~aim of~the~paper is~to~present the~influence of~the~dark sector on~the~collapse of~an~electrically charged scalar field. In~figure~\ref{fig:sf} we present the~Penrose diagrams of~spacetimes formed during the~gravitational evolution of~the~field without any additional factors as~a~reference for~further analyses. For~the~initial amplitude value $\as=0.2$ the~emerging curved spacetime is~non-singular, because the~evolving field disperses towards infinity (figure~\ref{fig:sf-02}). The~process with $\as=0.6$ results in~a~dynamical Reissner-Nordstr\"{o}m spacetime~\cite{OrenPiran2003-044013}, shown in~figure~\ref{fig:sf-06}.

The~obtained Reissner-Nordstr\"{o}m spacetime consists of~two parts. The~first one covers the~dynamical region of~spacetime and~spreads up to~about~$v=7.05$. The~second one refers to~$v>7.05$ and~is non-dynamical. This distinction is~based on~the~analysis of~the~behavior of~the~apparent horizon, which lies along the~$r\POv=0$ line. It~changes its position along~$u$ in~the~dynamical region and~becomes null as~$v\to\infty$. Its location there, along $u=0.84$, determines the~event horizon location in~the~spacetime~\cite{FrolovNovikov}. The~Cauchy horizon is~situated at~the~future null infinity, i.e.,~at~$v=\infty$. It~is~invisible within the~covered spacetime domain, but its existence is~confirmed by~the~fact that the~$r=const.$ lines tend towards constant~$u$ with increasing advanced time. The~line $r=0$ is~non-singular up to~$v=3.04$ and~becomes a~spacelike singularity for~bigger values of~advanced time.

One of~physical quantities, which plays a~significant role during the~interpretation of~the~results is~the~quasi-local Hawking mass~\cite{Hawking1968-598}. Its value calculated for~the~spherically symmetric spacetime with two coupled gauge fields $A_\mu$ and~$P_\mu$ is~provided by
\ben
m\left(u,v\right) = \frac{r}{2} \left( 1 + \frac{4fg}{a^2} + \frac{2Q^2 + T^2 + \alpha_{DM}TQ}{r^2} \right).
\label{haw}
\een
It~describes the~mass contained within a~sphere of~a~radius $r(u,v)$ on~a~spacelike hypersurface containing the~point~$(u,v)$. The~mass of~a~particular object is~a~value of~the~above expression at~the~event horizon in~the~non-dynamical region of~the~emerging spacetime, i.e.,~for~$v=v_f$.

\section{Electrically charged scalar field collapse with dark energy}
\label{sec:de}

An~interpretation of~the~dark energy impact on~the~examined collapse was made on~the~basis of~two distinct sets of~solutions describing the~outcomes of~the~process. The~first one corresponds to~the~varying amplitude of~the~phantom field~$\ak$ and~the~constant amplitude of~the~electrically charged scalar field~$\as=0.6$. The~other set was obtained during simulations with the~constant phantom scalar field amplitude $\ak=0.1$ and~the~varying charged scalar field amplitude~$\as$. Such an~approach allows to~investigate the~influence of~the~gravitational self-interaction strength of~a~particular field on~the~studied process. The~evolutions conducted for~the~selected values of~constant amplitudes, when the~accompanying field is~absent, lead to~dynamical spacetimes of~Reissner-Nordstr\"{o}m and~Schwarzschild types, respectively~\cite{NakoniecznaRogatkoModerski2012-044043}.

\subsection{Dynamical emergence of~wormhole-like structures and~naked singularities}
\label{sec:DEstructures}

\begin{figure}[tbp]
\subfigure[][]{\includegraphics[width=0.45\textwidth]{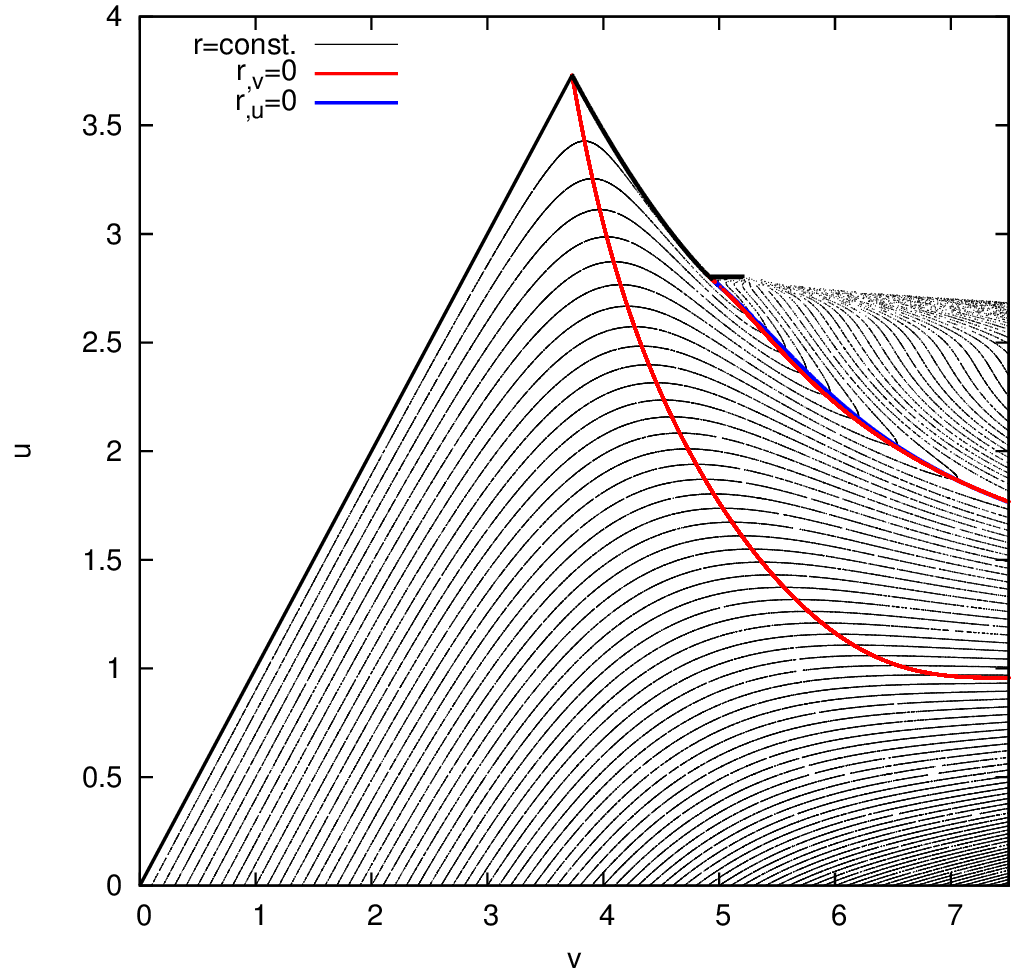}}
\hfill
\subfigure[][]{\includegraphics[width=0.45\textwidth]{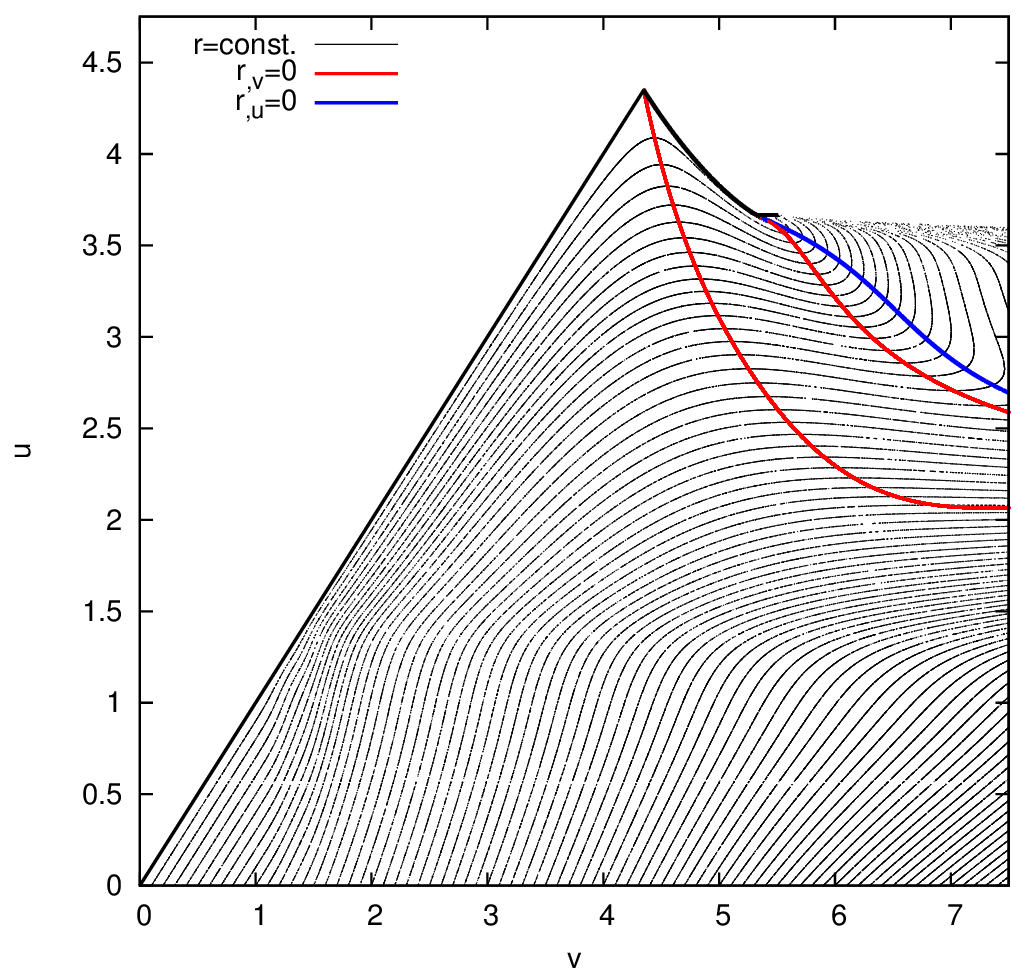}\label{fig:aD-b}}
\subfigure[][]{\includegraphics[width=0.45\textwidth]{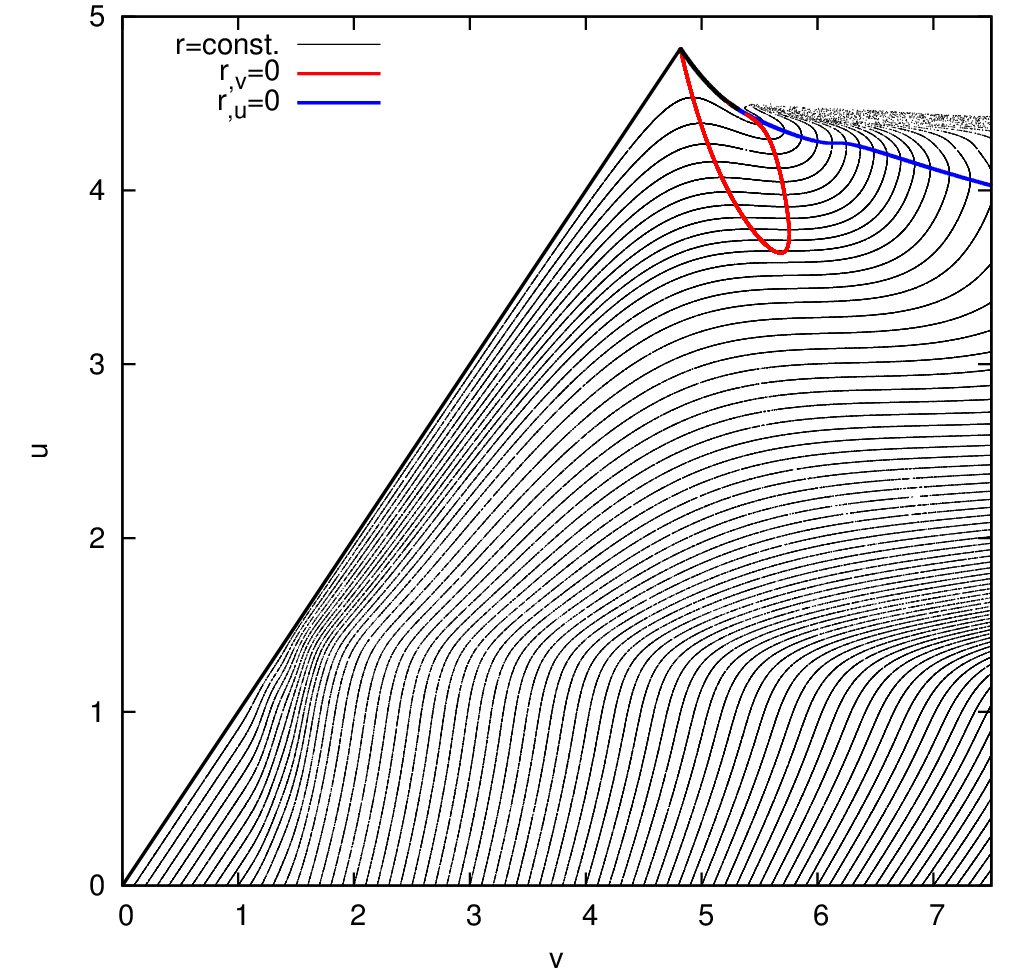}}
\hfill
\subfigure[][]{\includegraphics[width=0.45\textwidth]{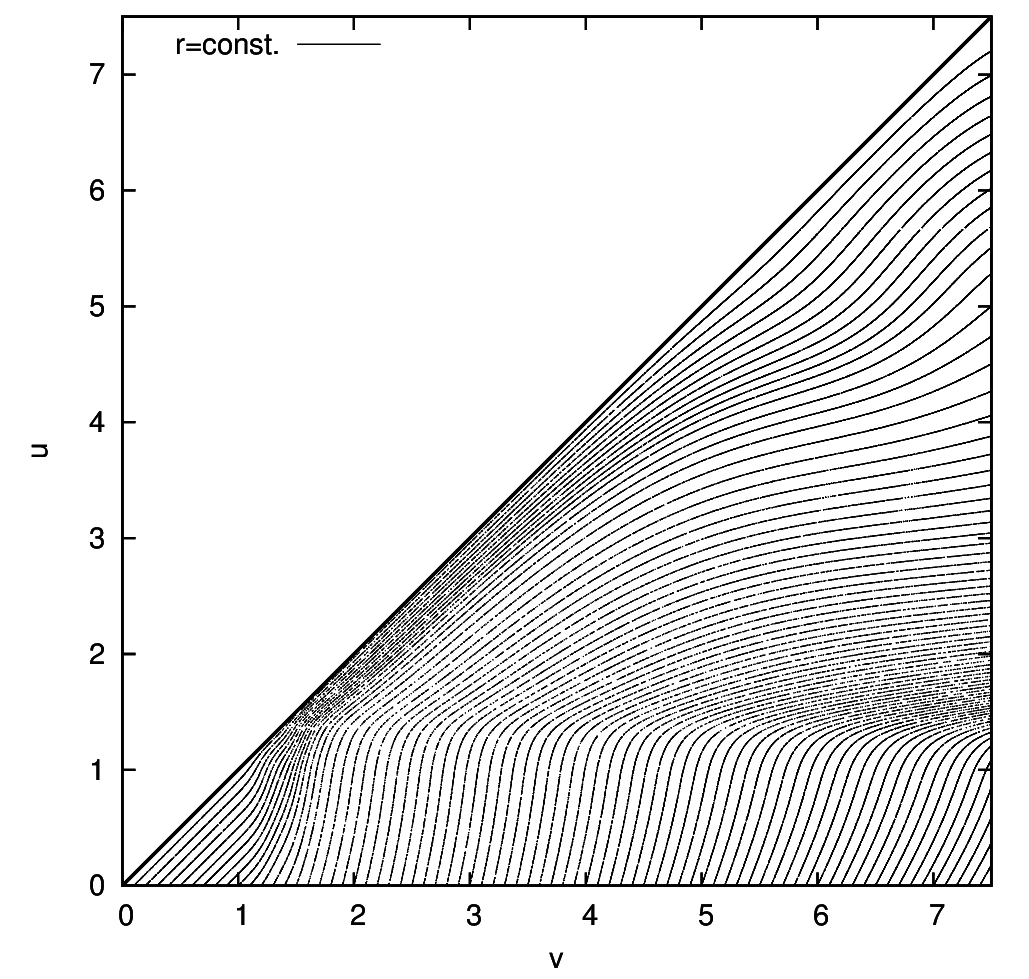}}
\caption{(color online) Penrose diagrams of~spacetimes emerging from the~$SF$--$DE$ evolution with the~varying phantom scalar field amplitude. The~family parameter for~the~electrically charged scalar field $\as=0.6$, while for~the~phantom scalar field $\ak$ is~equal to~(a)~$0.01$, (b)~$0.15$, (c)~$0.2$ and~(d)~$0.25$.}
\label{fig:aD}
\end{figure}

Dynamical spacetimes resulting from the~collapse proceeding for~the~constant electrically charged scalar field amplitude and~varying~$\ak$ are presented in~figure~\ref{fig:aD}. The~emerging spacetimes are singular for~small values of~$\ak$ and~become non-singular for~its bigger values. This results from the~fact that the~phantom field contributes negatively to~the~overall energy of~the~system. For~small values of~the~amplitude~$\ak$, up to~$0.22$, the~spacetimes are singular with both $r\POu=0$ and~$r\POv=0$ lines of~zero expansion visible in~the~domain of~integration. For~values of~the~altering amplitude not~exceeding $0.185$ there are two branches of~the~$r\POv=0$ line. The~outer one surrounds the~singular part of~$r=0$ (for explanation, see section~\ref{sec:particulars}) and~it becomes null as~$v\to\infty$. The~second branch of~the~line $r\POv=0$ lies inside the~first one and~its beginning coincides with the~end of~the~singular spacelike $r=0$ line. The~line of~$r\POu=0$ is~situated beyond the~second branch of~$r\POv=0$ and~begins at~the~same point. For~small values of~the~parameter~$\ak$ these two lines practically coincide.

The~structure described above, with two branches of~$r\POv=0$ lines and~the~line $r\POu=0$ beyond them, is~a~dynamical wormhole. The~strong flare-out condition $\theta_{i,i}>0$ with $i=u,v$~\cite{HochbergVisser1998-044021} is~fulfilled along the~$r\POu=0$ line and~the~inner branch of~$r\POv=0$. This fact proves that they are wormhole throats. Each of~the~throats is~related to~one travel direction. The~outgoing throat is~situated along $r\POu=0$ and~the~ingoing one along~$r\POv=0$. The~strong flare-out condition is~not satisfied along the~outer branch of~the~line $r\POv=0$, because the~relation $\theta_{v,v}<0$ holds there. This confirms that the~branch is~not a~throat, but an~apparent horizon in~the~spacetime, whose null course in~the~region $v\to\infty$ indicates the~location of~an event horizon.

For~values of~the~amplitude~$\ak$ larger than $0.185$, but not~exceeding $0.22$, the~two branches of~the~line $r\POv=0$ form a~loop, whose ends are joined to~the~ending points of~the~singular part of~the~$r=0$ line. The~line $r\POu=0$ remains outside the~loop. Such a~structure corresponds to~a~naked singularity in~the~spacetime~\cite{HwangYeom2011-064020}. For~values of~the~scalar field amplitude~$\ak$ larger than $0.22$, the~spacetime is~non-singular.

\begin{figure}[tbp]
\begin{minipage}{0.5\textwidth}
\subfigure[][]{\includegraphics[width=0.9\textwidth]{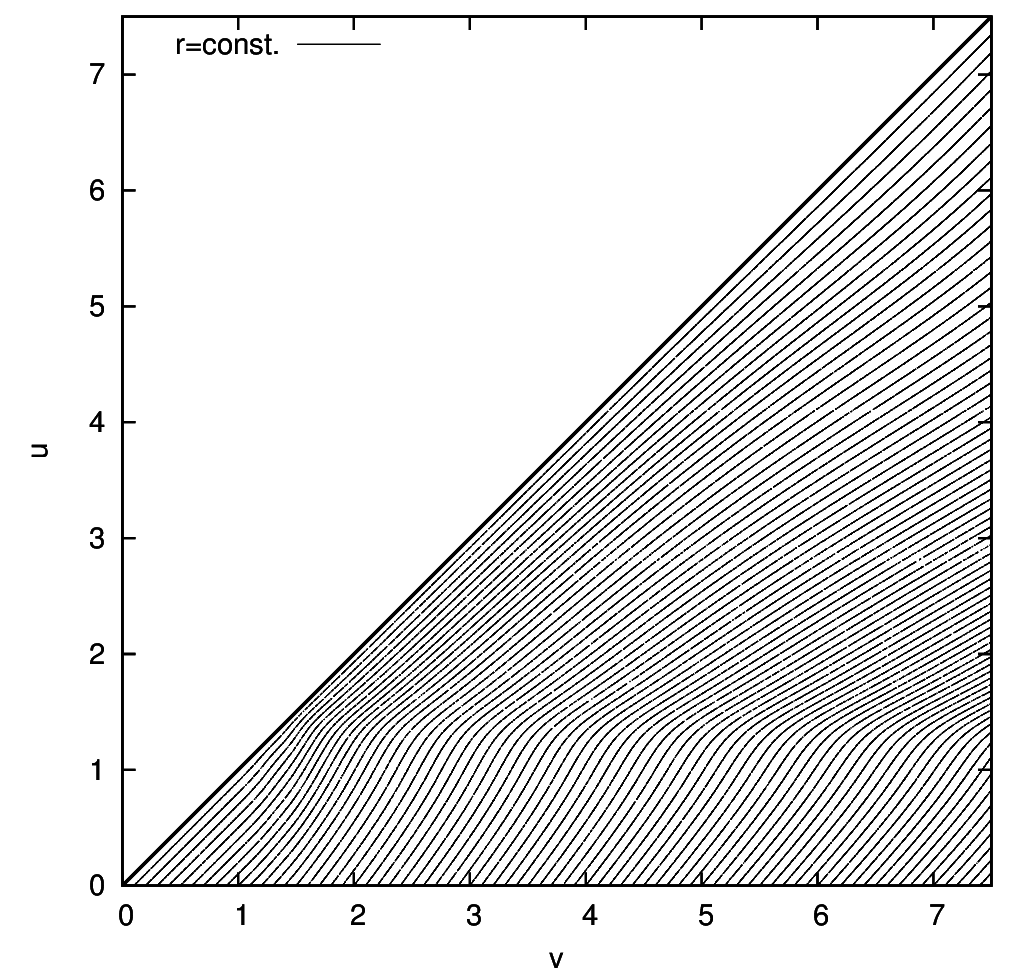}}
\end{minipage}
\hfill
\begin{minipage}{0.5\textwidth}
\subfigure[][]{\includegraphics[width=0.9\textwidth]{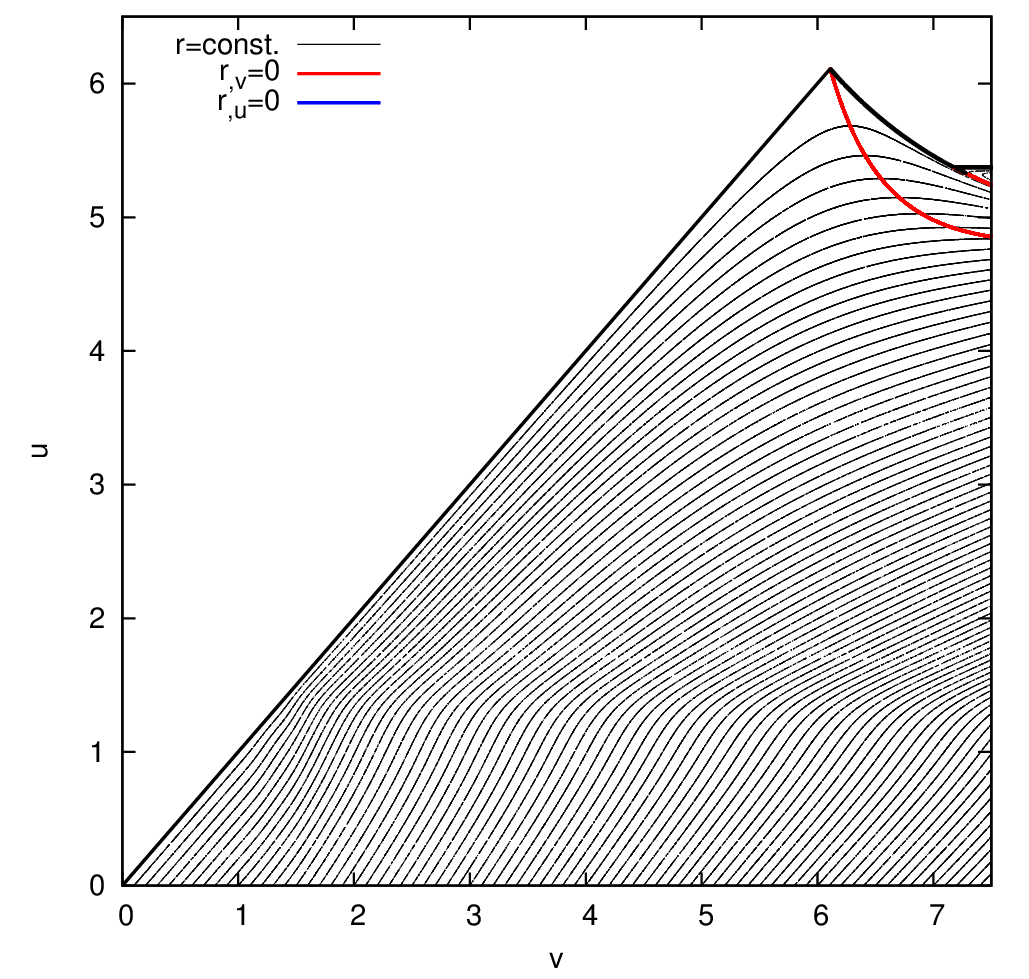}}
\end{minipage}
\begin{minipage}{0.5\textwidth}
\subfigure[][]{\includegraphics[width=0.9\textwidth]{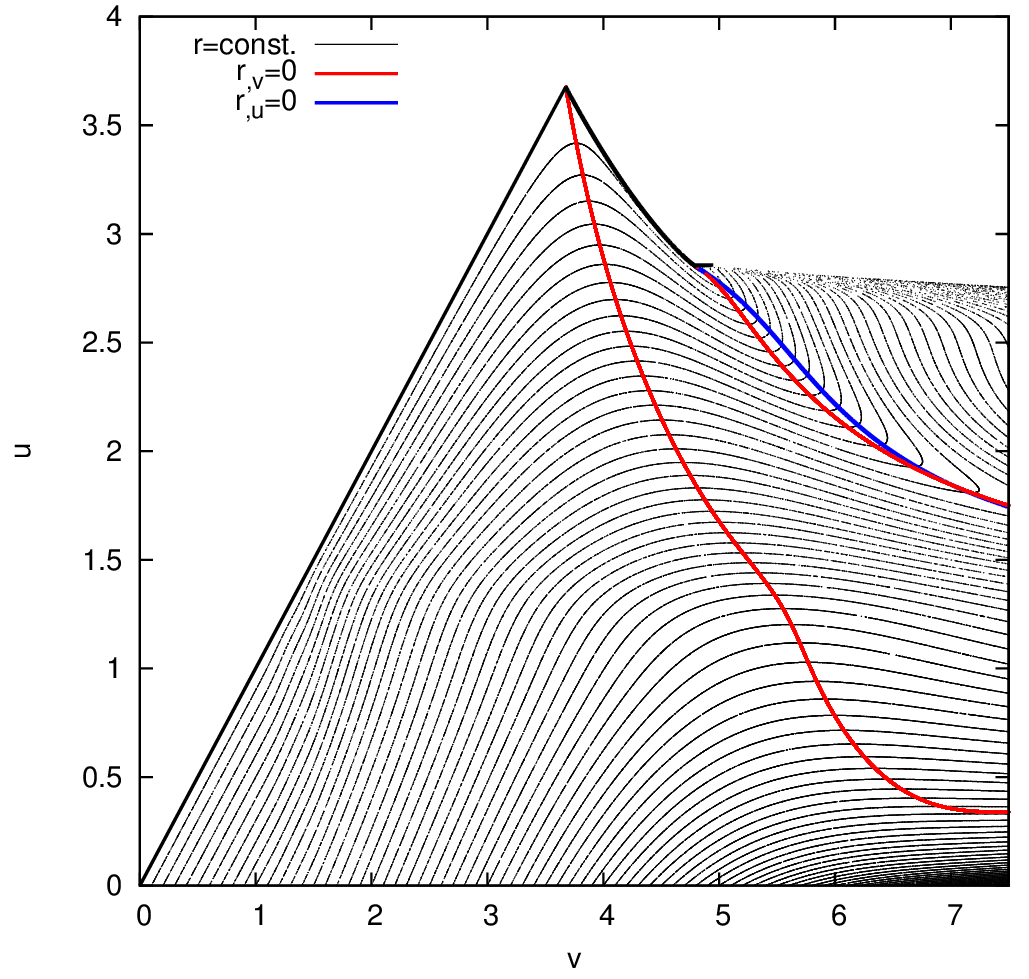}\label{fig:aM-c}}
\end{minipage}
\hfill
\begin{minipage}{0.475\textwidth}
\caption{(color online) Penrose diagrams of~spacetimes emerging from the~$SF$--$DE$ evolution with the~varying electrically charged scalar field amplitude. The~family parameter of~the~phantom scalar field $\ak=0.1$, while $\as$ is~equal to~(a)~$0.3$, (b)~$0.375$ and~(c)~$0.7$.}
\label{fig:aM}
\end{minipage}
\end{figure}

The~second set of~solutions involves spacetimes emerging from a~collapse when the~amplitude of~the~phantom scalar field is~constant and~the~amplitude of~the~electrically charged scalar field changes. In~figure~\ref{fig:aM} the~structures of~spacetimes which stem from the~collapse are depicted. As~in~the~previous set of~solutions, the~wormhole structures are observed. But~unlike the~preceding case, the~spacetime is~non-singular for~small values of~the~amplitude $\as$, not~exceeding $0.35$, and~contains a~wormhole for~its bigger values. The~wormhole structures are similar to~these presented above, that is~there are two branches of~the~$r\POv=0$ line, the~outer and~the~inner one, and~the~line $r\POu=0$ located within them. The~outer branch of~the~line $r\POv=0$ settles along a~constant value of~$u$-coordinate for~$v\to\infty$ indicating the~location of~an event horizon. Contrary to~the~previous case, no naked singularities were observed in~the~formed dynamical spacetimes.

\subsection{Vicinity of~wormhole throats}
\label{sec:de-vic}

As was explained in~the~previous section, due to~the~behavior of~the~expansion \eqref{eqn:expansion} and~its adequate derivatives, the~ingoing and~outgoing wormhole throats are located along the~inner branch of~$r\POv=0$ and~along the~$r\POu=0$ line, respectively. As was anticipated and~confirmed for~eternal wormholes, the~violation of~the~null energy condition~(NEC) 
\ben
T_{\mu \nu}n^\mu n^\nu \geqslant 0,
\label{eqn:nec}
\een
where $n^\mu$ denotes a~null vector, is~crucial for~their existence~\cite{HochbergVisser1998-044021,HochbergVisser1998-746}. 

In~spherical symmetry, taking into account the~line element \eqref{m}, the~stress-energy tensor can be written in~the~general form
\ben
T_{\mu\nu}dx^\mu dx^\nu = T_{uu}du^2 + T_{vv}dv^2 + 2T_{uv}dudv + T_{kk}r^2d\Omega^2,
\label{eqn:Tmn}
\een
where $T_{uu}$, $T_{vv}$, $T_{uv}$ and~$T_{kk}$ are its components, while $k=\Theta,\Phi$. The~matter described by~the~above stress-energy tensor fulfills the~null energy condition when diagonal components of~\eqref{eqn:Tmn} corresponding to~the~coordinates~$u$ and~$v$ are non-negative~\cite{MaedaHaradaCarr2009-044034}, i.e.,
\ben
T_{uu} \geqslant 0 \ \textrm{ and~} \ T_{vv} \geqslant 0.
\label{eqn:NECcond}
\een

\begin{figure}[tbp]
\subfigure[][]{\includegraphics[width=0.475\textwidth]{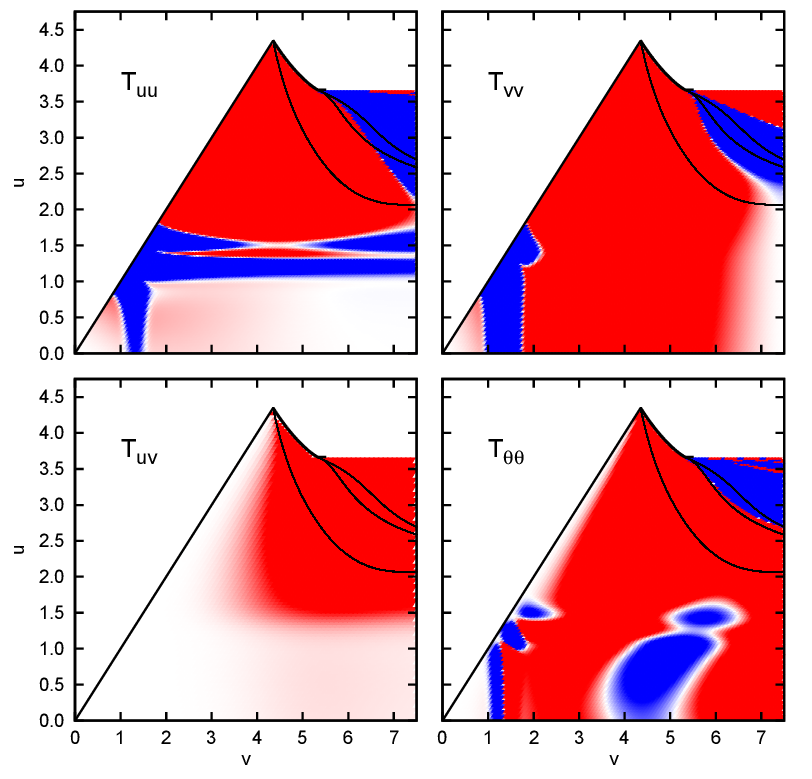}}
\hfill
\subfigure[][]{\includegraphics[width=0.475\textwidth]{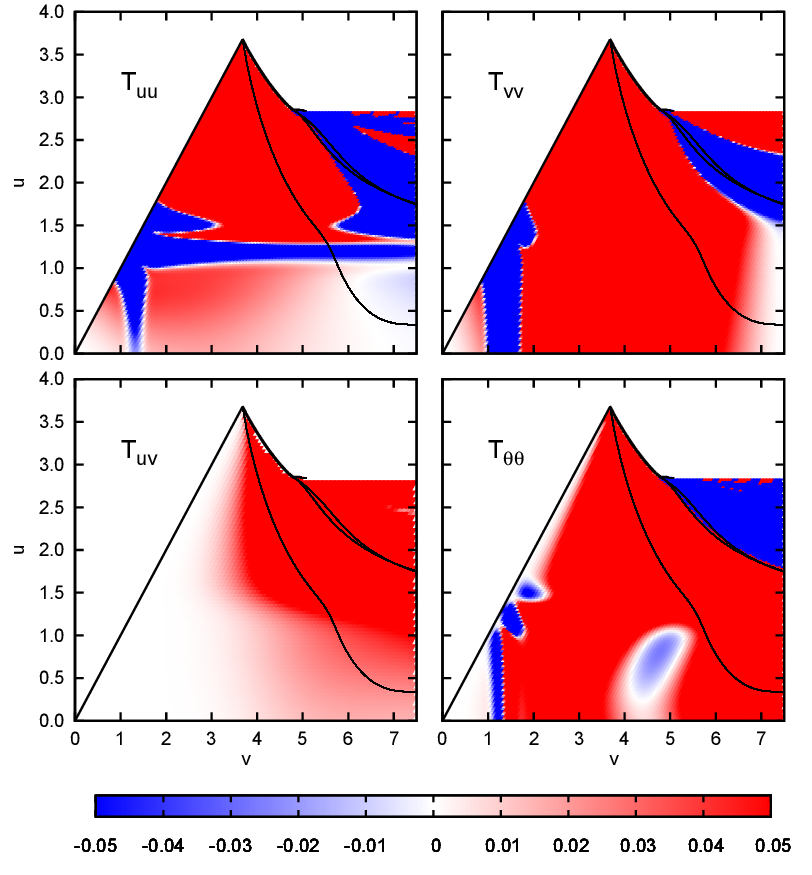}}
\caption{(color online) The~stress-energy tensor components \eqref{eqn:Tuu}--\eqref{eqn:Ttt} in~the~$(vu)$-plane for~evolutions carried out for~electrically charged scalar field and~phantom scalar field amplitudes set as~(a)~$\as=0.6$ and~$\ak=0.15$, (b)~$\as=0.7$ and~$\ak=0.1$.}
\label{fig:T}
\end{figure}

The~non-zero components of~the~stress-energy tensor \eqref{ten} were calculated for~the~wormhole spacetimes in~order to~examine the~fulfillment of~the~NEC in~the~dynamical case. Figure~\ref{fig:T} presents the~components \eqref{eqn:Tuu}--\eqref{eqn:Ttt} for~the~$SF$--$DE$ collapse with the~amplitudes of~the~electrically charged field and~the~phantom field equal to~$\as = 0.6$, $\ak = 0.15$ and~$\as = 0.7$, $\ak = 0.1$ (the~respective spacetime structures were presented in~figures~\ref{fig:aD-b} and~\ref{fig:aM-c}).

In~the~vicinity of~wormhole throats the~violation of~the~null energy condition is~observed, because the~requirement \eqref{eqn:NECcond} is~not satisfied. At~the~same time, this condition is~fulfilled nearby the~singular part of~$r=0$. The~violation of~NEC is~thus a~factor, which determines the~existence of~wormhole throats also in~the~dynamical case. Another area, in~which the~null energy condition is~violated, is~the~vicinity of~null lines $v\approx 1.3$ and~$u\approx 1.3$ in~both investigated wormhole spacetimes. This region is~related to~the~evolution of~the~phantom scalar field, during which it~neither falls into the~central singularity nor evolves through the~wormhole, but escapes to~null infinity as~$v\to\infty$. 

\begin{figure}[tbp]
\subfigure[][]{\includegraphics[width=0.45\textwidth]{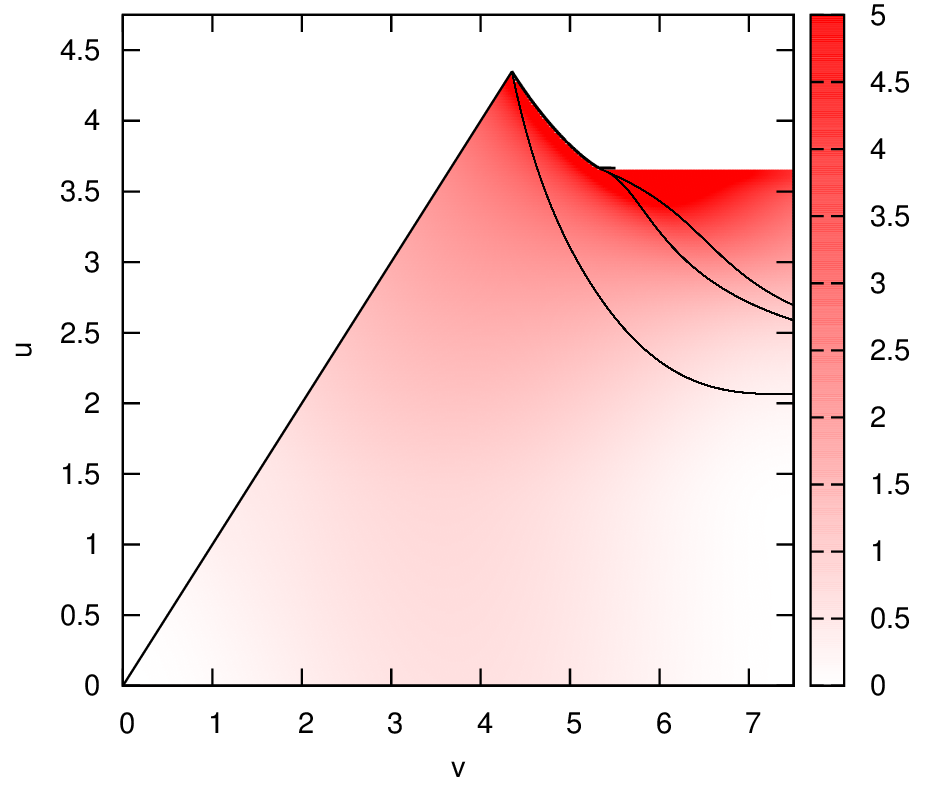}\label{fig:aDfields-a}}
\hfill
\subfigure[][]{\includegraphics[width=0.45\textwidth]{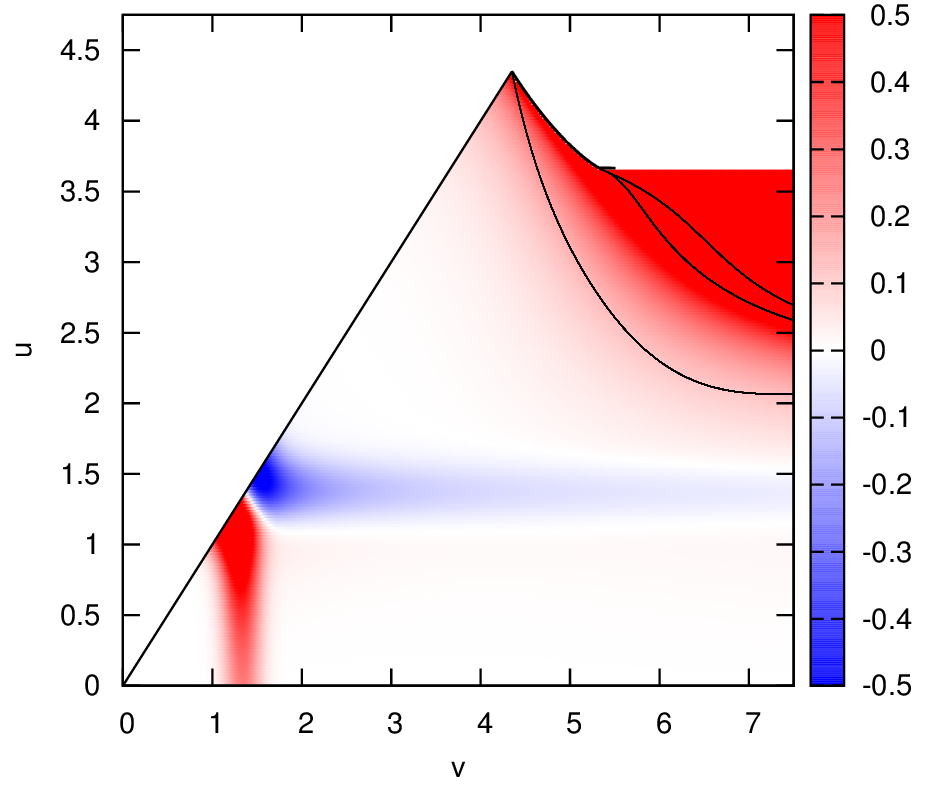}\label{fig:aDfields-b}}
\caption{(color online) The~$\left(vu\right)$-distribution of~(a)~the~modulus of~the~electrically charged scalar field, $|\psi|$, and~(b)~the~phantom scalar field, $\phi$, for~the~$SF$--$DE$ evolution characterized by~initial field amplitudes $\as=0.6$ and~$\ak=0.15$.}
\label{fig:aDfields}
\end{figure}

\begin{figure}[tbp]
\subfigure[][]{\includegraphics[width=0.45\textwidth]{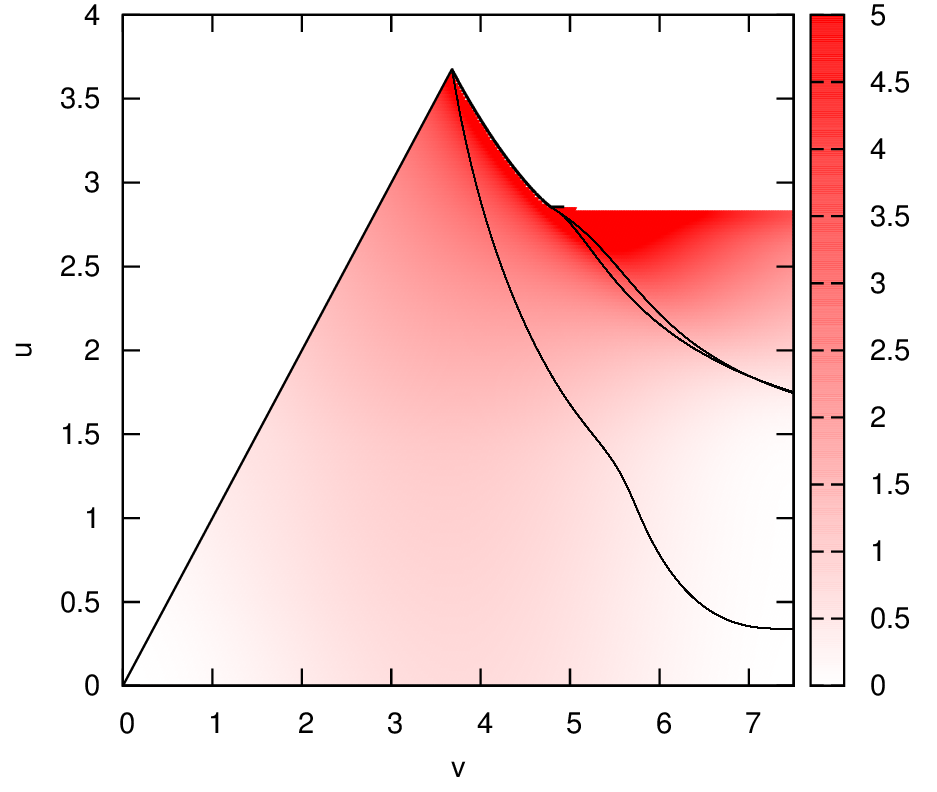}\label{fig:aMfields-a}}
\hfill
\subfigure[][]{\includegraphics[width=0.45\textwidth]{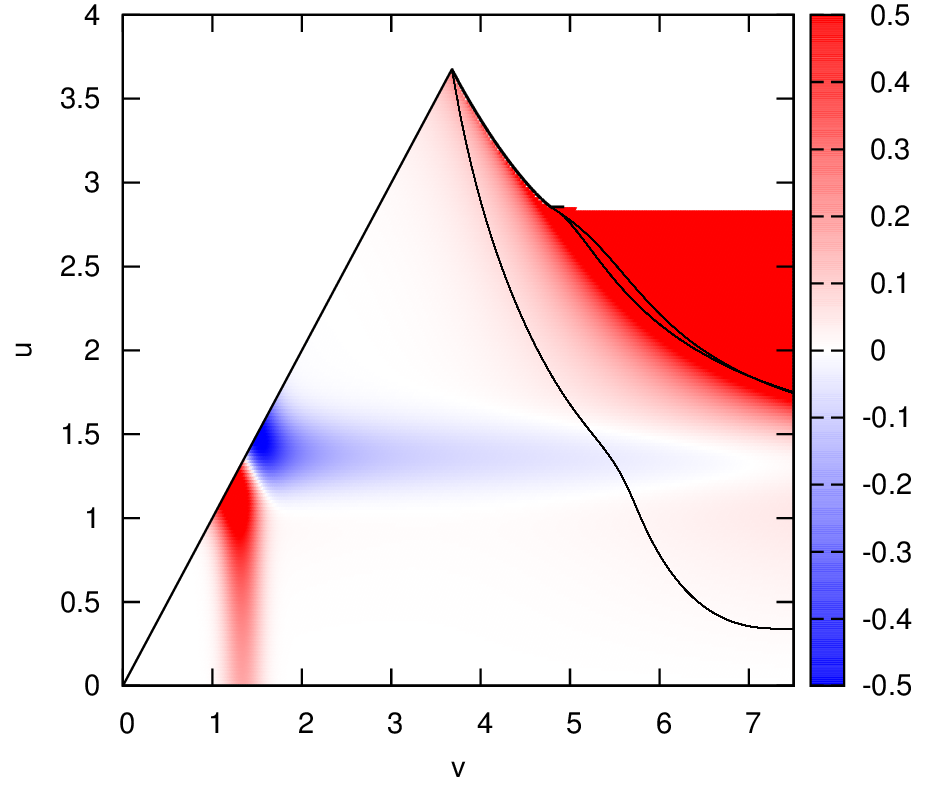}\label{fig:aMfields-b}}
\caption{(color online) The~$\left(vu\right)$-distribution of~(a)~$|\psi|$ and~(b)~$\phi$ for~the~$SF$--$DE$ evolution characterized by~initial field amplitudes $\as=0.7$ and~$\ak=0.1$.}
\label{fig:aMfields}
\end{figure}

Figures~\ref{fig:aDfields} and~\ref{fig:aMfields} depict the~modulus of~the~electrically charged complex scalar field and~the~phantom scalar field in~the~$\left(vu\right)$-plane for~the~same evolutions as~in~figure~\ref{fig:T}. The~modulus of~the~electrically charged scalar field increases with retarded time uniformly for~all values of~the~$v$-coordinate. It~is~hard to~distinguish any area, within which the~field accumulates and~the~biggest values are one order of~magnitude larger than the~smallest ones. The~value of~the~function related to~the~phantom field is~considerably higher in~the~region nearby wormhole throats, where it~is four orders of~magnitude bigger than in~the~surrounding spacetime region. It~can be interpreted as~a~phenomenon of~stabilization of~the~wormhole structure by~phantom matter~\cite{Sushkov2005-043520}. The~value of~the~phantom scalar field function is~also significantly bigger along the~hypersurfaces $v\approx 1.3$ and~then $u\approx 1.3$ indicating the~course of~the~dispersive evolution of~the~field, which was also noticed during the~above interpretation of~the~stress-energy tensor components.

\subsection{Black hole-wormhole duality}

A~possibility of~a~two-way interconversion between black holes and~wormholes was discussed in~\cite{Hayward1999-373,Hayward2009-124001,ShinkaiHayward2002-044005,KoyamaHaywardKim2003-084008,KoyamaHayward2004-084001}. It~was stated that a~black hole may turn into a~wormhole when irradiated by~matter with negative energy. On~the~other hand, a~wormhole stabilized by~such exotic matter shall turn into a~black hole when the~matter source disappears.

Dynamical wormhole spacetimes obtained in~our computations contain spacelike singularities for~large values of~retarded time and~the~existence of~event horizons is~due to~these late-$u$ singularities. A~singularity exists during the~dynamical stage of~the~collapse and~bifurcates into two wormhole throats when the~spacetime tends towards its stationary phase at~$v\to\infty$. Such a~course of~the~investigated process suggests that a~black hole, which emerges in~the~preliminary stage of~the~evolution, finally converts into a~wormhole. The~existence of~a~singularity at~large values of~retarded time is~strictly connected with the~fact that the~null energy condition is~fulfilled in~the~area (figure~\ref{fig:T}), what is~in~turn a~result of~a~considerably smaller value of~the~phantom field function in~this region (figures~\ref{fig:aDfields-b} and~\ref{fig:aMfields-b}). The~reason for~the~black hole-wormhole conversion is~a~substantial increase of~its values in~the~particular area, which in~consequence turns into a~tunnel outlined by~wormhole throats.

\subsection{Properties of~wormholes}

The~issue whether and~in~what manner do the~properties of~the~obtained spacetimes differ when one of~the~field amplitudes is~constant and~the~value of~the~other amplitude changes was also addressed. The~ranges of~varying amplitudes were chosen in~such a~way that in~the~case of~the~constant electrically charged scalar field amplitude the~biggest~$\ak$ is~approximately the~last one not~leading to~the~formation of~a~naked singularity. In~the~case of~a~constant phantom scalar field amplitude the~smallest~$\as$ refers to~the~first dynamical wormhole 
spacetime which appears instead of~a~non-singular spacetime.

\begin{figure}[tbp]
\subfigure[][]{\includegraphics[width=0.45\textwidth]{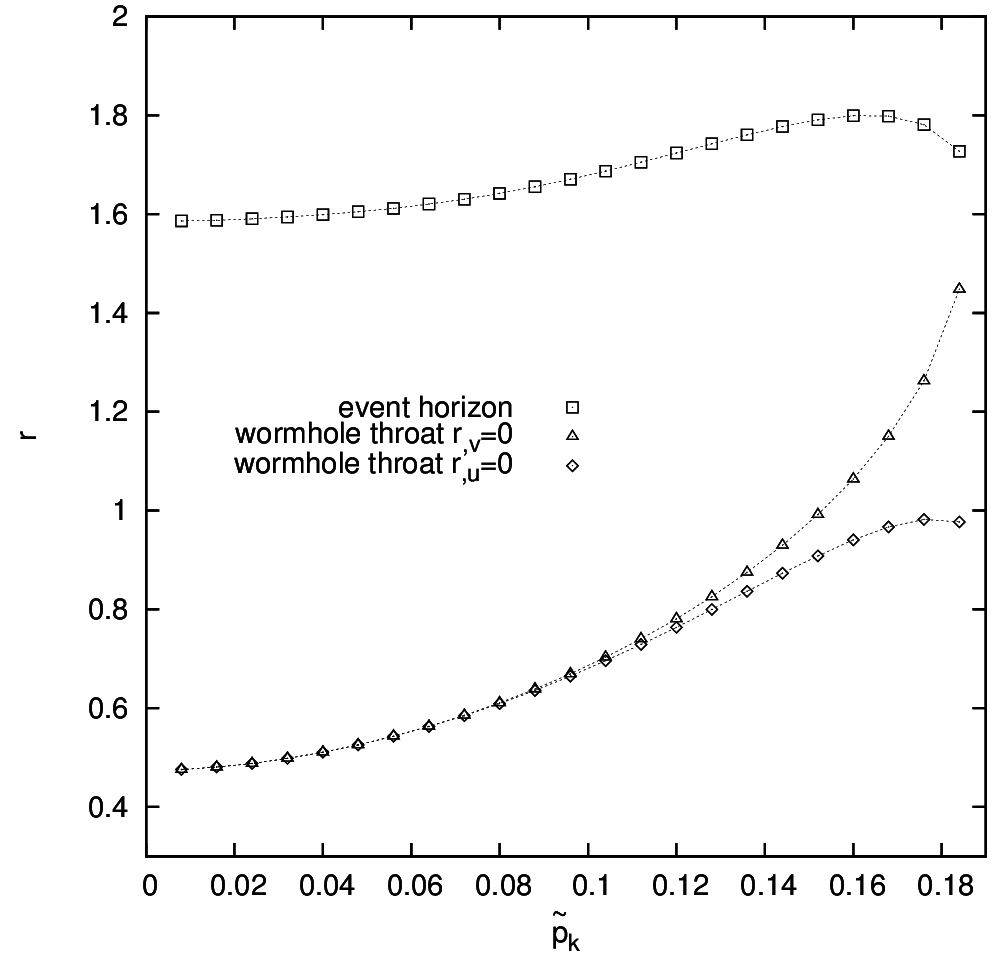}\label{fig:radiia}}
\hfill
\subfigure[][]{\includegraphics[width=0.45\textwidth]{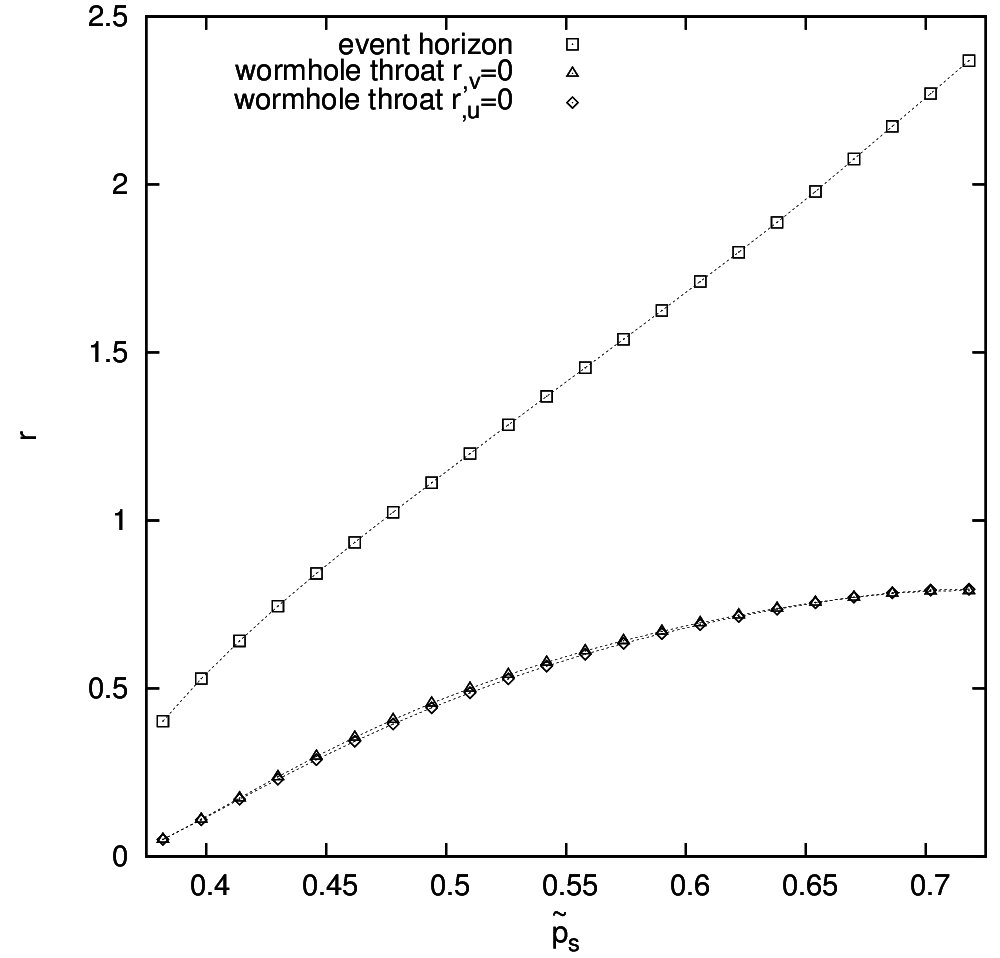}\label{fig:radiib}}
\caption{The~radii of~event horizons and~wormhole throats at~the~final value of~the~$v$-coordinate as~functions of~(a)~the~phantom scalar field amplitude~$\ak$ and~(b)~the~electrically charged scalar field amplitude~$\as$ for~the~$SF$--$DE$ collapse. The~constant free family parameters for~the~particular evolutions were chosen to~be equal to~$\as=0.6$ and~$\ak=0.1$, respectively.}
\label{fig:radii}
\end{figure}

In~figure~\ref{fig:radii} the~radii of~event horizons and~wormhole throats at~the~final value of~the~$v$-coordinate were plotted as~functions of~the~altering field amplitudes. The~chosen value of~advanced time $v_f$ corresponds to~the~well-established stationary phase of~the~evolution due to~the~fact that in~all considered cases the~apparent horizon $r\POv=0$ is~a~null hypersurface there. The~common feature of~both sets of~solutions is~that the~radii of~event horizons as~well as~wormhole throats increase when the~varying amplitude increases. The~radii of~event horizons for~the~biggest values of~$\ak$ are the~only exception (figure~\ref{fig:radiia}). In~this region two branches of~the~line $r\POv=0$ approach each other, which for~values of~$\ak$ exceeding $0.185$ results in~their connection and~naked singularity formation. Such a~tendency is~not visible for~the~other set of~solutions with the~varying amplitude~$\as$ (figure~\ref{fig:radiib}), hence a~naked singularity does not~form. The~changes of~radii of~event horizons are considerably more noticeable in~the~case of~a~varying amplitude of~the~electrically charged scalar field. On~the~contrary, the~radii of~wormhole throats vary more significantly when the~phantom scalar field amplitude alters. The~radii of~both wormhole throats are identical for~all values of~the~varying~$\as$. For~altering~$\ak$ they are equal up to~some point and~than they begin to~differ more and~more significantly.

\begin{figure}[tbp]
\subfigure[][]{\includegraphics[width=0.45\textwidth]{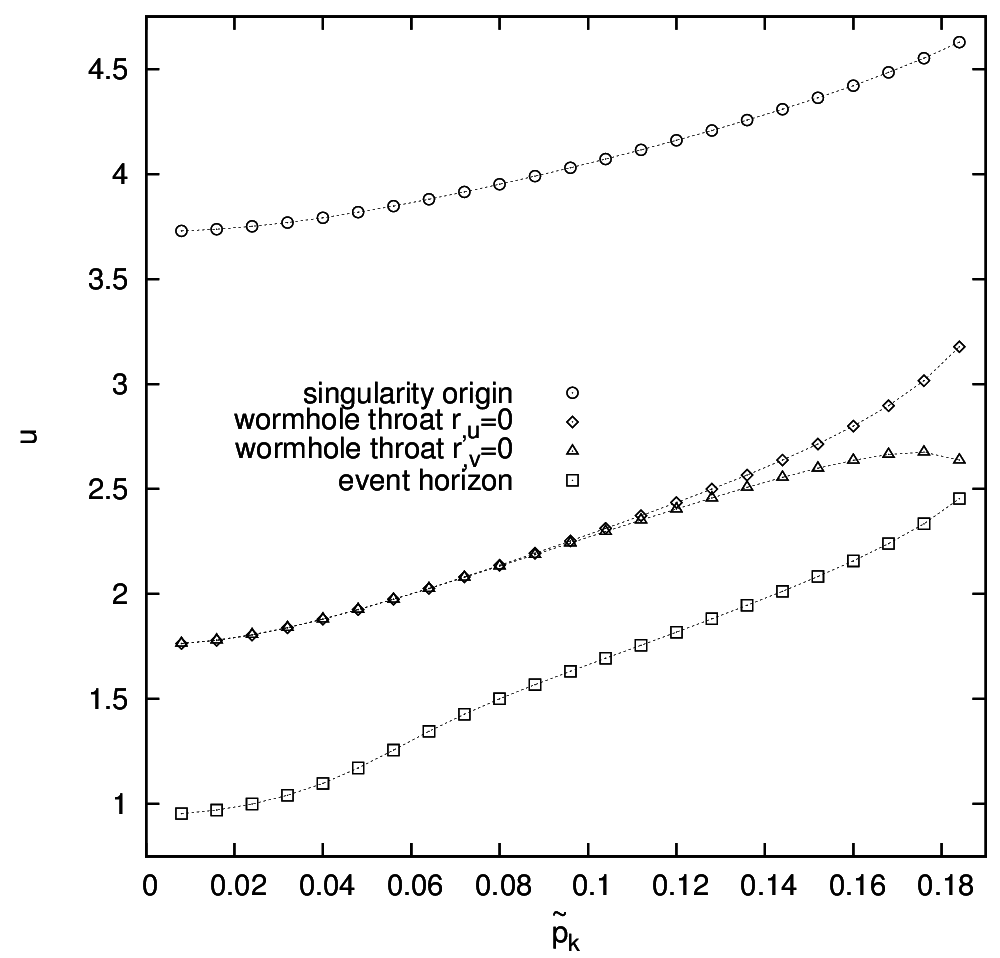}\label{fig:$u$-locationsa}}
\hfill
\subfigure[][]{\includegraphics[width=0.45\textwidth]{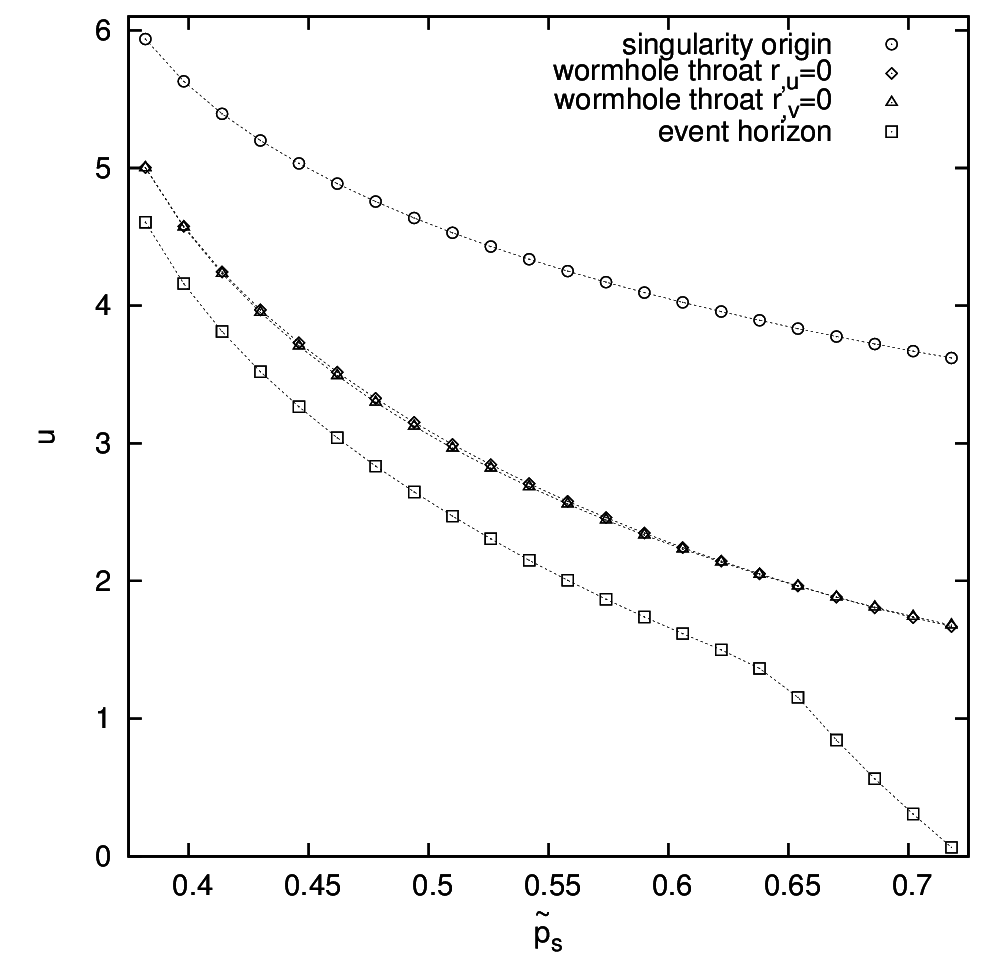}\label{fig:$u$-locationsb}}
\caption{The~$u$-locations of~event horizons, wormhole throats at~the~final value of~the~$v$-coordinate and~the~points, where the~line $r=0$ becomes singular, as~functions of~(a)~$\ak$ and~(b)~$\as$ for~the~$SF$--$DE$ collapse. The~respective constant free family parameters are the~same as~in~figure~\ref{fig:radii}.}
\label{fig:$u$-locations}
\end{figure}

Figure~\ref{fig:$u$-locations} depicts the~$u$-locations of~event horizons, wormhole throats at~the~final value of~advanced time and~the~points, where the~line $r=0$ becomes singular (singularity origins) for~both sets of~solutions. When the~phantom scalar field amplitude increases, the~$u$-locations of~all the~mentioned features increase (figure~\ref{fig:$u$-locationsa}), while in~the~other case the~opposite tendency is~observed (figure~\ref{fig:$u$-locationsb}). Similarly to~the~radii, the~$u$-locations of~wormhole throats are exactly the~same in~the~case of~altering $\as$ and~they bifurcate at~some point when~$\ak$ increases. It~is~connected with the~forthcoming naked singularity formation. The~changes in~$u$-locations are considerably bigger when the~phantom scalar field amplitude is~constant and~the~amplitude of~the~electrically charged scalar field varies, in~comparison to~the~other set of~solutions.

\begin{figure}[tbp]
\subfigure[][]{\includegraphics[width=0.475\textwidth]{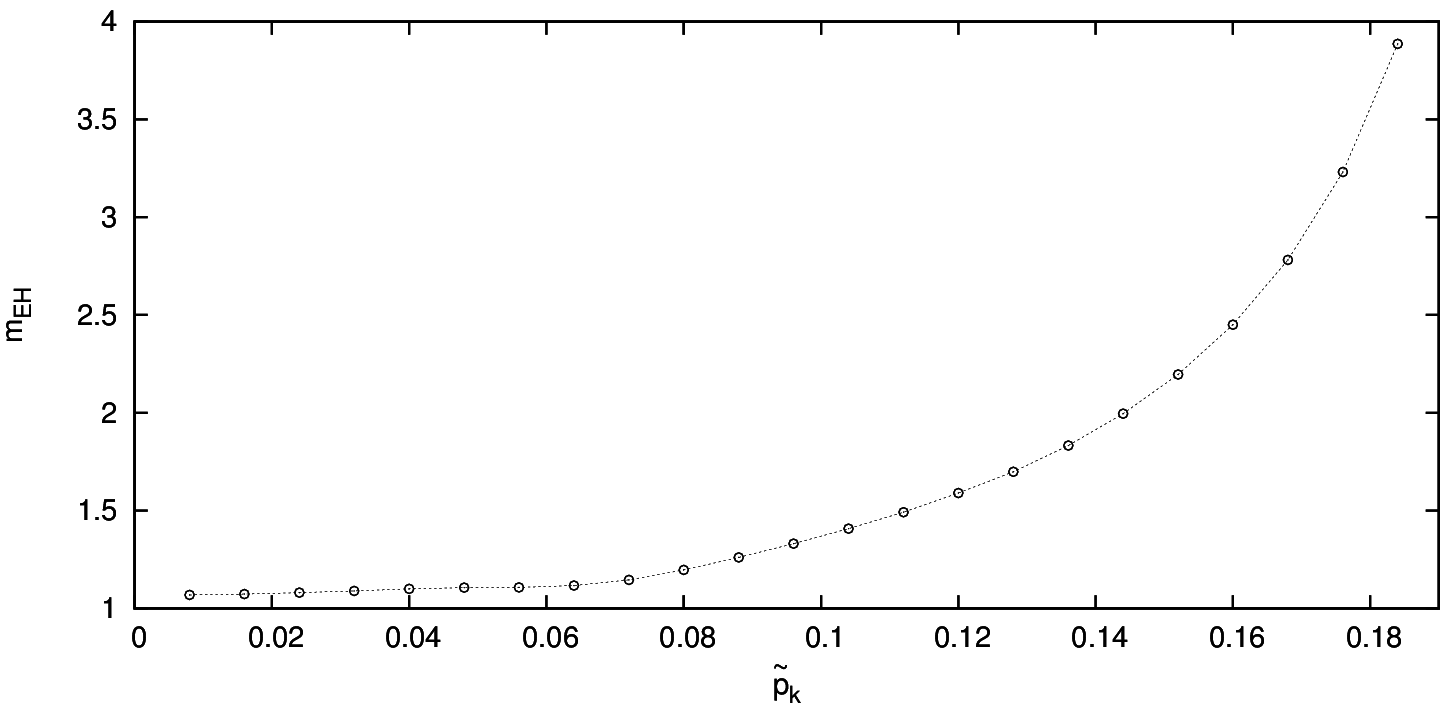}\label{fig:massesa}}
\hfill
\subfigure[][]{\includegraphics[width=0.475\textwidth]{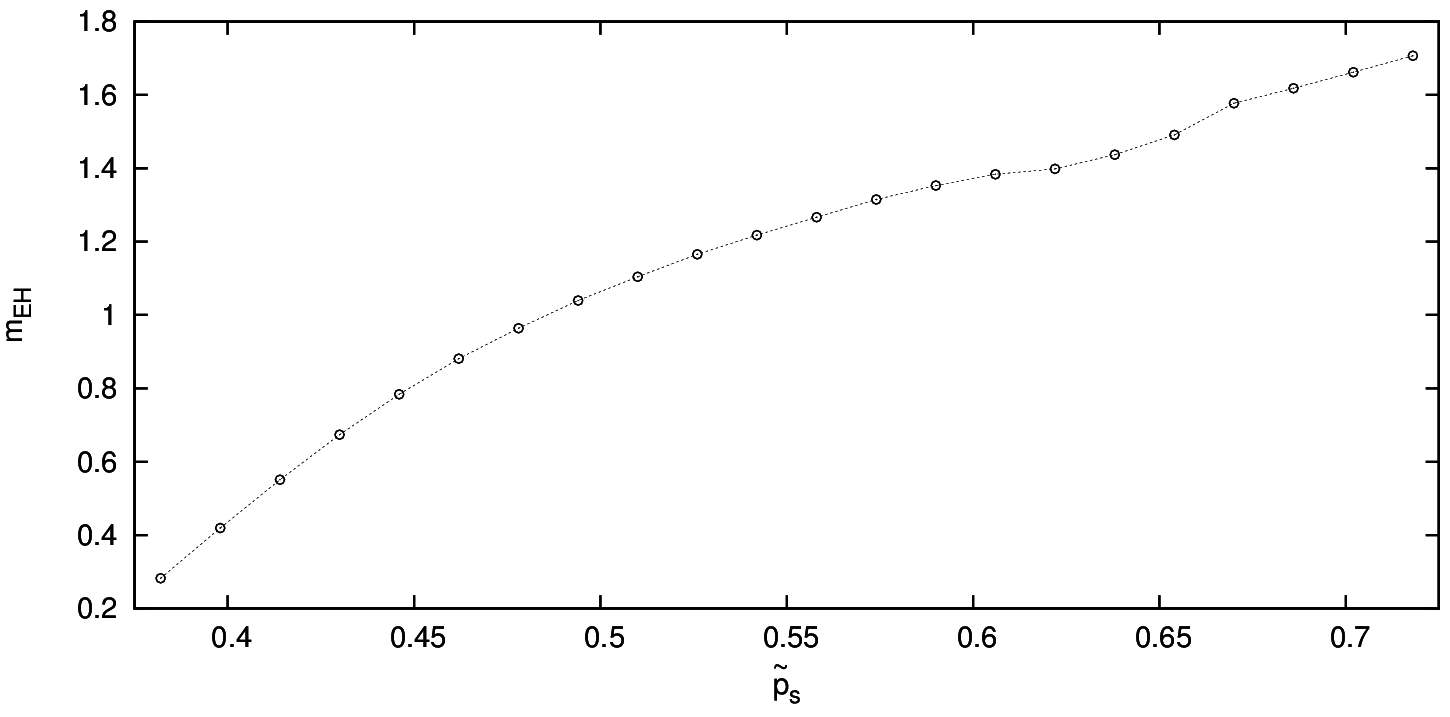}\label{fig:massesb}}
\caption{The~wormhole masses, $m_{EH}$, as~functions of~(a)~$\ak$ and~(b)~$\as$ for~the~$SF$--$DE$ collapse. The~respective constant free family parameters are the~same as~in~figure~\ref{fig:radii}.}
\label{fig:masses}
\end{figure}

The~masses of~wormholes versus the~phantom scalar field and~electrically charged scalar field amplitudes are presented in~figure~\ref{fig:masses}. In~both cases the~mass increases when the~amplitude raises.

\section{Influence of~dark matter on~electrically charged scalar field evolution}
\label{sec:dm}

During examining the~role of~dark matter in~the~studied collapse the~values of~parameters, which characterize the~considered dark matter model, i.e.,~$\alpha_{DM}$, $m^2$ and~$\lambda_{DM}$, were varied within the~permissible ranges outlined in~section~\ref{sec:model}. It~turned out that $\alpha_{DM}$ and~$\lambda_{DM}$ do~not~play a~significant role during the~process. The~latter result is~an~extension of~the~previous attempt to~include the~$\chi^4$-type interactions in~the~gravitational collapse mentioned in~\cite{OkawaCardosoPani2014-041502}, which was made for~the~evolution of~a~real scalar field without electric charge. The~results of~the~conducted simulations will be presented for~$\alpha_{DM}=10^{-3}$ and~$\lambda_{DM}=0.1$. The~course of~the~$SF$--$DM$ evolution was also interpreted regarding the~strength of~a~gravitational self-interaction of~the~evolving fields by~changing the~values of~their initial amplitudes $\ah$ and~$\as$.

\subsection{Spacetime structures}
\label{sec:DMstructures}

\begin{figure}[tbp]
\subfigure[][]{\includegraphics[width=0.24\textwidth]{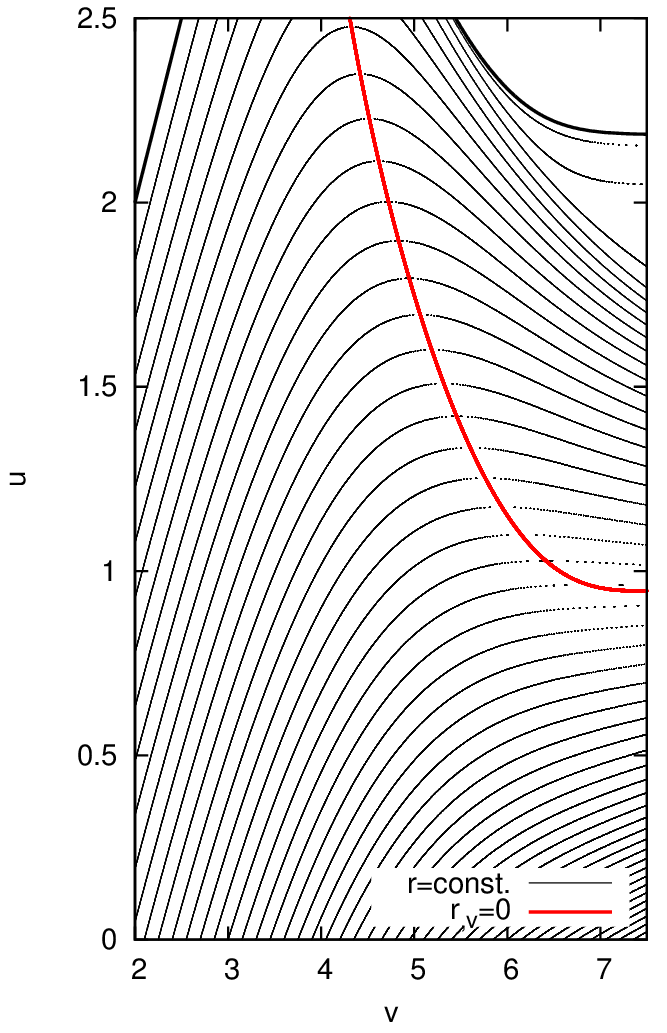}\label{fig:10a}}
\hfill
\subfigure[][]{\includegraphics[width=0.24\textwidth]{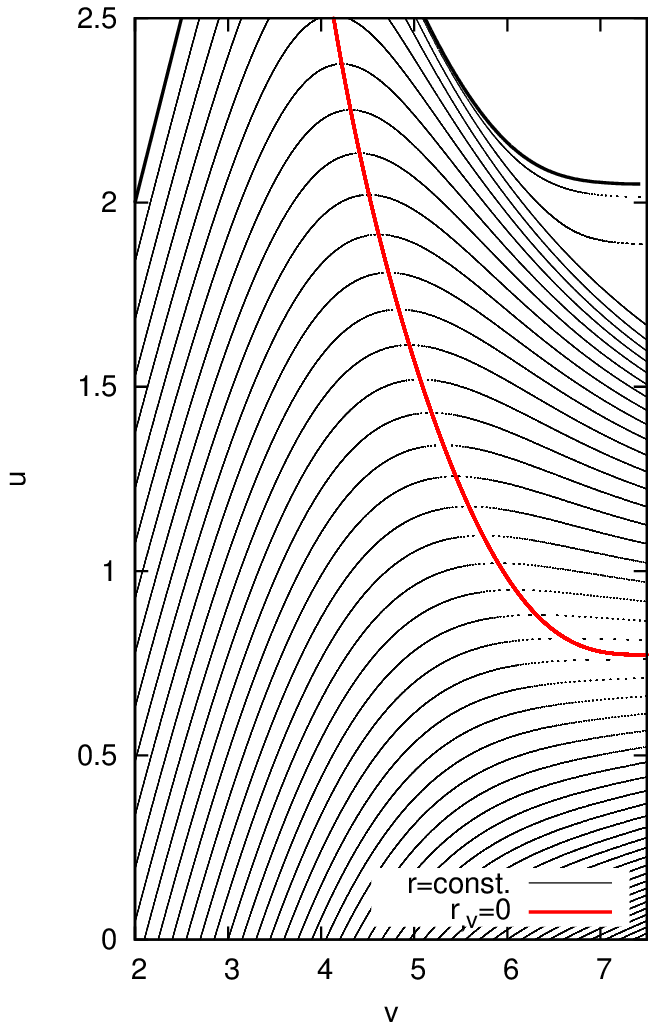}}
\hfill
\subfigure[][]{\includegraphics[width=0.24\textwidth]{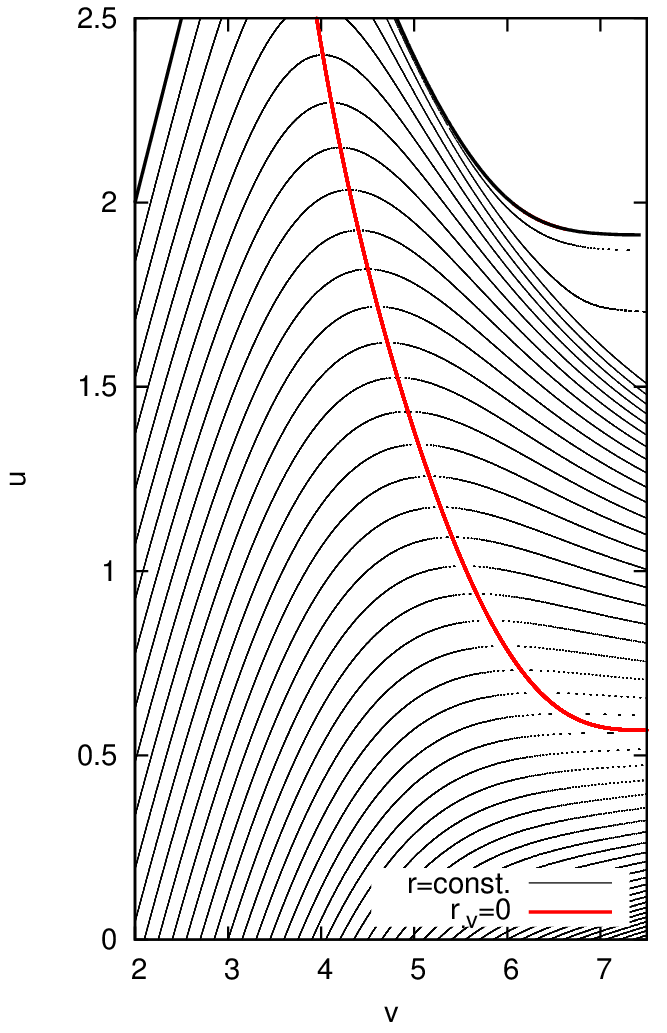}\label{fig:10c}}
\hfill
\subfigure[][]{\includegraphics[width=0.24\textwidth]{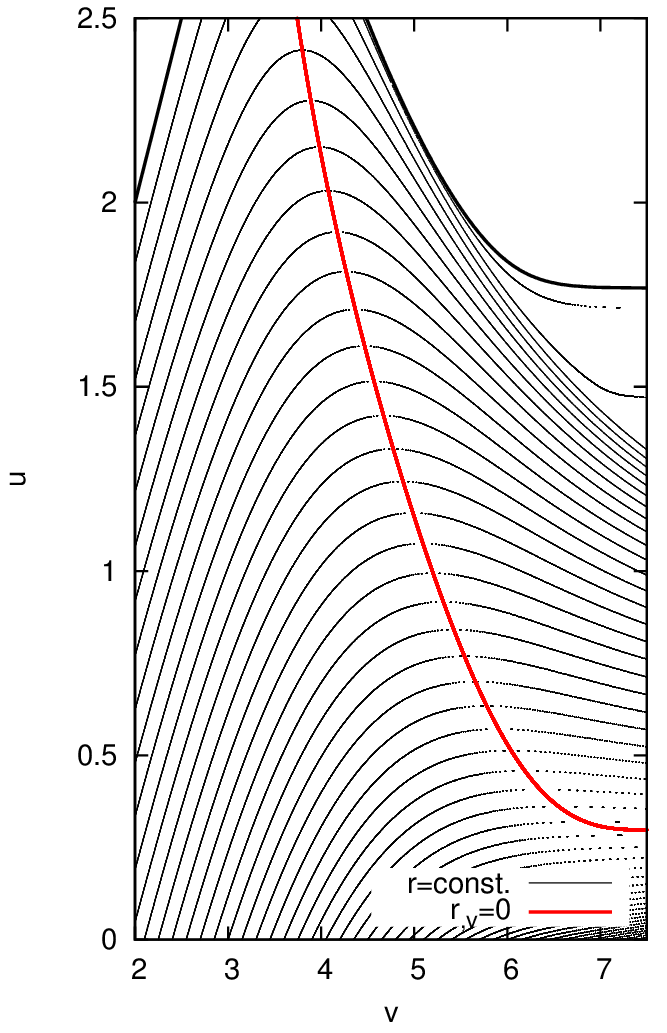}}
\caption{(color online) Penrose diagrams of~spacetimes formed during the~collapse of~the~electrically charged scalar field of~the~amplitude $\as=0.6$ in~the~presence of~dark matter with $m^2=0.1$ and~$\ah$ equal to~(a)~$10^{-3}$, (b)~$0.1$, (c)~$0.15$ and~(d)~$0.2$.}
\label{fig:SFDM-06varAh}
\end{figure}

\begin{figure}[tbp]
\begin{minipage}{0.5\textwidth}
\subfigure[][]{\includegraphics[width=0.48\textwidth]{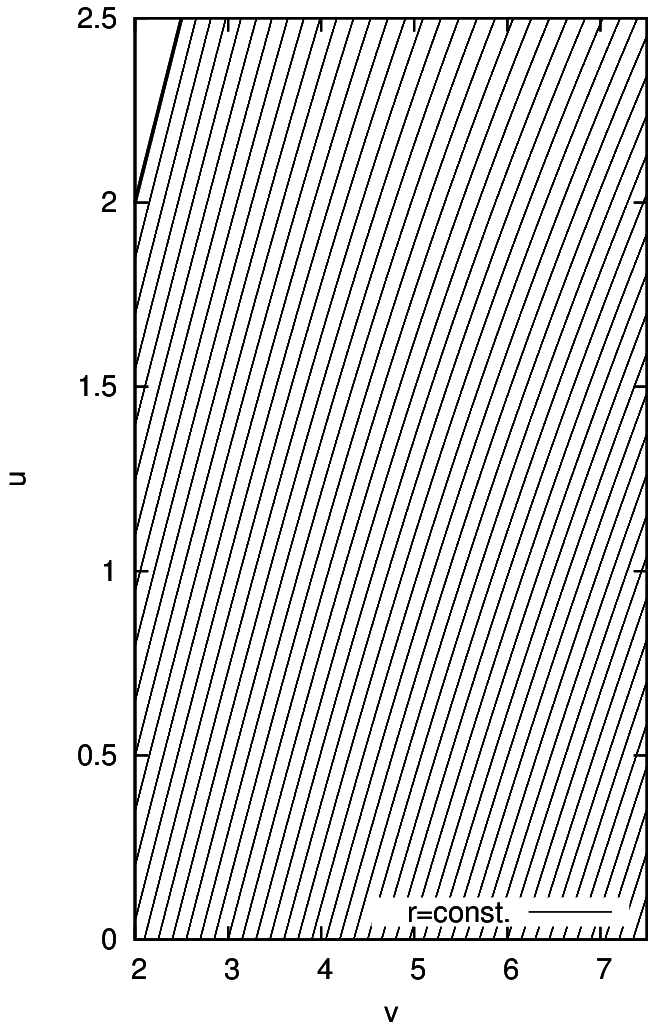}}
\hfill
\subfigure[][]{\includegraphics[width=0.48\textwidth]{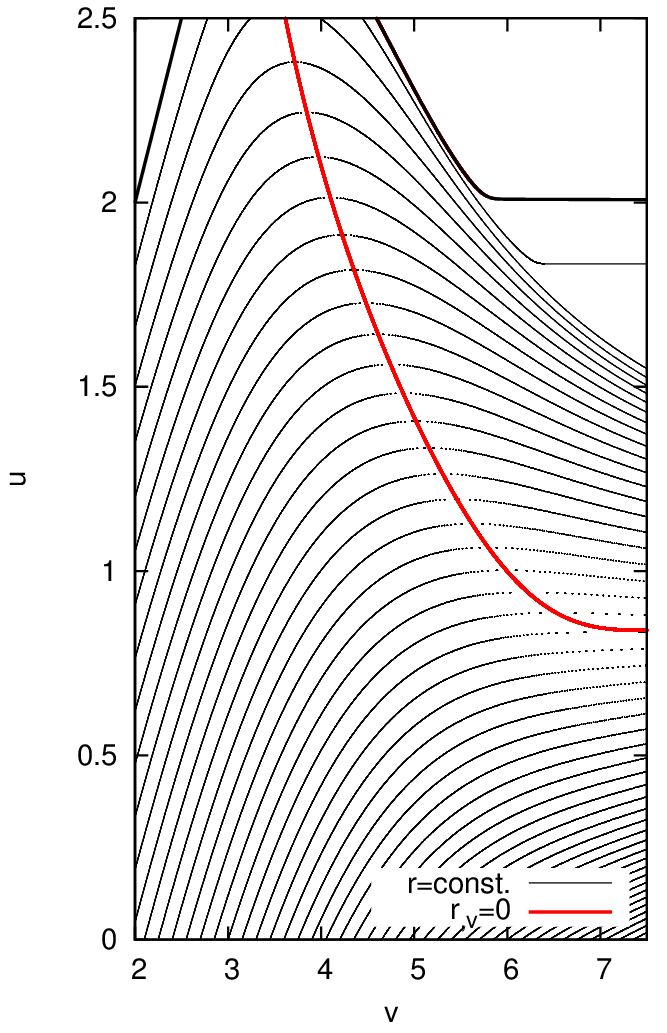}}
\end{minipage}
\hfill
\begin{minipage}{0.475\textwidth}
\caption{(color online) Penrose diagrams of~spacetimes formed during the~$SF$--$DM$ evolution with $\as=0.2$, $m^2=0.1$ and~$\ah$ equal to~(a)~$0.1$ and~(b)~$0.4$.}
\label{fig:SFDM-02varAh}
\end{minipage}
\end{figure}

The~Penrose diagrams of~spacetimes formed during the~$SF$--$DM$ collapse with a~varying amplitude $\ah$ are presented in~figures~\ref{fig:SFDM-06varAh} and~\ref{fig:SFDM-02varAh} for~$m^2=0.1$ and~the~electrically charged scalar field amplitude $\as$ equal to~$0.6$ and~$0.2$, respectively. The~structures of~spacetimes, which stem from the~process in~the~absence of~dark energy are shown in~figure~\ref{fig:sf}. The~former is~the~dynamical Reissner-Nordstr\"{o}m spacetime and~the~latter is~non-singular.

In~the~case of~$\as=0.6$, all the~resulting spacetimes are dynamical Reissner-Nordstr\"{o}m spacetimes for~all initial amplitudes of~the~complex field, which constitutes the~dark matter model. The~central spacelike singularity at~$r=0$ is~surrounded within each of~them by~a~single apparent horizon situated along the~$r\POv=0$ line, which settles down along an~event horizon as~$v\to\infty$. The~Cauchy horizon is~located at~future null infinity~$v~=~\infty$. An~increasing value of~$\ah$ results in~an~earlier formation of~a~black hole in~terms of~retarded time. When $\as$ equals $0.2$, the~obtained spacetime is~non-singular for~values of~$\ah$ not~exceeding $0.135$ or~contains a~dynamical Reissner-Nordstr\"{o}m black hole for~its larger values. This result is~particularly interesting due to~the~role played by~dark matter in~this case. Although the~self-interaction of~an~electrically charged scalar field is~too weak to~form a~black hole (figure~\ref{fig:sf-02}), the~presence of~dark matter enables the~emergence of~a~singular spacetime. The~appearance of~the~singularity is~clearly related to~the~dark matter action during the~collapse, but the~existence of~a~Cauchy horizon is~strictly connected with the~charge associated with the~scalar field coupled with the~Maxwell field.

\begin{figure}[tbp]
\subfigure[][]{\includegraphics[width=0.24\textwidth]{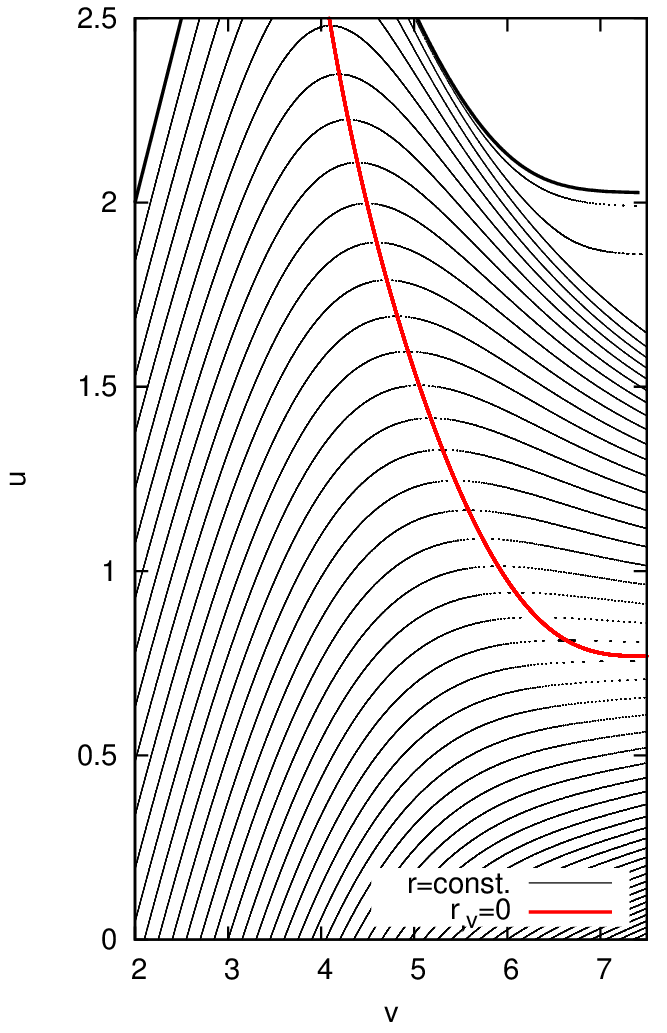}\label{fig:11a}}
\hfill
\subfigure[][]{\includegraphics[width=0.24\textwidth]{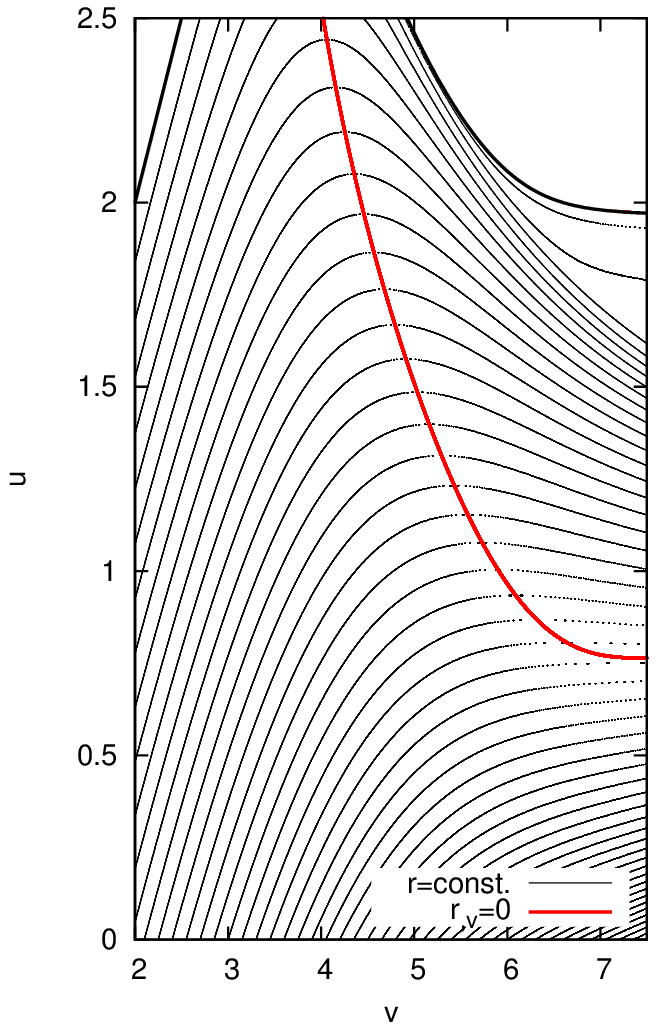}}
\hfill
\subfigure[][]{\includegraphics[width=0.24\textwidth]{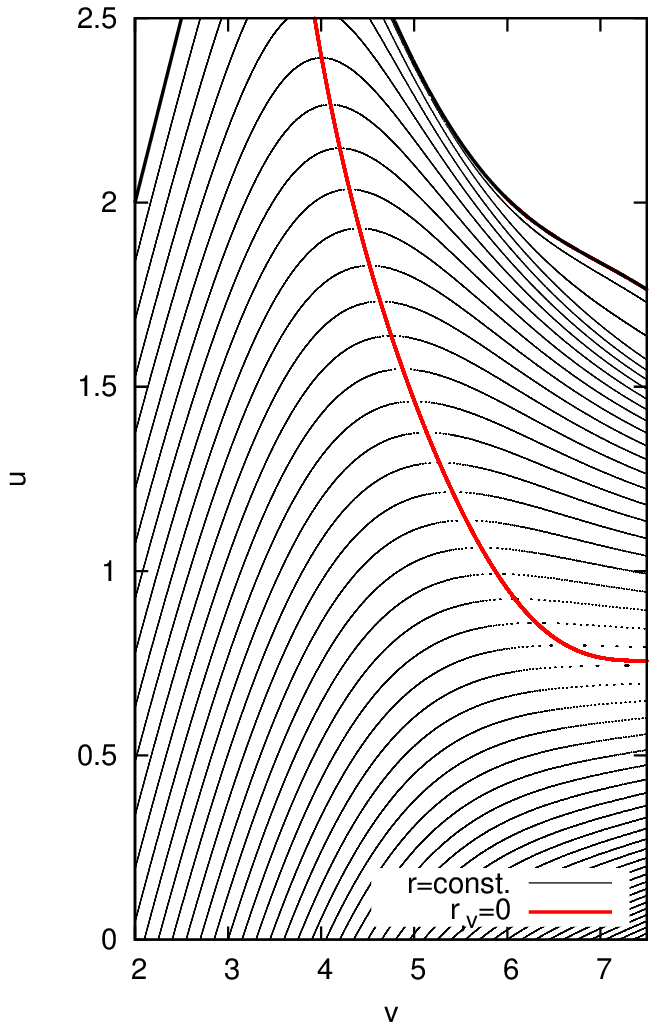}}
\hfill
\subfigure[][]{\includegraphics[width=0.24\textwidth]{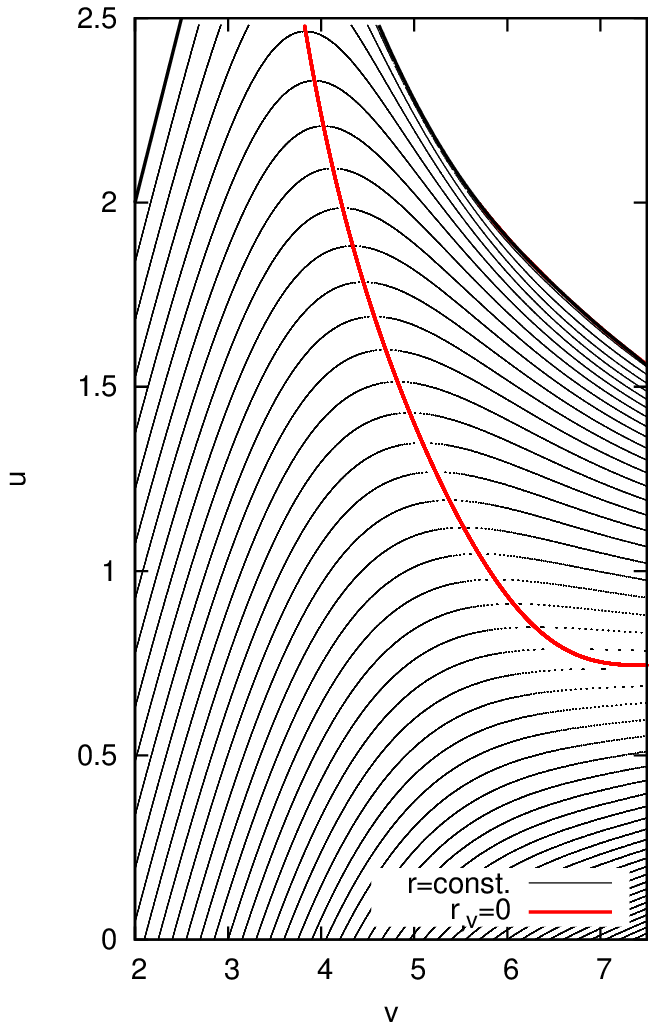}\label{fig:11d}}
\caption{(color online) Penrose diagrams of~spacetimes formed during the~$SF$--$DM$ collapse with $\as=0.6$, $\ah=0.1$ and~$m^2$ equal to~(a)~$-1$, (b)~$-2$, (c)~$-3$ and~(d)~$-4$.}
\label{fig:SFDM-06varm2}
\end{figure}

The~spacetimes formed during the~discussed collapse for~$\as=0.6$, $\ah=0.1$ and~several negative values of~$m^2$ are shown in~figure~\ref{fig:SFDM-06varm2}. For~$m^2$ larger than $-3.85$ the~dynamical Reissner-Nordstr\"{o}m spacetimes form, while for~its smaller values the~Cauchy horizon is~absent in~the~spacetime and~hence dynamical Schwarzschild black holes stem from the~evolution. In~the~latter case, the~central $r=0$ spacelike singularity is~surrounded by~an~apparent horizon $r\POv=0$, which coincides with an~event horizon at~$v\to\infty$. The~Cauchy horizon does not~exist, because the~$r=const.$ lines do~not~settle along $u=const.$ hypersurfaces as~$v\to\infty$. For~$m^2>0$ all the~spacetimes are of~the~Reissner-Nordstr\"{o}m type. The~non-zero vev of~the~dark matter complex scalar field, i.e.,~the~existence of~dark matter in~the~form of~a~massive gauge boson (called $Z^\prime$ or~dark photon) during the~course of~the~studied process for~$m^2<-3.85$, results in~the~disappearance of~a~Cauchy horizon and~favors the~formation of~a~simpler Schwarzschild-type spacetime.

\subsection{Fields behavior in~spacetimes}

\begin{figure}[tbp]
\subfigure[][]{\includegraphics[width=0.475\textwidth]{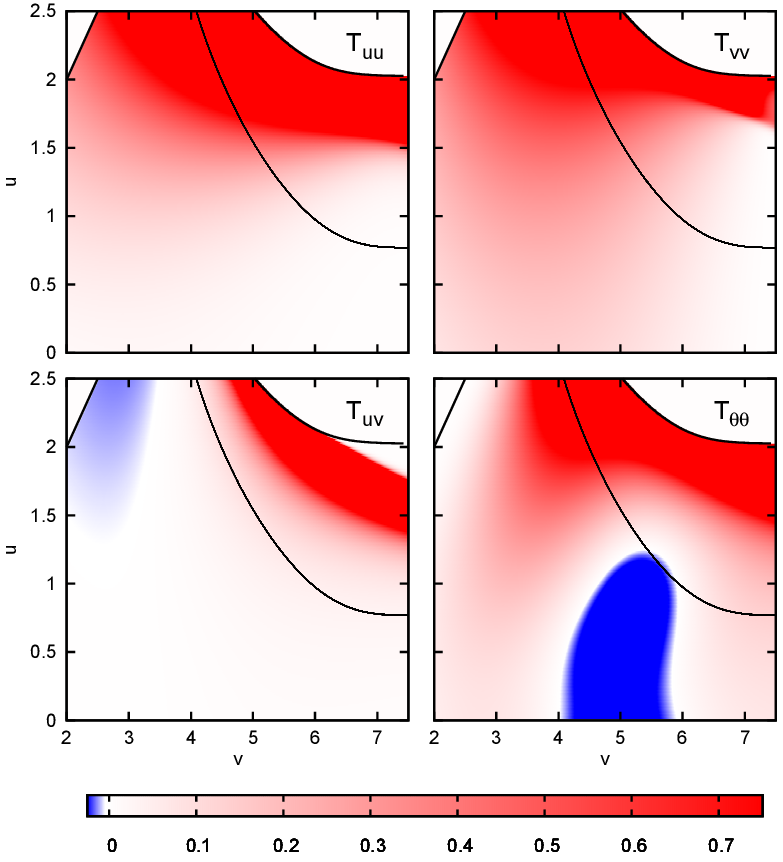}}
\hfill
\subfigure[][]{\includegraphics[width=0.475\textwidth]{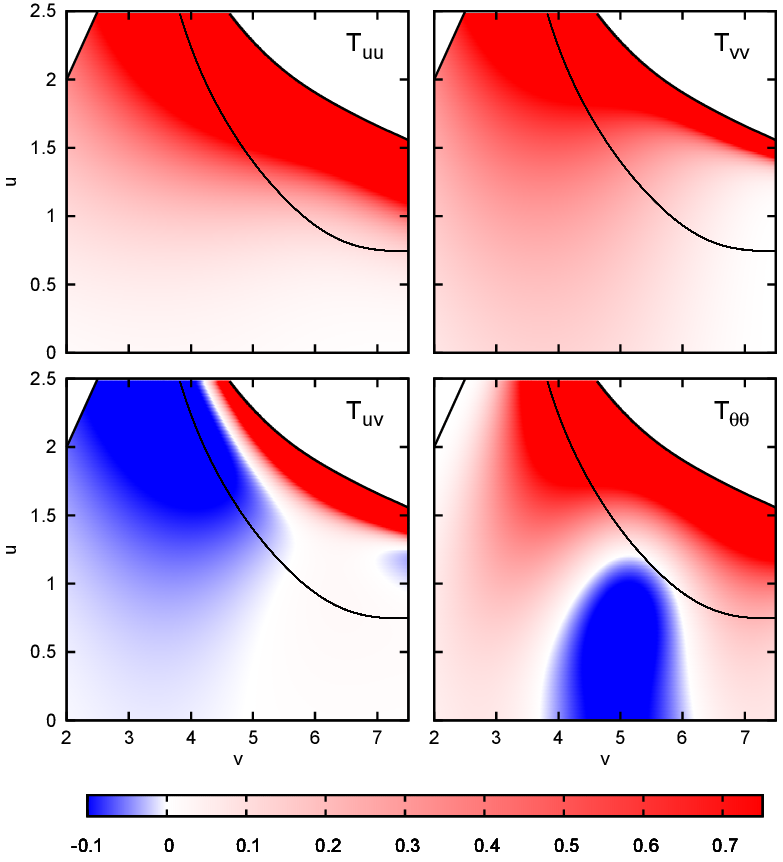}}
\caption{(color online) The~stress-energy tensor components \eqref{eqn:Tuu}--\eqref{eqn:Ttt} in~the~$(vu)$-plane for~$SF$--$DM$ evolutions with $\as=0.6$, $\ah=0.1$ and~$m^2$ equal to~(a)~$-1$ and~(b)~$-4$.}
\label{fig:SFDM-ten}
\end{figure}

\begin{figure}[tbp]
\begin{minipage}{0.475\textwidth}
\subfigure[][]{\includegraphics[width=0.48\textwidth]{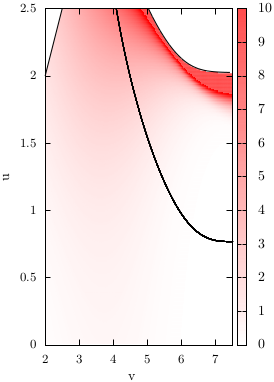}}
\hfill
\subfigure[][]{\includegraphics[width=0.48\textwidth]{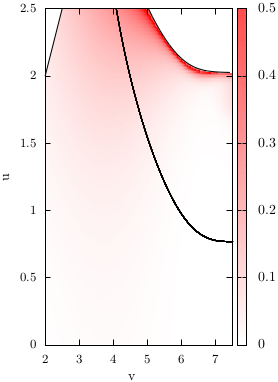}}
\caption{(color online) The~$\left(vu\right)$-dis\-tri\-bu\-tion of~the~moduli of~(a)~the~electrically charged scalar field, $|\psi|$, and~(b)~the~dark matter complex scalar field, $|\chi|$, for~the~$SF$--$DM$ evolution with $\as=0.6$, $\ah=0.1$ and~$m^2=-1$.}
\label{fig:SFDM-dis-1}
\end{minipage}
\hfill
\begin{minipage}{0.475\textwidth}
\subfigure[][]{\includegraphics[width=0.4625\textwidth]{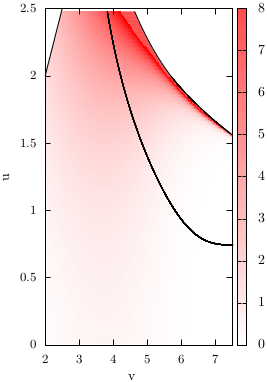}}
\hfill
\subfigure[][]{\includegraphics[width=0.48\textwidth]{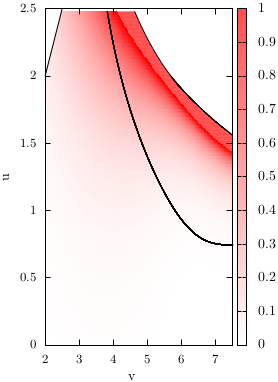}}
\caption{(color online) The~$\left(vu\right)$-dis\-tri\-bu\-tion of~(a)~$|\psi|$ and~(b)~$|\chi|$ for~the~$SF$--$DM$ evolution with $\as=0.6$, $\ah=0.1$ and~$m^2=-4$.}
\label{fig:SFDM-dis-4}
\end{minipage}
\end{figure}

The~components of~the~stress-energy tensor \eqref{ten} calculated for~the~spacetimes formed during the~$SF$--$DM$ collapse characterized by~the~amplitudes $\as=0.6$, $\ah=0.1$ and~$m^2$ equal to~$-1$ and~$-4$ are presented in~figure~\ref{fig:SFDM-ten} (the~respective spacetime structures were presented in~figures~\ref{fig:11a} and~\ref{fig:11d}). The~$\left(vu\right)$-distributions of~the~moduli of~the~electrically charged scalar field and~the~dark matter complex scalar field for~the~same evolutions are shown in~figures~\ref{fig:SFDM-dis-1} and~\ref{fig:SFDM-dis-4}.

Both $T_{uu}$ and~$T_{vv}$ components of~the~stress-energy tensors are positive nearby the~singular $r=0$ lines in~the~presented spacetimes. Hence, the~null energy condition \eqref{eqn:nec} is~fulfilled in~the~vicinity of~the~central singularities. They do~not~bifurcate into wormhole throats and~black holes form during the~$SF$--$DM$ evolution. Similar conclusions stem from the~analysis of~the~stress-energy tensor components for~singular spacetimes presented in~figures~\ref{fig:SFDM-06varAh} and~\ref{fig:SFDM-02varAh}.

An~inspection of~the~moduli of~the~field functions reveals that the~type of~the~forming spacetime depends on~their~values in~the~vicinity of~$r=0$ as~$v\to\infty$. A~significant value of~the~modulus of~an electrically charged scalar field in~this region, three orders of~magnitude bigger than in~the~remaining spacetime, accompanied by~a~small value of~the~other field modulus, results in~the~formation of~a~dynamical Reissner-Nordstr\"{o}m spacetime. It~is~connected with the~repulsive character of~the~scalar field due to~the~electric charge associated with~it. On~the~other hand, when the~modulus of~the~dark matter complex scalar field is~bigger at~$v\to\infty$ nearby $r=0$ (two orders of~magnitude in~relation to~the~rest of~the~spacetime in~the~considered case), the~importance of~the~other field is~reduced and~the~dynamical Schwarzschild spacetime emerges. Such behavior of~dark matter during the~gravitational collapse is~consistent with its action on~a~cosmological scale, where it~tends to~concentrate in~confined areas due to~the~attractive character of~its gravitational interaction. It~is~worth noting, that in~the~regime of~the~selected dark matter model, the~described behavior is~related to~the~change in~potential (the~value of~$m^2$), not~in~the~gravitational self-interaction strength of~the~complex field (the~amplitude~$\ah$).

\subsection{Characteristics of~the~formed black holes}

\begin{figure}[tbp]
\subfigure[][]{\includegraphics[width=0.5\textwidth]{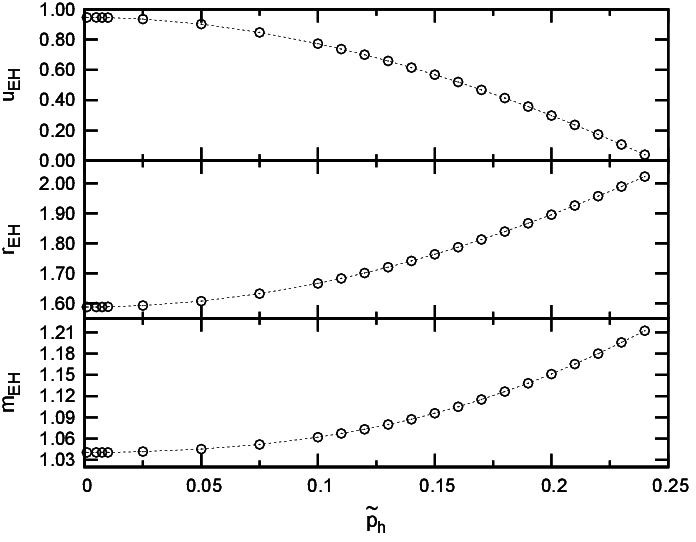}\label{fig:DMchar-ph}}
\hfill
\subfigure[][]{\includegraphics[width=0.49\textwidth]{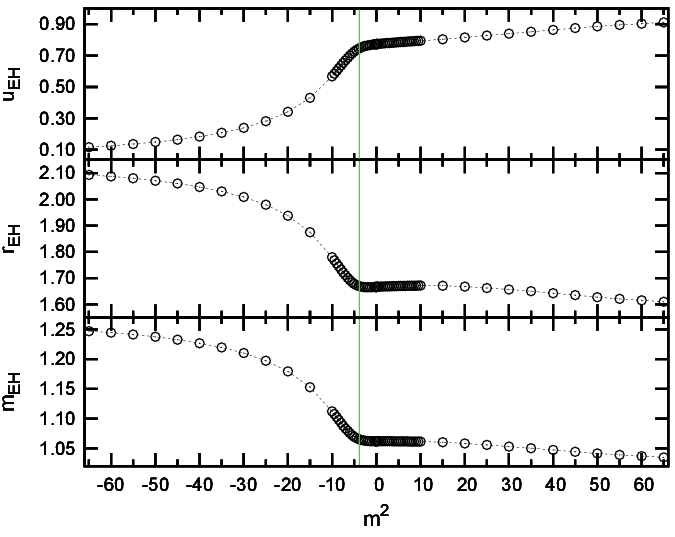}\label{fig:DMchar-m2}}
\caption{(color online) The~$u$-locations of~the~event horizons, $u_{EH}$, the~masses, $m_{EH}$, and~the~radii, $r_{EH}$, of~black holes formed during the~$SF$--$DM$ collapse as~functions of~(a)~$\ah$ with $m^2=0.1$ and~(b)~$m^2$ with $\ah=0.1$. The~green line indicates the~border between spacetimes of~the~Schwarzschild ($m^2<-3.85$) and~Reissner-Nordstr\"{o}m ($m^2\geqslant -3.85$) types.}
\label{fig:DMchar}
\end{figure}

The~dependence of~the~event horizon~$u$-locations, the~masses and~the~radii of~black holes formed during the~collapse of~an electrically charged scalar field accompanied by~dark matter on~$\ah$ and~$m^2$ is~depicted in~figure~\ref{fig:DMchar}. The~maximal value of~the~initial amplitude of~the~dark matter scalar field was selected so that for~all investigated cases the~collapse begins outside of~the~event horizon.

The~black holes form earlier in~terms of~retarded time and~their masses and~radii increase as~$\ah$ gets bigger (figure~\ref{fig:DMchar-ph}). When $m^2$ increases the~moment of~the~event horizon formation appears later in~terms of~$u$ and~the~black hole masses and~radii decrease with an~inflection nearby $m^2=-3.85$, which is~the~point of~the~change between the~Reissner-Nordstr\"{o}m and~Schwarzschild spacetime structures (figure~\ref{fig:DMchar-m2}).

\section{Collective effect of~dark components on~the~considered collapse}
\label{sec:dmde}

Similarly to~the~evolutions described in~section~\ref{sec:dm}, the~collapse of~an electrically charged scalar field with both dark energy and~dark matter will be presented for~the~values of~$\alpha_{DM}$ and~$\lambda_{DM}$ equal to~$10^{-3}$ and~$0.1$, respectively. The~altering of~the~initial amplitudes $\as$, $\ah$ and~$\ak$ is~equivalent to~changing the~strength of~the~particular field gravitational self-interaction within the~examined physical system.

\subsection{Spacetime structures}
\label{sec:DEDMstructures}

\begin{figure}[tbp]
\subfigure[][]{\includegraphics[width=0.24\textwidth]{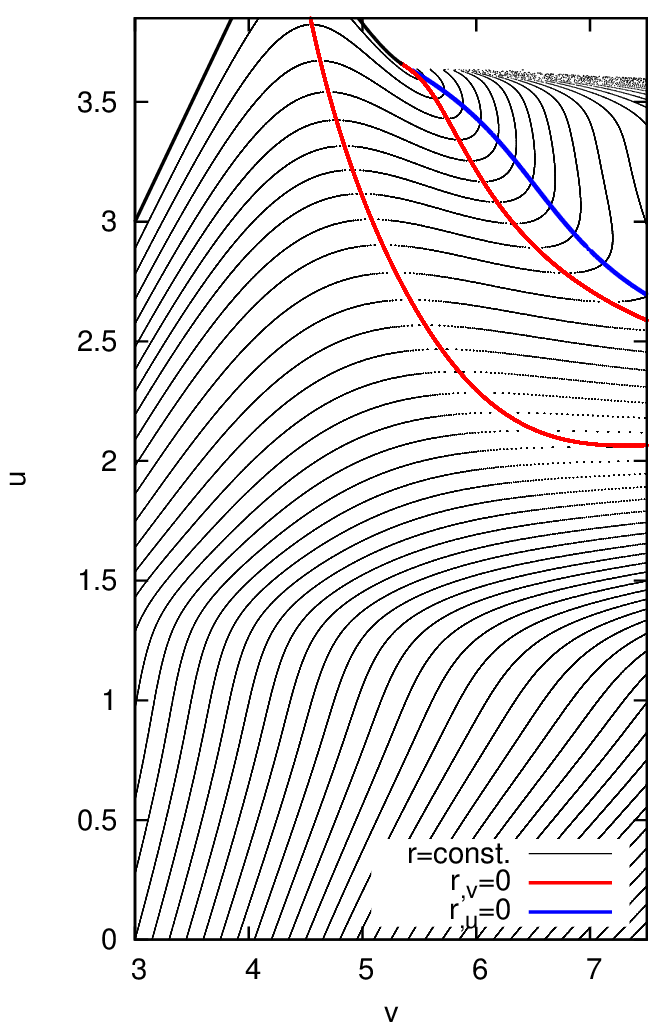}\label{fig:17a}}
\hfill
\subfigure[][]{\includegraphics[width=0.24\textwidth]{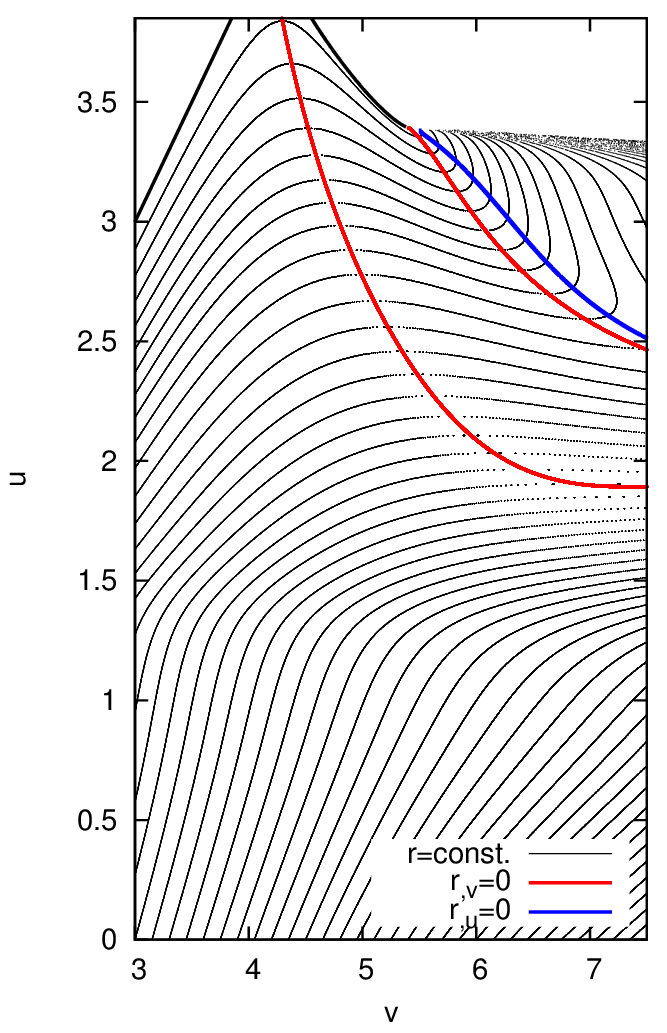}}
\hfill
\subfigure[][]{\includegraphics[width=0.24\textwidth]{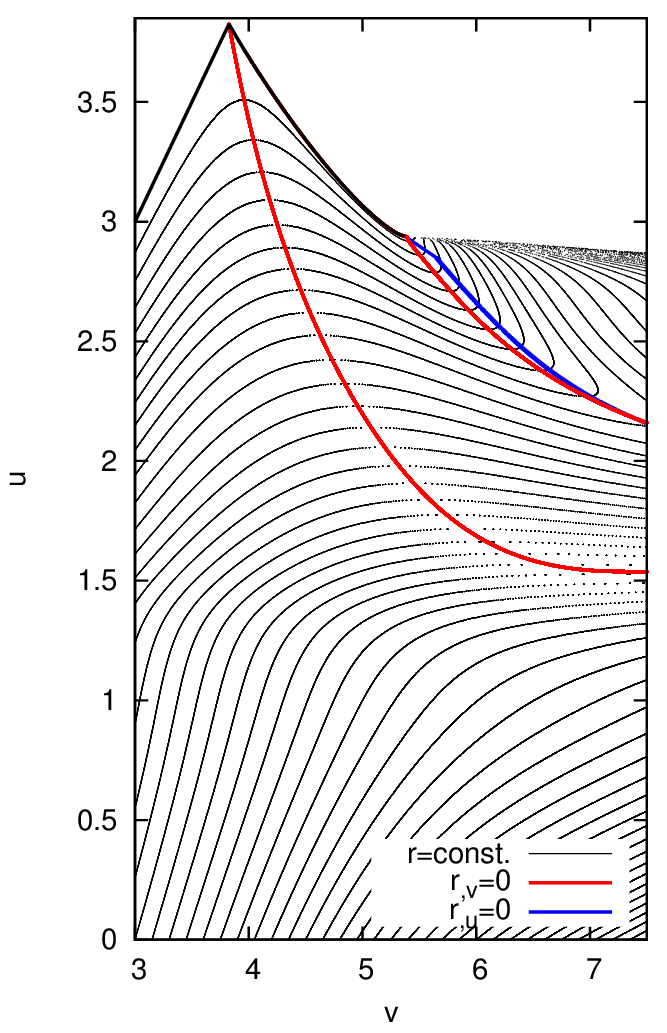}}
\hfill
\subfigure[][]{\includegraphics[width=0.24\textwidth]{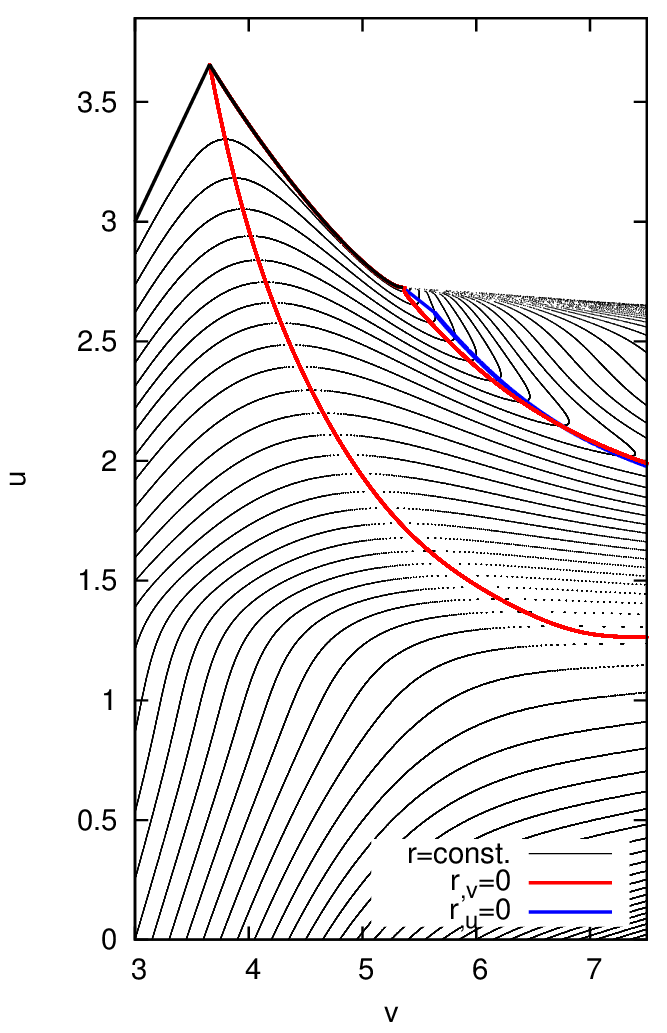}\label{fig:17d}}
\caption{(color online) Penrose diagrams of~spacetimes formed during the~gravitational evolution of~the~electrically charged scalar field of~the~amplitude $\as=0.6$ in~the~presence of~the~phantom scalar field with $\ak=0.15$ and~dark matter, whose model is~described by~the~parameters $m^2=0.1$ and~$\ah$ equal to~(a)~$10^{-3}$, (b)~$0.1$, (c)~$0.2$ and~(d)~$0.25$.}
\label{fig:SFDEDM-varAh}
\end{figure}

The~structures of~spacetimes formed in~the~course of~the~$SF$--$DE$--$DM$ collapse for~$\as=0.6$, $\ak=0.15$, $m^2=0.1$ and~the~varying amplitude of~the~dark matter complex scalar field are presented in~figure~\ref{fig:SFDEDM-varAh}. The~obtained spacetimes contain dynamical wormholes. The~central spacelike $r=0$ singularity bifurcates into two wormhole throats located along the~lines $r\POv=0$ and~$r\POu=0$ as~$v\to\infty$. The~whole is~surrounded by~an~apparent horizon $r\POv=0$, which coincides with the~event horizon in~the~region, where the~spacetime settles down after the~dynamical stage of~the~process. The~value of~$\ah$ does not~influence the~type of~the~emerging spacetime, but it~causes the~event horizon to~form earlier in~terms of~retarded time.

\begin{figure}[tbp]
\subfigure[][]{\includegraphics[width=0.24\textwidth]{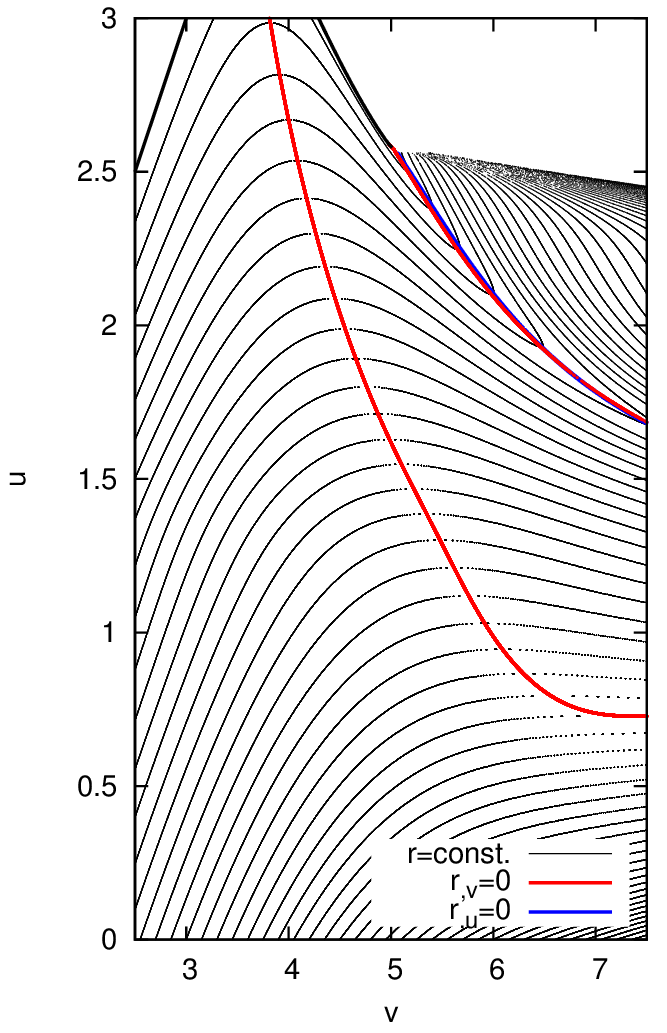}}
\hfill
\subfigure[][]{\includegraphics[width=0.24\textwidth]{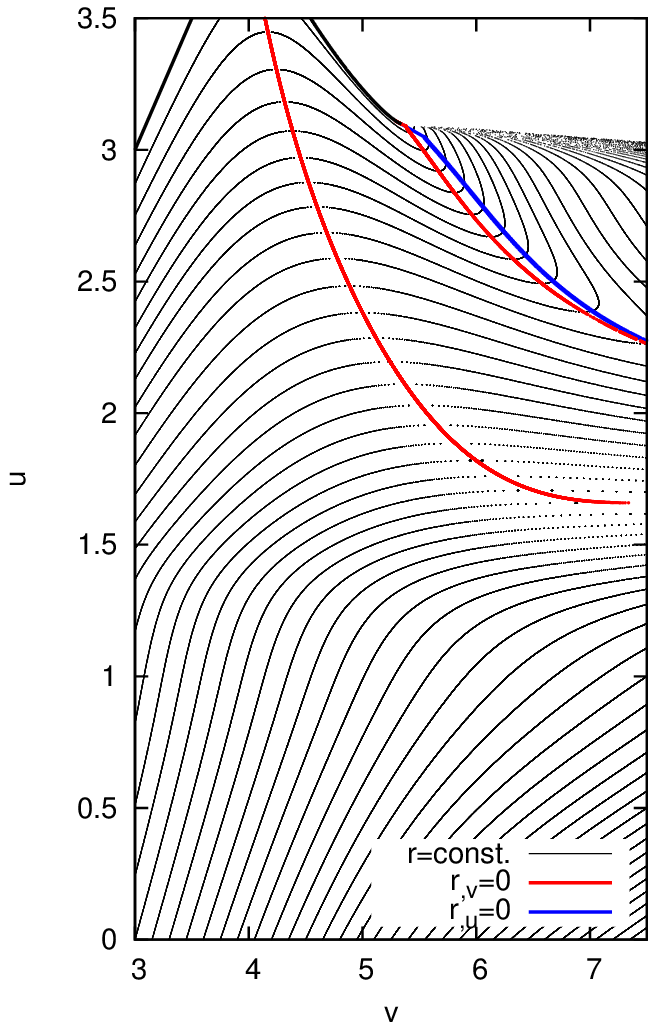}\label{fig:19b}}
\hfill
\subfigure[][]{\includegraphics[width=0.24\textwidth]{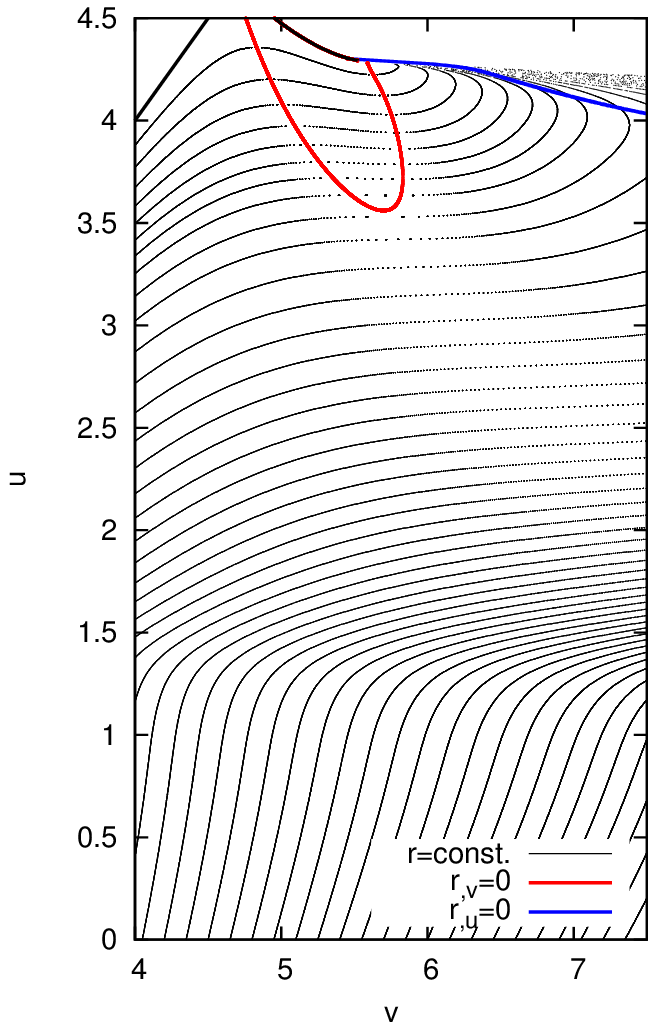}\label{fig:19c}}
\caption{(color online) Penrose diagrams of~spacetimes formed during the~collapse of~the~electrically charged scalar field of~the~amplitude $\as=0.6$ in~the~presence of~dark matter with $\ah=0.15$, $m^2=0.1$ and~the~phantom scalar field with $\ak$ equal to~(a)~$0.05$, (b)~$0.14$ and~(c)~$0.25$.}
\label{fig:SFDEDM-varAk}
\end{figure}

The~Penrose diagrams of~spacetimes which stem from the~studied collapse with a~varying phantom field amplitude with $\as=0.6$, $\ah=0.15$ and~$m^2=0.1$ are shown in~figure~\ref{fig:SFDEDM-varAk}. The~spacetimes contain either wormholes or~naked singularities. The~former emerge for~$\ak$ smaller than $0.23$ and~the~latter for~its larger values. The~$u$-location of~the~event horizon increases with the~value of~the~altering amplitude. The~gravitational self-interaction strength of~the~phantom field decides about the~type of~a~spacetime structure, which forms in~the~process.

\begin{figure}[tbp]
\subfigure[][]{\includegraphics[width=0.24\textwidth]{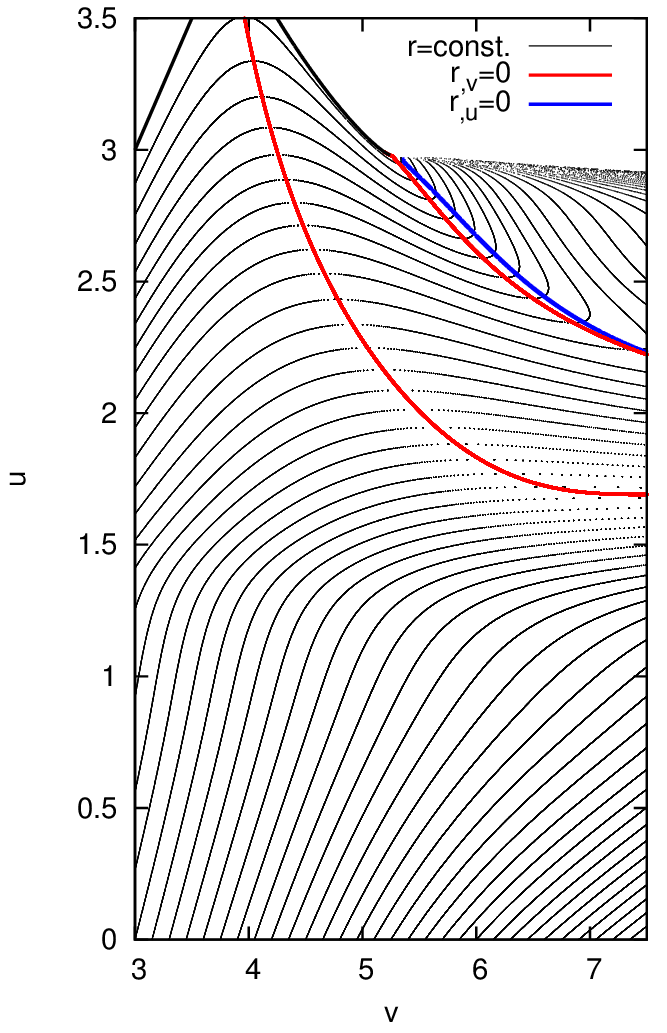}\label{fig:18a}}
\hfill
\subfigure[][]{\includegraphics[width=0.24\textwidth]{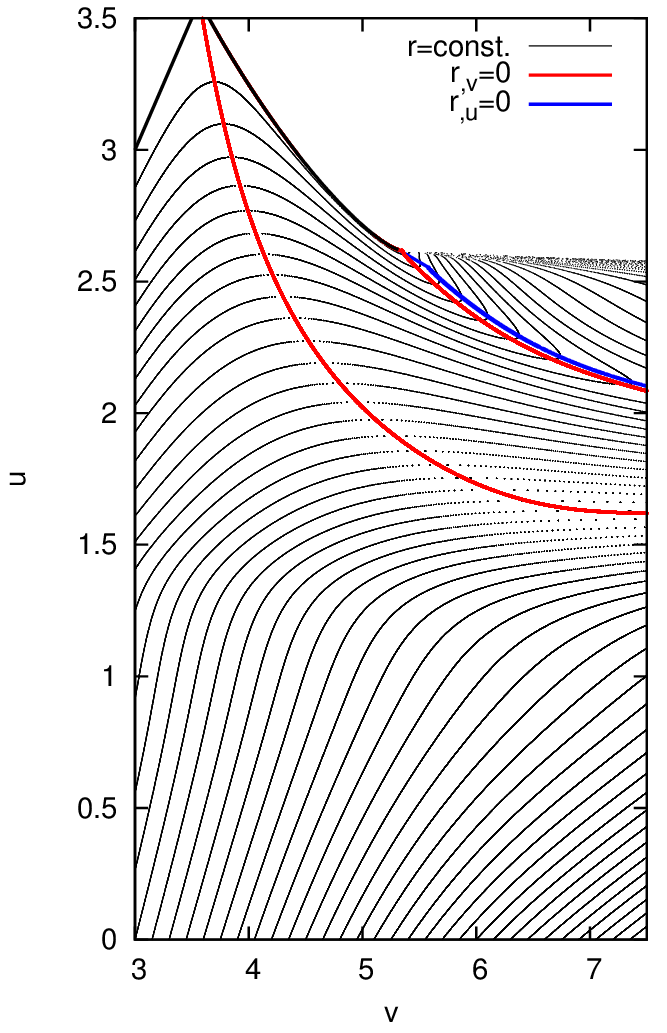}}
\hfill
\subfigure[][]{\includegraphics[width=0.24\textwidth]{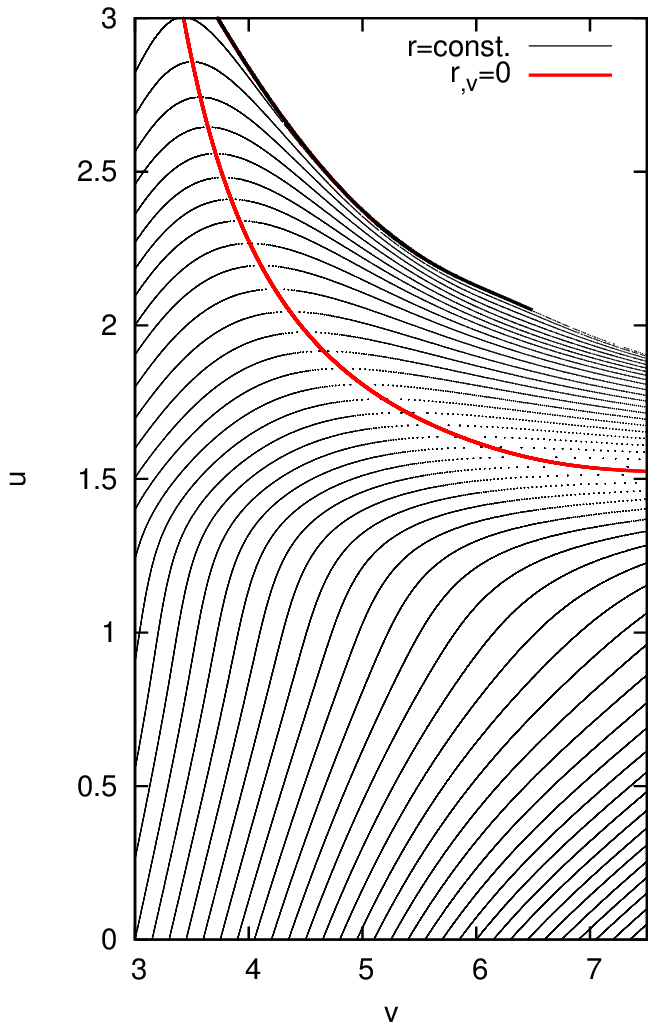}\label{fig:18c}}
\hfill
\subfigure[][]{\includegraphics[width=0.24\textwidth]{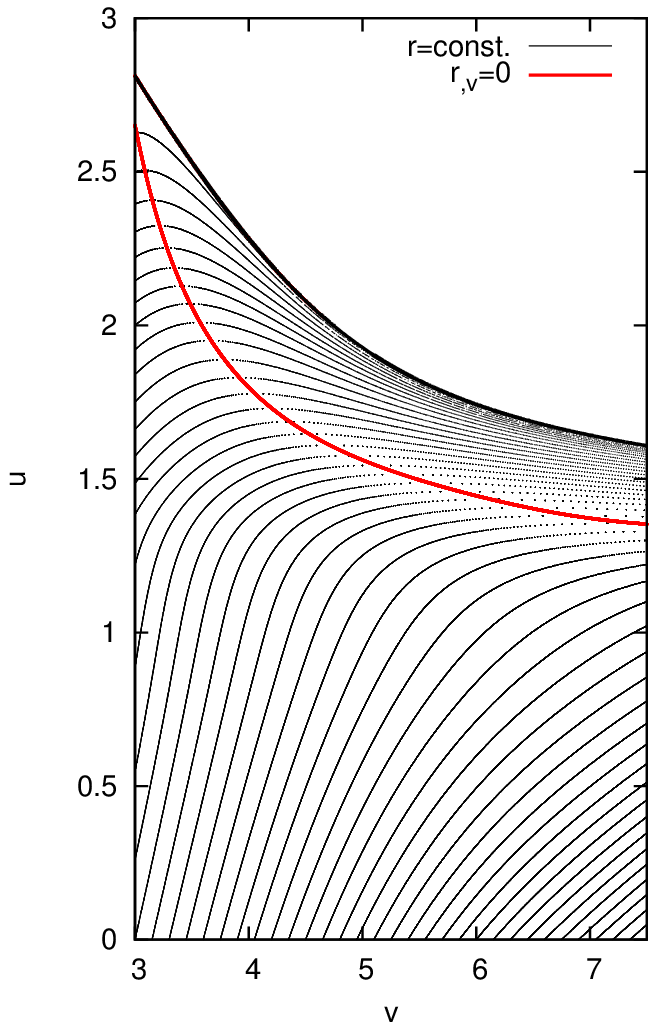}}
\caption{(color online) Penrose diagrams of~spacetimes formed during the~$SF$--$DE$--$DM$ collapse with $\as=0.6$, $\ak=0.15$, $\ah=0.15$ and~$m^2$ equal to~(a)~$-1$, (b)~$-2$, (c)~$-3$ and~(d)~$-5$.}
\label{fig:SFDEDM-varm2}
\end{figure}

Figure~\ref{fig:SFDEDM-varm2} presents the~results of~the~considered evolution for~different negative values of~the~parameter $m^2$, when the~initial field amplitudes are $\as=0.6$, $\ak=0.15$ and~$\ah=0.15$. There exist dynamical wormholes in~the~emerging spacetimes for~$m^2\geqslant -2.1$ and~Schwarzschild black holes for~its smaller values. The~collapse with positive $m^2$ always results in~a~spacetime, which contains a~dynamical wormhole.

The~above results indicate that when the~field gravitational self-interaction strength is~concerned, the~role of~dark energy coexists with the~influence of~dark matter. The~dark energy component decides what type of~an object forms. Dark matter does not~interfere with the~structure type, but regulates the~amount of~time which is~needed for~the~intrinsic object to~form. When the~impact of~the~dark matter potential is~considered, it~turns out that the~value of~$m^2$ influences the~observed structure by~controlling whether the~wormhole throats remain open in~the~spacetime. Their absence results in~the~formation of~a~Schwarzschild spacetime and~the~existence of~a~central singularity even after the~dynamical phase of~the~collapse is~finished. It~should be emphasized that the~dark sector suppresses the~natural tendency of~an~electrically charged scalar field to~form a~dynamical Reissner-Nordstr\"{o}m spacetime in~the~course of~the~gravitational collapse.

\subsection{Fields behavior in~spacetimes}

\begin{figure}[tbp]
\subfigure[][]{\includegraphics[width=0.475\textwidth]{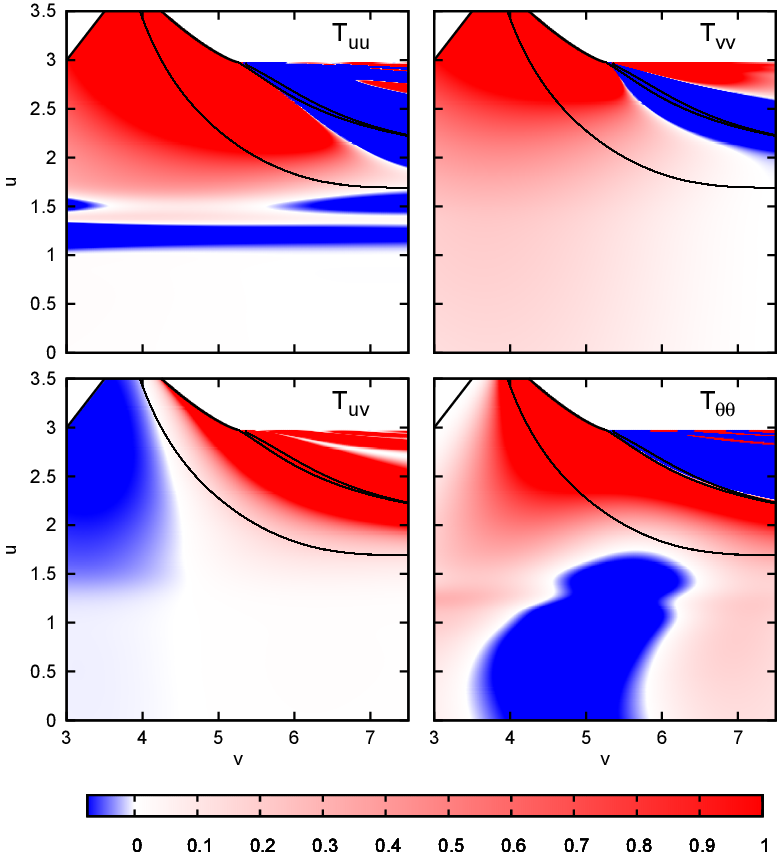}}
\hfill
\subfigure[][]{\includegraphics[width=0.475\textwidth]{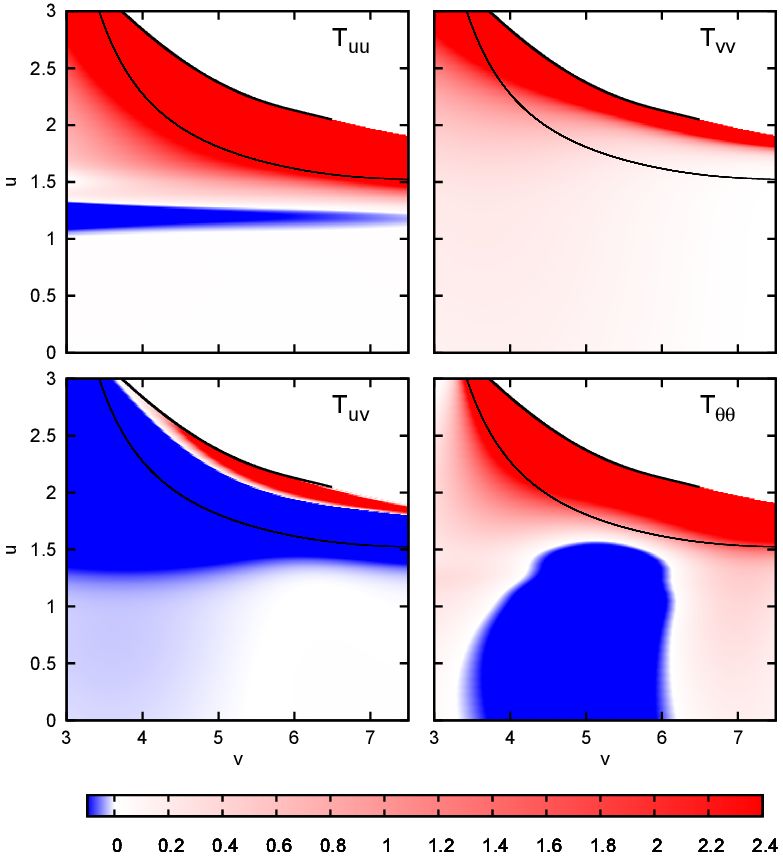}}
\caption{(color online) The~stress-energy tensor components \eqref{eqn:Tuu}--\eqref{eqn:Ttt} in~the~$(vu)$-plane for~$SF$--$DE$--$DM$ evolutions with $\as=0.6$, $\ak=0.15$, $\ah=0.15$ and~$m^2$ equal to~(a)~$-1$ and~(b)~$-3$.}
\label{fig:SFDEDM-ten}
\end{figure}

\begin{figure}[tbp]
\subfigure[][]{\includegraphics[width=0.24\textwidth]{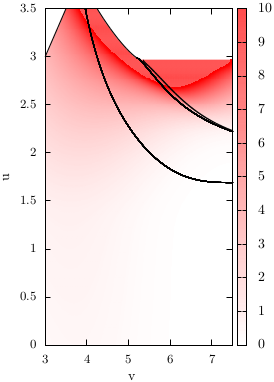}}
\hfill
\subfigure[][]{\includegraphics[width=0.24\textwidth]{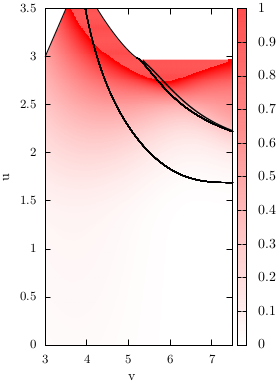}}
\hfill
\subfigure[][]{\includegraphics[width=0.24\textwidth]{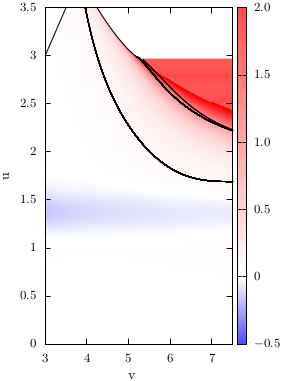}}
\caption{(color online) The~$\left(vu\right)$-distribution of~(a)~the~moduli of~the~electrically charged scalar field, $|\psi|$, (b)~the~moduli of~the~dark matter complex scalar field, $|\chi|$, and~(c)~the~phantom scalar field, $\phi$, for~the~$SF$--$DE$--$DM$ evolution with $\as=0.6$, $\ak=0.15$, $\ah=0.15$ and~$m^2=-1$.}
\label{fig:SFDEDM-dis-1}
\end{figure}

\begin{figure}[tbp]
\subfigure[][]{\includegraphics[width=0.24\textwidth]{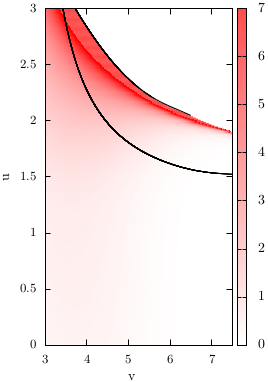}}
\hfill
\subfigure[][]{\includegraphics[width=0.24\textwidth]{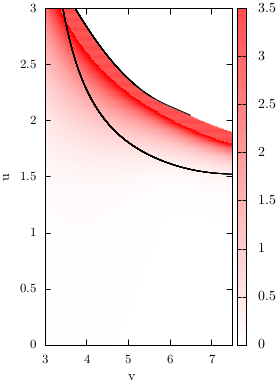}}
\hfill
\subfigure[][]{\includegraphics[width=0.24\textwidth]{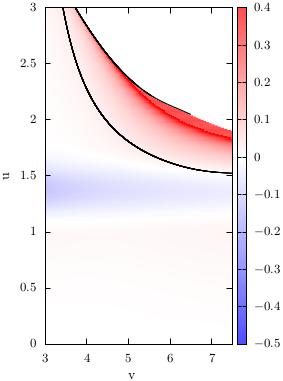}}
\caption{(color online) The~$\left(vu\right)$-distribution of~(a)~$|\psi|$, (b)~$|\chi|$ and~(c)~$\phi$ for~the~$SF$--$DE$--$DM$ evolution with $\as=0.6$, $\ak=0.15$, $\ah=0.15$ and~$m^2=-3$.}
\label{fig:SFDEDM-dis-3}
\end{figure}

The~$\left(vu\right)$-distributions of~the~stress-energy tensor components \eqref{eqn:Tuu}--\eqref{eqn:Ttt} for~the~dynamical spacetimes emerging from the~$SF$--$DE$--$DM$ evolutions characterized by~the~amplitudes $\as=0.6$, $\ak=0.15$, $\ah=0.15$ and~$m^2$ equal to~$-1$ and~$-3$ are presented in~figure~\ref{fig:SFDEDM-ten} (the~respective spacetime structures were presented in~figures~\ref{fig:18a} and~\ref{fig:18c}). The~spacetime distributions of~the~moduli of~the~electrically charged scalar field, the~dark matter complex scalar field and~the~phantom scalar field for~the~same cases are shown in~figures~\ref{fig:SFDEDM-dis-1} and~\ref{fig:SFDEDM-dis-3}. 

The~$T_{uu}$ and~$T_{vv}$ components of~the~stress-energy tensors are negative in~the~vicinity of~wormhole throats when~$m^2=-1$. The~null energy condition \eqref{eqn:nec} is~violated there. In~the~case of~$m^2=-3$ the~components are positive nearby $r=0$ and~hence a~black hole, not~a~dynamical wormhole arises in~the~spacetime. The~obtained result strengthens the~conclusion presented in~section~\ref{sec:de-vic} that the~violation of~NEC is~needed for~the~formation of~wormholes during dynamical evolutions. Clearly, this statement is~valid even for~complicated matter-geometry systems evolutions, in~which several types of~interacting matter are involved.

Similarly to~the~$SF$--$DM$ case, the~obtained spacetime type depends on~the~values of~the~evolving field functions in~the~vicinity of~$r=0$. The~formation of~a~dynamical wormhole stems from a~significant value of~the~modulus of~the~phantom scalar field in~this region as~$v\to\infty$ (four orders of~magnitude bigger than in~the~remaining spacetime) accompanied by~a~small value of~the~dark matter field modulus. When the~modulus of~the~dark matter complex scalar field is~bigger at~$v\to\infty$ nearby $r=0$ (three orders of~magnitude in~relation to~the~rest of~the~spacetime in~the~considered case), the~importance of~the~remaining fields is~reduced and~the~dynamical Schwarzschild spacetime emerges. The~structure type and~the~fields behavior in~a~spacetime depend on~the~value of~the~parameter $m^2$, not~the~initial amplitude of~either of~the~collapsing fields.

\subsection{Properties of~the~emerging objects}

\begin{figure}[tbp]
\begin{minipage}{0.5\textwidth}
\subfigure[][]{\includegraphics[width=0.95\textwidth]{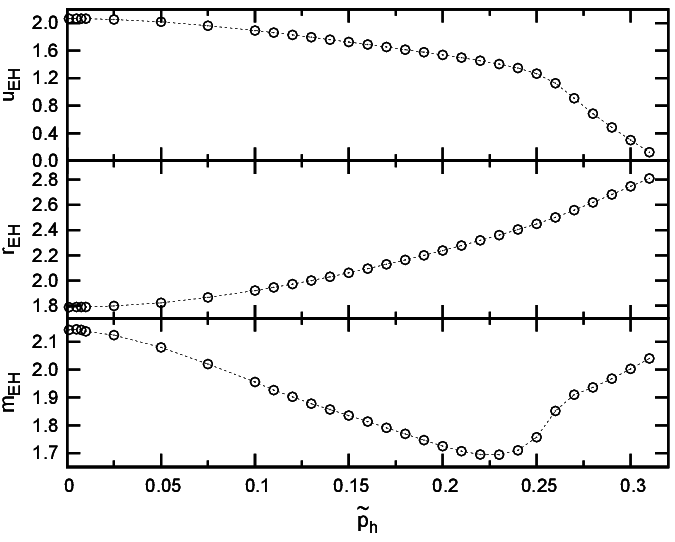}\label{fig:DEDMchar-ph}}
\end{minipage}
\hfill
\begin{minipage}{0.5\textwidth}
\subfigure[][]{\includegraphics[width=0.95\textwidth]{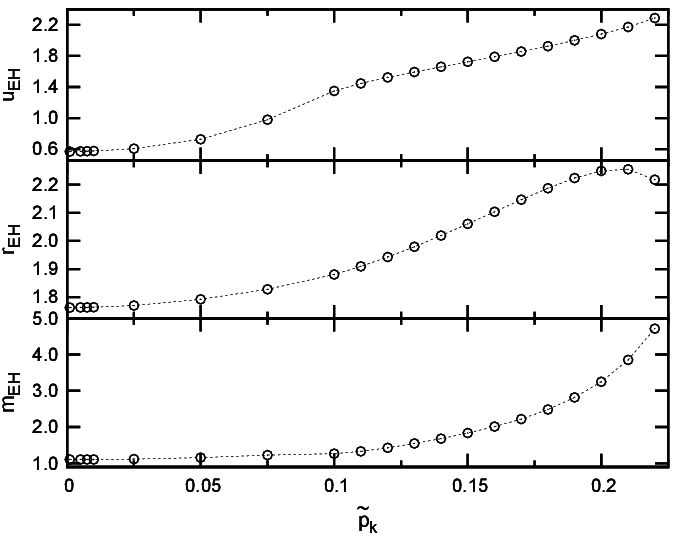}\label{fig:DEDMchar-pk}}
\end{minipage}
\begin{minipage}{0.5\textwidth}
\subfigure[][]{\includegraphics[width=0.95\textwidth]{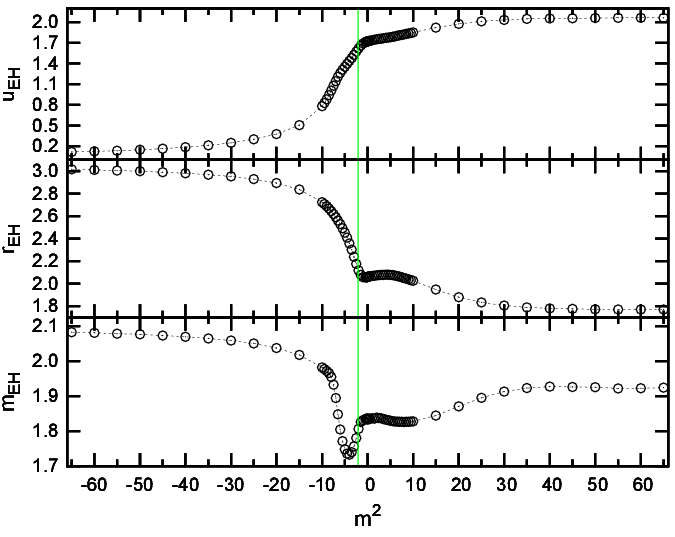}\label{fig:DEDMchar-m2}}
\end{minipage}
\hfill
\begin{minipage}{0.475\textwidth}
\caption{(color online) The~$u$-locations of~the~event horizons, $u_{EH}$, the~masses, $m_{EH}$, and~the~radii, $r_{EH}$, of~the~objects formed during the~$SF$--$DE$--$DM$ collapse as~functions of~(a)~$\ah$ with $\ak=0.15$, $m^2=0.1$, (b)~$\ak$ with $\ah=0.15$, $m^2=0.1$ and~(c)~$m^2$ with $\ak=0.15$, $\ah=0.15$. The~green line indicates the~border between spacetimes containing black holes of~the~Schwarzschild type ($m^2<-2.1$) and~wormholes ($m^2\geqslant -2.1$).}
\label{fig:DEDMchar}
\end{minipage}
\end{figure}

The~influence of~$\ah$, $\ak$ and~$m^2$ on~the~event horizon~$u$-locations, the~masses and~the~radii of~objects formed during the~gravitational evolution of~an electrically charged scalar field accompanied by~dark matter and~dark energy is~shown in~figure~\ref{fig:DEDMchar}. The~maximal value of~the~amplitude $\ah$ was determined by~the~event horizon location in~the~spacetime so~as~to~begin the~collapse outside~it. The~maximal considered value of~the~phantom scalar field amplitude was the~biggest value not~leading to~the~formation of~a~naked singularity in~spacetime.

As~the~dark matter complex scalar field amplitude increases, the~wormholes appear earlier in~$u$, their radii increase and~masses decrease up~to~$\ah=0.23$ and~increase for~larger~$\ah$ (figure~\ref{fig:DEDMchar-ph}). In~the~case of~an~increasing $\ak$, the~values describing all the~properties increase, apart from $r_{EH}$ for~values of~the~amplitude exceeding $0.21$ (figure~\ref{fig:DEDMchar-pk}). Just~as~in~the~case of~the~$SF$--$DE$ system, such behavior indicates the~upcoming formation of~a~naked singularity in~the~spacetime. When~$m^2$~increases the~event horizon forms later in~terms of~retarded time and~the~wormhole radii decrease with an~inflection nearby $m^2=-2.1$, which is~the~point of~the~change between the~wormhole and~Schwarzschild spacetime structures (figure~\ref{fig:DEDMchar-m2}). The~behavior of~masses is~more complicated, as~apart from the~inflection around $m^2=-2.1$, there exists a~minimal mass for~$m^2=-3.9$. The~non-monotonic behavior of~the~described characteristics is~related to~the~fact that the~investigated system consists of~multiple components and~the~observed phenomena are a~compromise among the~tendencies characteristic for~particular evolving matter types.

\section{Conclusions}
\label{sec:conclusions}

The~influence of~dark matter and~dark energy on~the~electrically charged scalar field collapse was investigated. Dark matter was described by~a~complex scalar field with a~quartic potential, coupled to~a~$U(1)$-gauge field. Dark energy was modeled by~a~scalar field coupled to~gravity in~a~phantom manner. The~exponential relation of~the~scalar field with dark energy originated from the~low-energy limit of~the~string theory. Dark matter was coupled with the~electrically charged scalar field through a~kinetic coupling between the~two gauge fields present in~the~system.

The~existence of~phantom matter in~spacetime can result in~the~formation of~naked singularities or~dynamical wormholes during the~gravitational collapse of~an electrically charged scalar field. The~latter objects have two throats related to~two travel directions and~are surrounded by~event horizons. The~values of~the~phantom field are considerably bigger in~the~spacetime region situated nearby wormhole throats, in~comparison to~the~surrounding area. This accumulation leads to~the~violation of~the~null energy condition, which is~essential for~the~dynamical wormhole formation and~the~stabilization of~its throats. The~masses of~wormholes, the~radii of~event horizons and~wormhole throats increase when the~amplitudes of~both phantom and~electrically charged scalar fields increase. The~event horizon, wormhole throats and~singularity origin appear later in~terms of~retarded time as~$\ak$ increases while~$\as$ is~constant and~at smaller values of~the~$u$-coordinate in~the~opposite case.

In~the~obtained wormhole spacetimes spacelike singularities situated along $r=0$ exist during the~dynamical stage of~the~evolution. They turn into wormhole throats as~the~spacetime tends towards the~final stage of~the~collapse. This suggests that the~considered spacetime shortcuts ensue from the~process, during which a~black hole gains exotic matter of~negative energy. Thus the~existence of~duality between black holes and~wormholes was confirmed in~the~course of~fully non-linear computations.

The~spacetimes resulting from the~examined collapse in~the~presence of~dark matter are either non-singular or~contain black holes, depending on~the~gravitational self-interaction strengths of~both evolving fields, that is~their initial amplitudes~$\as$ and~$\ah$. The~type of~an~emerging object depends on~the~value of~the~square of~mass parameter from the~dark matter model. In~general, the~non-zero vev of~the~dark matter complex scalar field favors the~formation of~dynamical Schwarzschild spacetimes for~$m^2<-3.85$, while for~the~vanishing vev the~dynamical Reissner-Nordstr\"{o}m black holes appear.

The~null energy condition is~fulfilled in~the~vicinity of~singularities in~the~emerging spacetimes, hence the~stabilization of~wormhole throats is~impossible and~such objects do~not~form. The~values of~the~moduli of~the~electrically charged scalar field and~the~dark matter complex scalar field are considerably bigger nearby $r=0$, where $v\to\infty$, than in~the~remaining regions in~spacetimes of~Reissner-Nordstr\"{o}m and~Schwarzschild types, respectively. This indicates a~repulsive character of~the~former and~an~attractive character of~the~latter field.

The~moment of~a~black hole formation, i.e.,~an~appearance of~an~event horizon in~terms of~retarded time, as~well as~the~black hole masses and~radii are controlled by~the~values of~$\ah$ and~$m^2$. The~event horizon $u$-locations increase and~the~remaining characteristics decrease as~the~dark matter complex scalar field initial amplitude raises. The~opposite tendency is~observed in~the~case of~an~increasing parameter~$m^2$.

The~dark sector suppresses the~tendency of~a~self-interacting electrically charged scalar field to~form a~dynamical Reissner-Nordstr\"{o}m black hole during the~gravitational evolution. In~the~case of~its overall influence on~the~investigated process, the~strength of~the~dark energy gravitational self-interaction prevails over the~dark matter gravitational self-interaction strength. The~dark energy component decides on~the~type of~an object, while dark matter regulates the~amount of~time which is~needed for~it~to form. The~parameter $m^2$ of~the~dark matter potential influences the~observed structures by~controlling whether the~wormhole throats appear in~the~spacetime. Their absence results in~the~formation of~a~Schwarzschild-type black hole, for~the~non-zero vev of~the~dark matter complex scalar field and~$m^2<-2.1$.

In~the~spacetimes resulting from the~$SF$--$DE$--$DM$ collapse the~null energy condition is~violated in~the~vicinity of~wormhole throats and~fulfilled nearby the~singularities, just as~in~the~previous cases. The~formation of~a~wormhole requires values of~a~phantom scalar field considerably bigger nearby the~throats in~comparison to~the~rest of~a~spacetime. Large values of~the~modulus of~the~dark matter complex scalar field result in~the~emergence of~a~Schwarzschild-type object.

The~behavior of~$u$-locations, masses and~radii as~functions of~$\ah$, $\ak$ and~$m^2$ are complicated due to~the~fact that the~investigated system consists of~multiple components, each of~which displays a~different tendency during the~evolution. The~electrically charged scalar field favors the~Reissner-Nordstr\"{o}m black hole formation, dark energy supports the~emergence of~wormholes or~naked singularities and~dark matter tends to~form the~Schwarzschild-type objects. The~observed phenomena are a~compromise among these tendencies characteristic for~the~involved matter types.

The~changes introduced by~the~dark sector to~the~electrically charged scalar field gravitational collapse are presented in~table~\ref{tab:summary}. It~summarizes the~obtained results in~the~context of~the~role of~particular parameters in~the~studied evolution for~the~initial amplitude of~an~electrically charged scalar field, which solely leads to~the~formation of~a~black hole. The~obtained outcomes confirm the~previous conclusions regarding the~role of~dark energy during the~gravitational collapse. It~does not prevent the~formation of~singular spacetimes during the~dynamical evolution when it~is~accompanied by~other types of~matter, in~particular dark matter~\cite{CaiWang2006-063005,RudraDebnath2014-2668}. The~current studies also widened previous findings in~this regard. They allowed us to~describe the~structures of~objects existing in~these singular spacetimes, which turned out to~be wormholes, naked singularities or~black holes of~Schwarzschild or~Reissner-Nordstr\"{o}m types.

\begin{table}[tbp]
\centering
\begin{tabular}{|c|c|c|c|}
\hline
\multirow{2}{*}{Evolution} & \multirow{2}{*}{$\tpe_k\nearrow$} & \multirow{2}{*}{$\tpe_h\nearrow$} & \multirow{2}{*}{$m^2\nearrow$}\\
& & &\\
\hline
\multirow{3}{*}{$SF$--$DE$} & $WH\to NS\to ns$~\cite{NakoniecznaRogatkoModerski2012-044043} & \multirow{3}{*}{$-$} & \multirow{3}{*}{$-$}\\
& $u_{EH}\nearrow$& &\\
& $r_{EH}\nearrow\ m_{EH}\nearrow$ & &\\
\hline
\multirow{3}{*}{$SF$--$DM$} & \multirow{3}{*}{$-$} & $RN$ & $RN\to S$\\
& & $u_{EH}\searrow$ & $u_{EH}\nearrow$\\
& & $r_{EH}\nearrow\ m_{EH}\nearrow$ & $r_{EH}\searrow\ m_{EH}\searrow$\\
\hline
\multirow{3}{*}{$SF$--$DE$--$DM$} & $WH\to NS$ & $WH$ & $WH\to S$\\
& $u_{EH}\nearrow$ & $u_{EH}\searrow$ & $u_{EH}\nearrow$\\
& $r_{EH}\nearrow\searrow\ m_{EH}\nearrow$ & $r_{EH}\nearrow\ m_{EH}\searrow\nearrow$ & $r_{EH}\searrow\ m_{EH}\searrow\nearrow$\\
\hline
\end{tabular}
\caption{\label{tab:summary} Spacetime structures formed during the~gravitational collapse of~an~electrically charged scalar field with $\as=0.6$ accompanied by~dark energy and~dark matter and~the~selected characteristics of~the~emerging objects. Symbols: $ns$ -- non-singular spacetime, $S$ -- Schwarzschild-type spacetime, $RN$ -- Reissner-Nordstr\"{o}m-type spacetime, $WH$ -- wormhole, $NS$ -- naked singularity, $u_{EH}$ -- $u$-location of~an~event horizon, $r_{EH}$ and~$m_{EH}$ -- radius and~mass of~the~intrinsic object, respectively. Arrows: $\nearrow$/$\searrow$ -- an~increase/decrease of~a~value of~the~particular characteristic or~amplitude, $\rightarrow$ -- the~change of~the~spacetime structure while the~particular parameter varies.}
\end{table}

\begin{figure}[tbp]
\begin{minipage}{0.5\textwidth}
\centering
\subfigure[][]{\includegraphics[width=0.55\textwidth]{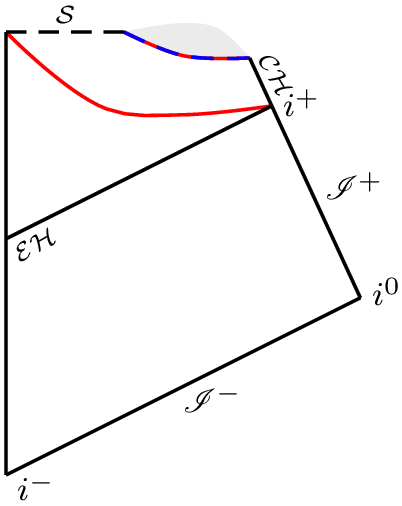}}
\end{minipage}
\begin{minipage}{0.5\textwidth}
\centering
\subfigure[][]{\includegraphics[width=0.55\textwidth]{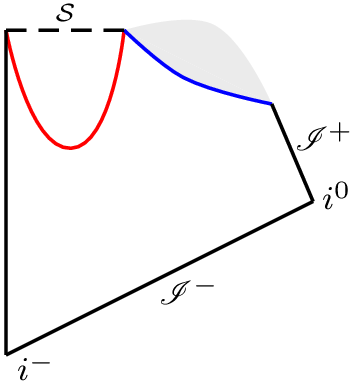}}
\end{minipage}
\caption{(color online) The~Carter-Penrose diagrams of~dynamical spacetimes, which contain (a)~a~wormhole and~(b)~a~naked singularity. The~red  and~blue solid lines are the~$r\POv=0$ and~$r\POu=0$ horizons, respectively. The~red and~blue dashed line symbolizes wormhole throats. The~meaning of~the~symbols is~the~same as~in~figure~\ref{fig:domain}. The~causal structure of~the~spacetime regions marked gray on~the~diagrams is~impossible to~interpret without any doubts due to~the~limitations of~numerical computations in~these areas.}
\label{fig:CPdiag}
\end{figure}

The~Penrose diagrams of~spacetimes, on~which the~spacetime structures were presented in~the~current paper, are related to~the~Carter-Penrose diagrams via conformal transformation, which preserves the~causal structure of~a~spa\-ce\-ti\-me. The~Carter-Penrose diagrams of~both static and~dynamical Schwarzschild and~Reissner-Nordstr\"{o}m spacetimes can be found, e.g.,~in~\cite{MisnerThorneWheeler,OrenPiran2003-044013,HongHwangStewartYeom2010-045014}. In~figure~\ref{fig:CPdiag} we present such diagrams for~the~dynamical wormhole and~naked singularity spacetimes, which stem from the~performed analyses of~the~dynamical formation of~these objects. 

Apart from investigations of~spacetime structures formed during the~collapse in~the~presence of~the~dark sector, other tantalizing questions are the~mechanism of~black hole formation and~the~behavior of~matter accreting onto a~nascent black hole taking the~cosmological evolution into account. It~was claimed that primordial black holes can accrete the~surrounding dark energy, phantom energy or~ghost condensate very effectively \cite{HaradaMaedaCarr2006-024024,CarrKohriSendoudaYokoyama2010-104019}. However, the~conducted analyses neglected cosmic expansion.

Until now, there have been several attempts to~find black hole solutions embedded in~the~expanding Universe, starting from the~Einstein-Straus~\cite{EinsteinStraus1945-120} and~McVittie~\cite{McVittie1933-325} solutions to~the~multi black hole Kastor-Traschen cosmological solution~\cite{KastorTrashen1993-5370}, which is a~time-dependent generalization of~the~Majumdar-Papapetrou one. An~exact analytical black hole solution in~the~expanding Universe was derived within the~Einstein-scalar-Maxwell system with two $U(1)$-gauge fields and~an~exponential potential of~the~scalar field~\cite{GibbonsMaeda2010-131101}. The~dilaton black hole on~a~thick brane and~a~cosmological brane black hole solutions were studied in~\cite{Rogatko2001-064014,Rogatko2004-044022}, while the~time-dependent solution from compactification of~intersecting branes in~higher-dimensional unified theories was found in~\cite{MaedaNozawa2010-044017}. As~was already stated in~the~Introduction, the~dynamical solutions may behave quite differently in~comparison to~their stationary cousins. Some realistic attempts to~analytically describe the~accretion of~a~matter field onto an~evolving black hole were conducted in~\cite{ChadburnGregory2014-195006}, where a~time-dependent scalar field and~a~time-dependent black hole were considered. In~order to~correctly account for the~scalar field accretion, the~expansion in~terms of~a~slow roll parameter was implemented. The~only initial assumptions about the~geometry were its $SO(3)$ symmetry and~the~time and~radial dependence of~the~line element. Hence, the~presented method did not require a~direct assumption about the~form of~the~underlying black hole metric.

The~issues introduced above to~be studied in~the~fully non-linear dynamical context are beyond of~the~scope of~the~presented research, as~they require different methodology than the~one employed in~the~paper. We~hope to~return to~the~announced problems in~relation to~the~cosmological evolution of~the~Universe in~the~future researches.

\appendix
\section{Numerical computations}
\label{sec:appendix}

The~numerical simulations of~the~investigated process were conducted with the~use of~a~modified version of~the~code, the~specifics of~which were presented in~\cite{BorkowskaRogatkoModerski2011-084007}. The~scheme was broadened and~adapted for~the~presently considered problem by~including the~phantom constant~$\xi$ and~the~additional sector \eqref{eqn:HR}--\eqref{eqn:C2}, which refers to~dark matter. Due~to~the~coupling between the~$P_\mu$ and~$A_\mu$ fields, the~module covering the~Maxwell field evolution \eqref{eqn:M2} was also modified. The~module governing the~geometry dynamics \eqref{eqn:E1}--\eqref{eqn:E4} was adjusted to~the~presently examined physical system.

The~inclusion of~the~additional sector in~the~code required posing adequate initial and~boundary conditions for~functions $z_1$, $z_2$, $h_1$, $h_2$, $w_1$, $w_2$, $T$, $\gamma$, $\kappa_1$ and~$\kappa_2$. The~first two evolved along~$u$ according to~equations $H_{_{\left(Re\right)}}$ and~$H_{_{\left(Im\right)}}$. The~dynamics of~$h_1$, $h_2$, $w_1$, $w_2$, $T$ and~$\gamma$ along~$v$ was governed by~equations $P7$, $P8$, $H_{_{\left(Re\right)}}$, $H_{_{\left(Im\right)}}$, $C2$ and~$C1$, respectively. The~values of~the~remaining two quantities were calculated on~the~basis of~their definitions included in~\eqref{defdef}.

Initial conditions were posed on~a~null hypersurface denoted as~$u=0$. The~profiles of~functions $h_1$ and~$h_2$ were assigned according to~\eqref{psichi-prof}. The~values of~$z_1$ and~$z_2$ were calculated analytically using the~relations $P7$ and~$P8$. The~behavior of~the~rest of~the~quantities, i.e.,~$w_1$, $w_2$, $T$ and~$\gamma$, at~the~initial spacetime slice was resolved with the~use of~the~three-point Simpson method and~the~Newton method at~the~first point.

The~line $u=v$ was the~line of~boundary conditions. Because of~the~fact that the~collapsing physical system consists of~a~matter shell, the~values of~$T$ and~$\gamma$ were equal to~zero at~the~boundary, which refers to~the~non-singular $r=0$ line. The~quantities $\kappa_1$ and~$\kappa_2$ also vanished due to~the~equations $H_{_{\left(Re\right)}}$ and~$H_{_{\left(Im\right)}}$. The~flattening of~the~functions $h_1$ and~$h_2$ was imposed on~the~boundary, i.e.,~$h_{1,r}$ and~$h_{2,r}$ were equal to~zero there. The~values of~the~quantities $w_1$ and~$w_2$ were the~same as~$z_1$ and~$z_2$, respectively.

The~accuracy of~the~numerical code was checked indirectly, due to~the~fact that no~analytical solutions exist for~the~investigated process. The~tests were performed for~the~evolutions initiated with parameters $\as=0.6$, $\ah=0.15$, $m^2=0.1$, $\ak=0.14$ and~$\as=0.6$, $\ah=0.15$, $m^2=-3$, $\ak=0.15$. The~respective spacetime structures are presented in~figures~\ref{fig:19b} and~\ref{fig:18c}. For~the~purpose of~testing the~numerical code, these two cases will be referred to~as evolutions~1 and~2, respectively. The~analyses were done for~spacetimes obtained within the~$SF$--$DE$--$DM$ physical system, which was the~most comprehensive among those discussed in~the~paper. Hence, the~performed code accuracy checks covered the~operation of~all the~constituent modules of~the~code.

The~first test was based on~checking the~convergence of~the~obtained results. In~order to~monitor the~convergence, the~computations for~evolutions 1 and~2 were conducted on~four grids with integration steps being multiples of~$10^{-4}$. A~step of~a~particular grid was twice the~size of~a~denser one. The~convergence was examined on~$u=const.$ hypersurfaces chosen for~each case arbitrarily. The~selected hypersurfaces were situated close to~the~emerging horizon, but in~the~region where the~adaptive grid was yet inactive, which was necessary for~performing a~proper comparison of~the~outcomes.

\begin{figure}[tbp]
\subfigure[][]{\includegraphics[width=0.475\textwidth]{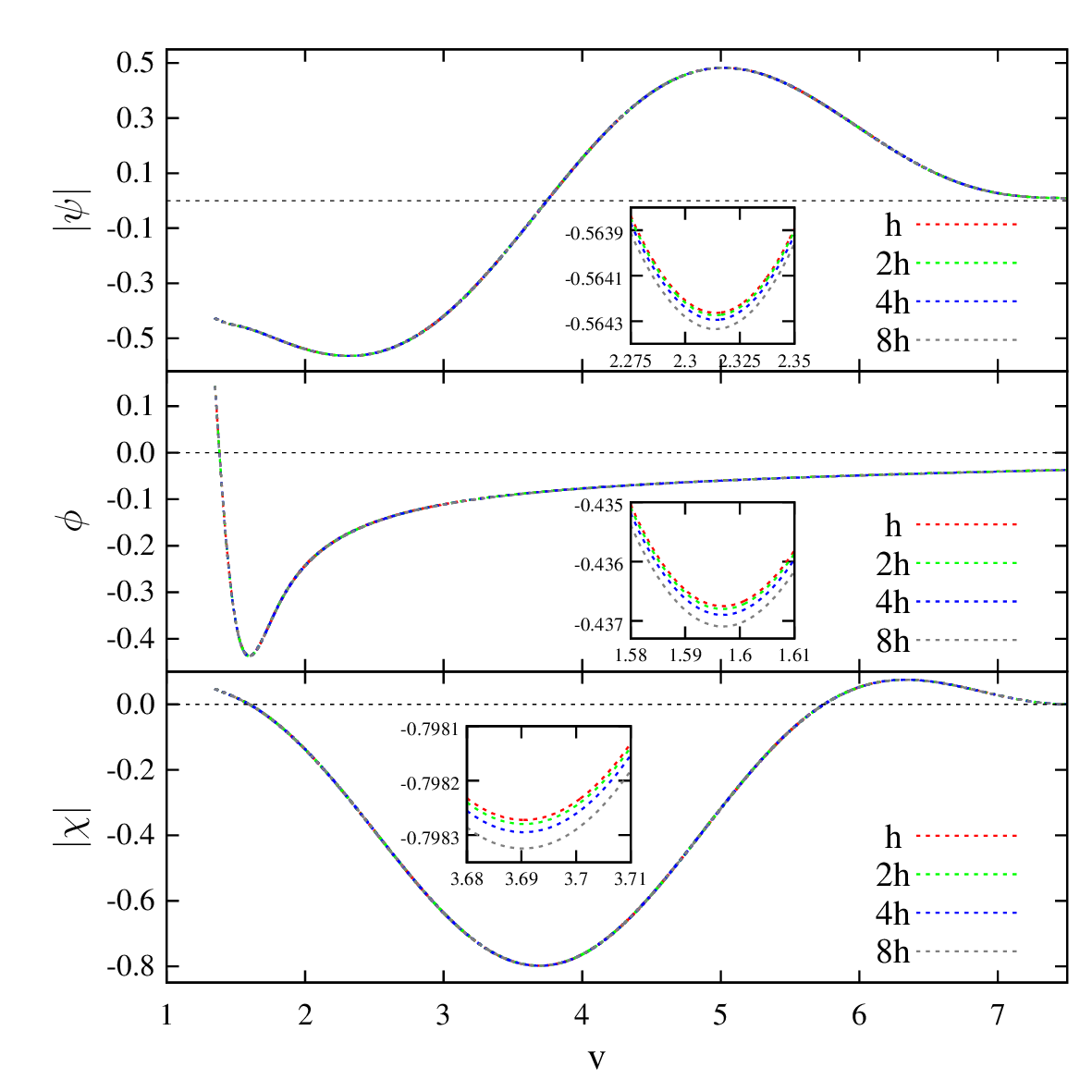}}
\hfill
\subfigure[][]{\includegraphics[width=0.475\textwidth]{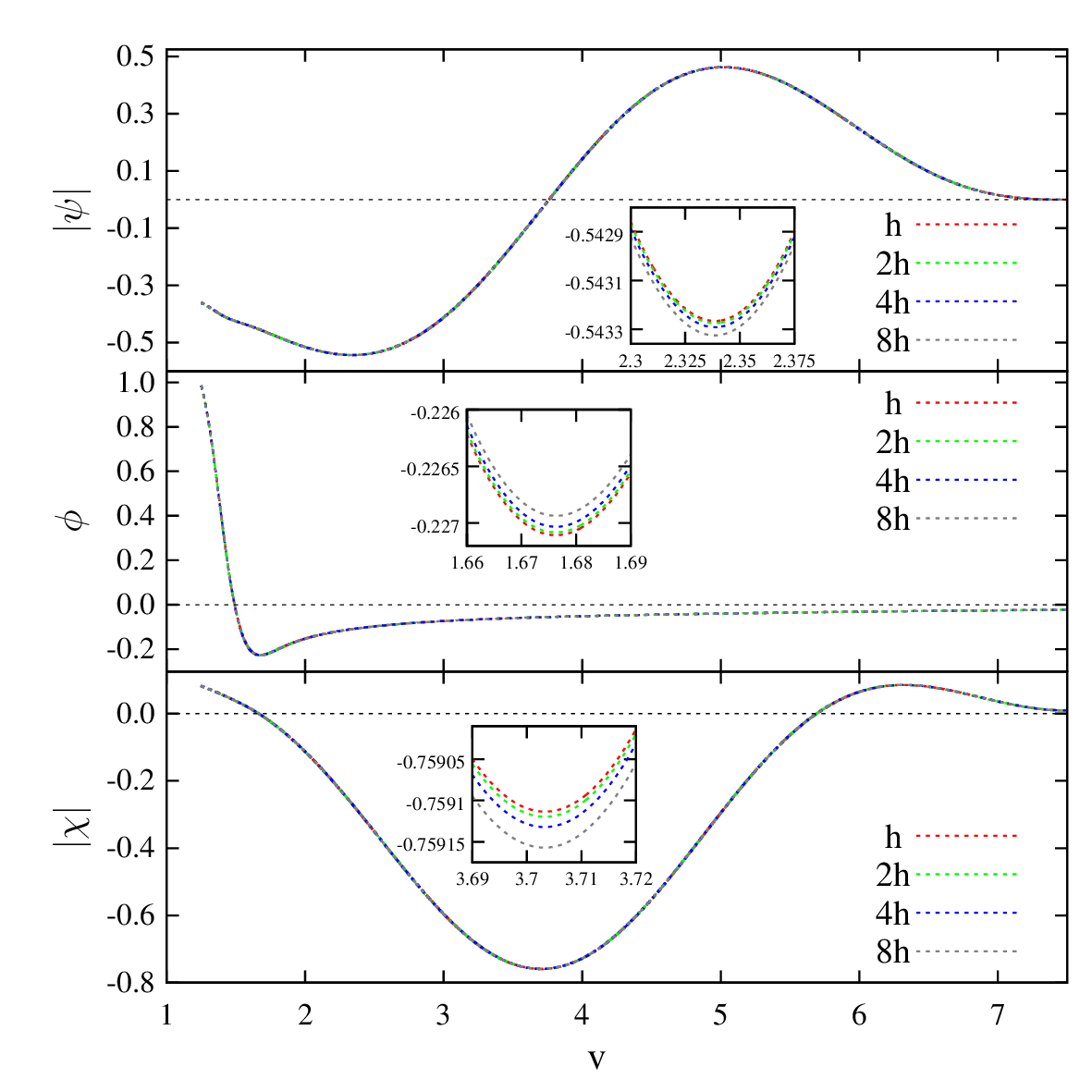}}
\caption{(color online) The~convergence of~field functions. The~phantom field, $\phi$, and~the~moduli of~complex scalar fields, $|\psi|$ and~$|\chi|$, were plotted versus $v$ for~evolutions conducted with integration steps, which were multiples of~\mbox{$h=10^{-4}$}, along hypersurfaces of~constant~$u$ equal to~(a)~$1.3504$ for~evolution~1 and~(b)~$1.2504$ for~evolution~2.}
\label{fig:Conv1}
\end{figure}

\begin{figure}[tbp]
\subfigure[][]{\includegraphics[width=0.475\textwidth]{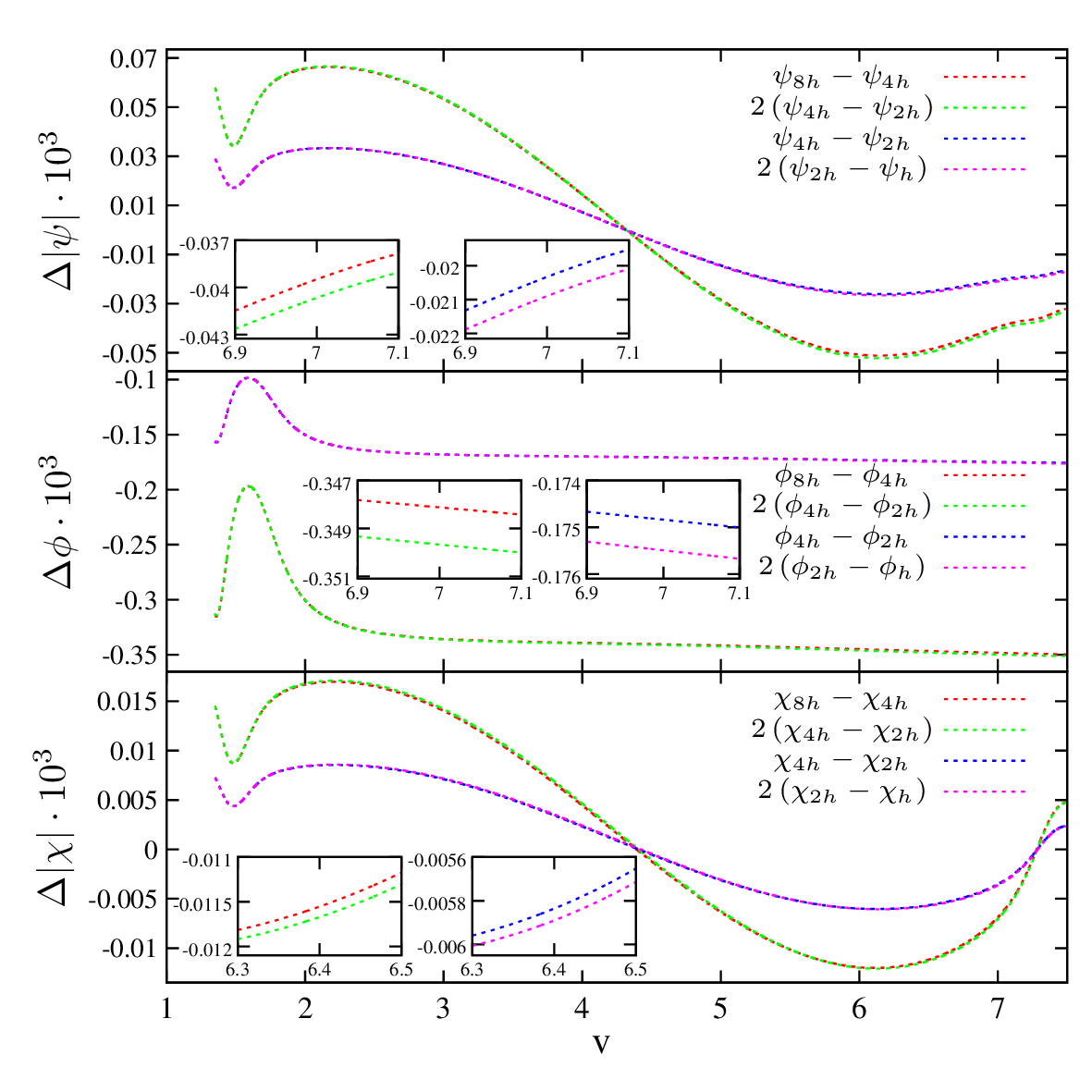}}
\hfill
\subfigure[][]{\includegraphics[width=0.475\textwidth]{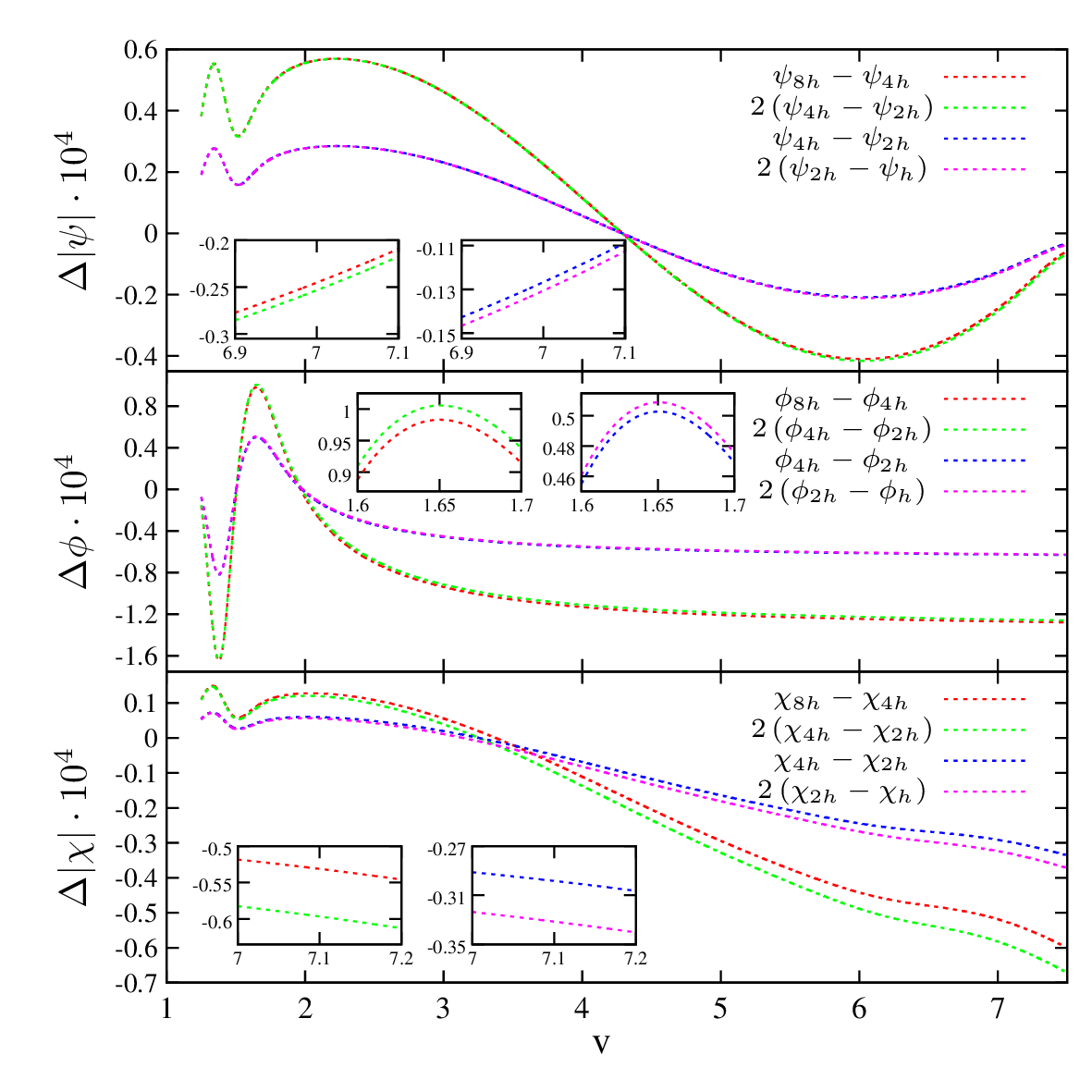}}
\caption{(color online) The~linear convergence of~the~code. The~differences between the~phantom field functions, $\Delta\phi$, and~the~moduli of~complex scalar fields, $\Delta|\psi|$ and~$\Delta|\chi|$, calculated on~grids with different integration steps (multiples of~\mbox{$h=10^{-4}$}) and~their doubles were obtained along the~same hypersurfaces of~constant $u$ as~in~figure~\ref{fig:Conv1} for~(a)~evolution~1 and~(b)~evolution~2.}
\label{fig:Conv2}
\end{figure}

The~field functions along the~selected hypersurfaces of~constant~$u$ are shown in~figure~\ref{fig:Conv1}. The~areas in~which the~differences among functions obtained on~various grids were most significant were magnified. The~maximum observed discrepancy between the~finest and~coarsest grids was equal to~$1.5\%$. Figure~\ref{fig:Conv2} presents the~linear convergence of~the~numerical code. The~maximal divergence between the~field profiles obtained on~two grids with a~quotient of~integration steps equal to~$2$ and~their respective doubles was~$5.7\%$. As~expected, the~errors became smaller linearly as~the~grid density increased. The~overall convergence analysis revealed that both the~algorithm and~the~numerical code were adequate for~solving the~system of~equations \eqref{eqn:P1-2}--\eqref{eqn:C2}.

\begin{figure}[tbp]
\centering
\includegraphics[width=0.475\textwidth]{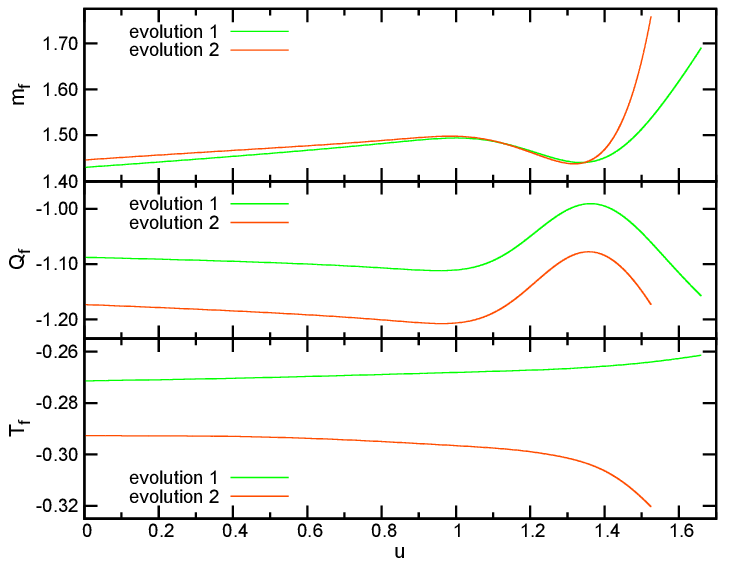}
\caption{(color online) Hawking masses, $m_f$, and~charges related to~the~$U(1)$-gauge fields, $Q_f$ and~$T_f$, calculated along $v_f=7.5$ as~functions of~retarded time, $u$, for~both tested evolutions.}
\label{fig:Cons}
\end{figure}

The~second test of~the~numerical code was based on~checking the~mass and~charge conservation in~the~spacetime. The~Hawking mass \eqref{haw} and~the~charges related to~both complex scalar fields coupled to~the~$U(1)$-gauge fields \eqref{charge} and~\eqref{charge-dm} as~functions of~retarded time along the~line $v=7.5$, which was a~maximal value of~advanced time achieved numerically, are presented in~figure~\ref{fig:Cons} for~the~investigated evolutions. Since during the~process the~matter was scattered by~the~gravitational potential barrier when the~collapsing shell approached its gravitational radius, the~physical quantities were not~conserved during the~whole evolution. The~effect of~the~outgoing flux was negligible except for~the~vicinity of~the~event horizon. The~deviation from the~constancy increased with advanced time, as~the~horizon was approached. The~maximal percentage deviations from the~particular quantity conservation up to~$u=1.25$ were equal to~$3.3\%$, $8.1\%$ and~$1.1\%$ for~the~mass, the~electric charge and~the~charge associated with the~dark matter sector, respectively. The~analysis of~mass and~charge conservation in~spacetime led to~the~conclusion that the~behavior of~matter investigated numerically was correct within the~domain of~integration.

\acknowledgments

A.~N. was supported by~the~Polish National Science Centre under doctoral scholarship ETIUDA DEC-2013/08/T/ST2/00007
and partially by~the~Polish National Science Centre grant no.~DEC-2014/15/B/ST2/00089. M.~R. was partially supported by~the~Polish National Science Centre grant no.~DEC-2014/15/B/ST2/00089. {\L}.~N. was supported by~the~Polish National Science Centre under postdoctoral scholarship FUGA\linebreak DEC-2014/12/S/ST2/00332.





\bibliographystyle{JHEP}
\bibliography{darksectorcollapseJHEP.bib}

\providecommand{\href}[2]{#2}\begingroup\raggedright\begin{thebibliography}{100}

\bibitem{Tolman1939-364}
R.~C. Tolman, {\it {Static solutions of~Einstein's field equations for~spheres
  of~fluid}},  {\em Phys.~Rev.} {\bf 55} (1939) 364.

\bibitem{OppenheimerVolkoff1939-374}
J.~R. Oppenheimer and G.~M. Volkoff, {\it {On massive neutron cores}},  {\em
  Phys.~Rev.} {\bf 55} (1939) 374.

\bibitem{SorkinPiran2001-084006}
E.~Sorkin and T.~Piran, {\it {The effects of~pair creation on~charged
  gravitational collapse}},  {\em Phys.~Rev.~D} {\bf 63} (2001) 084006.

\bibitem{SorkinPiran2001-124024}
E.~Sorkin and T.~Piran, {\it {Formation and evaporation of~charged black
  holes}},  {\em Phys.~Rev.~D} {\bf 63} (2001) 124024.

\bibitem{HodPiran1998-1554}
S.~Hod and T.~Piran, {\it {Mass inflation in~dynamical gravitational collapse
  of~a charged scalar field}},  {\em Phys.~Rev.~Lett.} {\bf 81} (1998) 1554.

\bibitem{OrenPiran2003-044013}
Y.~Oren and T.~Piran, {\it {Collapse of~charged scalar fields}},  {\em
  Phys.~Rev.~D} {\bf 68} (2003) 044013.

\bibitem{PoissonIsrael1990-1796}
E.~Poisson and W.~Israel, {\it {Internal structure of~black holes}},  {\em
  Phys.~Rev.~D} {\bf 41} (1990) 1796.

\bibitem{HongHwangStewartYeom2010-045014}
S.~E. Hong, D.~Hwang, E.~D. Stewart, and D.~Yeom, {\it {The causal structure
  of~dynamical charged black holes}},  {\em Class.~Quant.~Grav.} {\bf 27}
  (2010) 045014.

\bibitem{HwangYeom2011-064020}
D.~Hwang and D.~Yeom, {\it {Internal structure of~charged black holes}},  {\em
  Phys.~Rev.~D} {\bf 84} (2011) 064020.

\bibitem{HwangYeom2010-205002}
D.~Hwang and D.~Yeom, {\it {Responses of~the Brans-Dicke field due to
  gravitational collapses}},  {\em Class.~Quant.~Grav.} {\bf 27} (2010) 205002.

\bibitem{HwangLeeYeom2011-006}
D.~Hwang, B.~Lee, and D.~Yeom, {\it {Mass inflation in~f(R) gravity: A
  conjecture on~the~resolution of~the mass inflation singularity}},  {\em JCAP}
  {\bf 1112} (2011) 006.

\bibitem{HansenYeom2014-040}
J.~Hansen and D.~Yeom, {\it {Charged black holes in~string-inspired gravity: I.
  Causal structures and responses of~the Brans-Dicke field}},  {\em JHEP} {\bf
  10} (2014) 040.

\bibitem{HansenYeom2015-arxiv}
J.~Hansen and D.~Yeom, {\it {Charged black holes in~string-inspired gravity:
  II. Mass inflation and dependence on~parameters and potentials}},  {\em
  arXiv:1506.05689} (2015).

\bibitem{BorkowskaRogatkoModerski2011-084007}
A.~Borkowska, M.~Rogatko, and R.~Moderski, {\it {Collapse of~charged scalar
  field in~dilaton gravity}},  {\em Phys.~Rev.~D} {\bf 83} (2011) 084007.

\bibitem{NakoniecznaRogatko2012-3175}
A.~Nakonieczna and M.~Rogatko, {\it {Dilatons and the~dynamical collapse
  of~charged scalar field}},  {\em Gen.~Rel.~Grav.} {\bf 44} (2012) 3175.

\bibitem{NakoniecznaRogatkoModerski2012-044043}
A.~Nakonieczna, M.~Rogatko, and R.~Moderski, {\it {Dynamical collapse
  of~charged scalar field in~phantom gravity}},  {\em Phys.~Rev.~D} {\bf 86}
  (2012) 044043.

\bibitem{Ade2014-A16}
P.~A.~R. Ade et~al., {\it {Planck 2013 results. XVI. Cosmological parameters}},
   {\em Astron. Astrophys.} {\bf 571} (2014) A16.

\bibitem{CaiWang2006-063005}
R.-G. Cai and A.~Wang, {\it {Black hole formation from collapsing dust fluid in
  a background of dark energy}},  {\em Phys.~Rev.~D} {\bf 73} (2006) 063005.

\bibitem{ChakrabortyBandyopadhyay2010-151}
S.~Chakraborty and T.~Bandyopadhyay, {\it {Collapse dynamics of a star of dark
  matter and dark energy}},  {\em Grav.~Cosmol.} {\bf 16} (2010) 151.

\bibitem{RudraDebnath2014-2668}
P.~Rudra and U.~Debnath, {\it {Gravitational collapse with dark energy and dark
  matter in Ho\v{r}ava-Lifshitz gravity}},  {\em Int.~J.~Theor.~Phys.} {\bf 53}
  (2014) 2668.

\bibitem{WangFan2009-123012}
Q.~Wang and Z.~Fan, {\it {Dynamical evolution of quintessence dark energy in
  collapsing dark matter halos}},  {\em Phys.~Rev.~D} {\bf 79} (2009) 123012.

\bibitem{MotaBruck2004-71}
{D. F. Mota} and {C. van de Bruck}, {\it {On~the~spherical collapse model
  in~dark energy cosmologies}},  {\em A\&A} {\bf 421} (2004) 71.

\bibitem{DelliouBarreiro2013-037}
M.~L. Delliou and T.~Barreiro, {\it {Interacting dark energy collapse with
  matter components separation}},  {\em JCAP} {\bf 02} (2013) 037.

\bibitem{CaramesFabrisVelten2014-083533}
T.~R.~P. Caram{\^e}s, J.~C. Fabris, and H.~E.~S. Velten, {\it {Spherical
  collapse for unified dark matter models}},  {\em Phys.~Rev.~D} {\bf 89}
  (2014) 083533.

\bibitem{FullerOtt2015-L71}
J.~Fuller and C.~D. Ott, {\it {Dark matter-induced collapse of~neutron stars:
  a~possible link between fast radio bursts and the~missing pulsar problem}},
  {\em Mon.~Not.~Roy.~Astron.~Soc.} {\bf 450} (2015) L71.

\bibitem{LorimerEtAl2007-777}
D.~R. Lorimer, M.~Bailes, M.~A. McLaughlin, D.~J. Narkevic, and F.~Crawford,
  {\it {A bright millisecond radio burst of extragalactic origin}},  {\em
  Science} {\bf 318} (2007) 777.

\bibitem{JohnstonEtAl2006-L6}
S.~Johnston, M.~Kramer, D.~R. Lorimer, A.~G. Lyne, M.~McLaughlin, B.~Klein, and
  R.~N. Manchester, {\it {Discovery of two pulsars towards the Galactic
  Centre}},  {\em Mon.~Not.~Roy.~Astron.~Soc.} {\bf 373} (2006) L6.

\bibitem{ChoiShlosmanBegelman2015-4411}
J.-H. Choi, I.~Shlosman, and M.~C. Begelman, {\it {Supermassive black hole
  formation at high redshifts via direct collapse in a cosmological context}},
  {\em Mon.~Not.~Roy.~Astron.~Soc.} {\bf 450} (2015) 4411.

\bibitem{ShlosmanChoiBegelmanNagamine2015-arxiv}
I.~Shlosman, J.-H. Choi, M.~C. Begelman, and K.~Nagamine, {\it {Supermassive
  black hole seed formation at high redshifts: long-term evolution of the
  direct collapse}},  {\em arXiv:1508.05098}.

\bibitem{Zwicky1933-110}
F.~Zwicky, {\it {The redshift of~extragalactic nebulae}},  {\em
  Helv.~Phys.~Acta} {\bf 6} (1933) 110.

\bibitem{RubinFord1970-379}
V.~C. Rubin and W.~K. Ford, {\it {Rotation of~the andromeda nebula from
  a~spectroscopic survey of~emission regions}},  {\em Astrophys.~J.} {\bf 159}
  (1970) 379.

\bibitem{Ade2014-A17}
P.~A.~R. Ade et~al., {\it {Planck 2013 results. XVII. Gravitational lensing
  by~large-scale structure}},  {\em Astron. Astrophys.} {\bf 571} (2014) A17.

\bibitem{RandallMarkevitchCloweGonzalezBradac2008-1173}
S.~W. Randall, M.~Markevitch, D.~Clowe, A.~H. Gonzalez, and M.~Bradac, {\it
  {Constraints on~the~self-interaction cross section of~dark matter from
  numerical simulations of~the~merging galaxy cluster 1E 0657--56}},  {\em
  Astrophys.~J.} {\bf 679} (2008) 1173.

\bibitem{massey2015-science}
P.~Harvey, R.~Massey, T.~Kitching, A.~Taylor, and E.~Tittley, {\it {The
  nongravitational interactions of~dark matter in~colliding galaxy clusters}},
  {\em Science} {\bf 347} (2015) 1462.

\bibitem{massey2015-mnras}
R.~Massey et~al., {\it {The behaviour of~dark matter associated with four
  bright cluster galaxies in~the~10 kpc core of~the Abell 3827}},  {\em
  Mon.~Not.~Roy.~Astron.~Soc.} {\bf 449} (2015) 3393.

\bibitem{HooperGoodenough2011-412}
D.~Hooper and L.~Goodenough, {\it {Dark matter annihilation in~the~galactic
  center as seen by~the~Fermi gamma ray space telescope}},  {\em Phys.~Lett.~B}
  {\bf 697} (2011) 412.

\bibitem{AbazajianKaplinghat2012-083511}
K.~N. Abazajian and M.~Kaplinghat, {\it {Detection of~a gamma-ray source
  in~the~galactic center consistent with extended emission from dark matter
  annihilation and concentrated astrophysical emission}},  {\em Phys.~Rev.~D}
  {\bf 86} (2012) 083511.

\bibitem{GordonMacias2013-083521}
C.~Gordon and O.~Macias, {\it {Dark matter and pulsar model constraints from
  Galactic Center Fermi-LAT gamma-ray observations}},  {\em Phys.~Rev.~D} {\bf
  88} (2013) 083521.

\bibitem{BulbulEtAl2014-13}
E.~Bulbul et~al., {\it {Detection of~an unindentified emission line
  in~the~stacked X-ray spectrum of~galaxy clusters}},  {\em Astrophys.~J.} {\bf
  789} (2014) 13.

\bibitem{BoyarskyRuchayskiyIakubovskyiFranse2014-251301}
A.~Boyarsky, O.~Ruchayskiy, D.~Iakubovskyi, and J.~Franse, {\it {An
  unidentified line in~X-ray spectra of~the Andromeda galaxy and Perseus galaxy
  cluster}},  {\em Phys.~Rev.~Lett.} {\bf 113} (2014) 251301.

\bibitem{Mitsou2013-1330052}
V.~A. Mitsou, {\it {Shedding light on~dark matter at colliders}},  {\em
  Int.~J.~Mod.~Phys.~A.} {\bf 28} (2013) 1330052.

\bibitem{Abercrombie2015-arxiv}
D.~Abercrombie et~al., {\it {Dark matter benchmark models for~early LHC run-2
  searches: report of~the ATLAS/CMS dark matter forum}},  {\em
  arXiv:1507.00966} (2015).

\bibitem{BramanteLinden2014-191301}
J.~Bramante and T.~Linden, {\it {Detecting dark matter with imploding pulsars
  in~the~galactic center}},  {\em Phys.~Rev.~Lett.} {\bf 113} (2014) 191301.

\bibitem{LopesSilk2014-25}
I.~Lopes and J.~Silk, {\it {A particle dark matter footprint on~the~first
  generation of~stars}},  {\em Astrophys.~J.} {\bf 786} (2014) 25.

\bibitem{ChangMaYuan2014-054}
C.-F. Chang, E.~Ma, and T.-C. Yuan, {\it {Multilepton Higgs decays through
  the~dark portal}},  {\em JHEP} {\bf 03} (2014) 054.

\bibitem{HeavyPhotonSearchExperiment}
{\em
  {https://confluence.slac.stanford.edu/display/hpsg/Heavy+Photon+Search+Experiment}}.

\bibitem{LeesEtAl2014-201801}
J.~P. Lees et~al., {\it {Search for~a~dark photon in~e$^+$e$^-$ collisions at
  BaBar}},  {\em Phys.~Rev.~Lett.} {\bf 113} (2014) 201801.

\bibitem{BaekKoPark2013-013}
S.~Baek, P.~Ko, and W.-I. Park, {\it {Singlet portal extensions of~the~standard
  seesaw models to~a~dark sector with~local dark symmetry}},  {\em JHEP} {\bf
  07} (2013) 013.

\bibitem{BaekKoPark2014-73}
S.~Baek, P.~Ko, and W.-I. Park, {\it {An alternative to the~standard model}},
  {\em AIP~Conf.~Proc.} {\bf 1604} (2014) 73.

\bibitem{BaekKoPark2015-255}
S.~Baek, P.~Ko, and W.-I. Park, {\it {Local Z$_2$ scalar dark matter model
  confronting galactic GeV-scale $\gamma$-ray}},  {\em Phys.~Lett.~B} {\bf 747}
  (2015) 255.

\bibitem{NakoniecznyRogatko2014-106004}
{\L}.~Nakonieczny and M.~Rogatko, {\it {Analytic study on~backreacting
  holographic superconductors with dark matter sector}},  {\em Phys.~Rev.~D}
  {\bf 90} (2014) 106004.

\bibitem{NakoniecznyRogatkoWysokinski2015-046007}
{\L}.~Nakonieczny, M.~Rogatko, and K.~I. Wysoki{\'n}ski, {\it {Magnetic field
  in~holographic superconductor with dark matter sector}},  {\em Phys.~Rev.~D}
  {\bf 91} (2015) 046007.

\bibitem{NakoniecznyRogatkoWysokinski2015-066008}
{\L}.~Nakonieczny, M.~Rogatko, and K.~I. Wysoki{\'n}ski, {\it {Analytic
  investigation of~holographic phase transitions influenced by~dark matter
  sector}},  {\em Phys.~Rev.~D} {\bf 92} (2015) 066008.

\bibitem{PerlmutterEtAl1999-565}
S.~Perlmutter et~al., {\it {Measurements of~$\Omega$ and $\Lambda$ from 42
  high-redshift supernovae}},  {\em Astrophys.~J.} {\bf 517} (1999) 565.

\bibitem{RiessEtAl1998-1009}
A.~G. Riess et~al., {\it {Observational evidence from supernovae
  for~an~accelerating universe and~a~cosmological constant}},  {\em
  Astrophys.~J.} {\bf 116} (1998) 1009.

\bibitem{SchmidtEtA1998l-46}
B.~P. Schmidt et~al., {\it {The high-z supernova search: Measuring cosmic
  deceleration and global curvature of~the Universe using type Ia supernovae}},
   {\em Astrophys.~J.} {\bf 507} (1998) 46.

\bibitem{TonryEtAl2003-1}
J.~L. Tonry et~al., {\it {Cosmological results from high-z supernovae}},  {\em
  Astrophys.~J.} {\bf 594} (2003) 1.

\bibitem{CaldwellKamionkowski2009-397}
R.~R. Caldwell and M.~Kamionkowski, {\it {The physics of~cosmic acceleration}},
   {\em Ann.~Rev.~Nucl.~Part.~Sci.} {\bf 59} (2009) 397.

\bibitem{Caldwell2002-23}
R.~R. Caldwell, {\it {A Phantom Menace? Cosmological consequences of~a dark
  energy component with super-negative equation of~state}},  {\em
  Phys.~Lett.~B} {\bf 545} (2002) 23.

\bibitem{GibbonsRasheed1996-515}
G.~W. Gibbons and D.~A. Rasheed, {\it {Dyson pairs and zero-mass black holes}},
   {\em Nucl.~Phys.~B} {\bf 476} (1996) 515.

\bibitem{ClementFabrisRodrigues2009-064021}
G.~Cl{\'e}ment, J.~C. Fabris, and M.~E. Rodrigues, {\it {Phantom black holes
  in~Einstein-Maxwell-dilaton theory}},  {\em Phys.~Rev.~D} {\bf 79} (2009)
  064021.

\bibitem{AzregAinouClementFabrisRodrigues2011-124001}
M.~Azreg-A{\"i}nou, G.~Cl{\'e}ment, J.~C. Fabris, and M.~E. Rodrigues, {\it
  {Phantom black holes and~sigma models}},  {\em Phys.~Rev.~D} {\bf 83} (2011)
  124001.

\bibitem{WheelerGeometrodynamics}
J.~A. Wheeler, {\em {Geometrodynamics}}.
\newblock Academic Press, New York, 1963.

\bibitem{MorrisThorne1988-395}
M.~S. Morris and K.~S. Thorne, {\it {Wormholes in~spacetime and their use
  for~interstellar travel: A tool for~teaching general relativity}},  {\em
  Am.~J.~Phys.} {\bf 56} (1988) 395.

\bibitem{LoboParsaelRiazi2013-084030}
F.~S.~N. Lobo, F.~Parsael, and N.~Riazi, {\it {New asymptotically flat phantom
  wormhole solutions}},  {\em Phys.~Rev.~D} {\bf 87} (2013) 084030.

\bibitem{LemosLoboOliveira2003-064004}
J.~P.~S. Lemos, F.~S.~N. Lobo, and S.~Q. de~Oliveira, {\it {Morris-Thorne
  wormholes with a~cosmological constant}},  {\em Phys.~Rev.~D} {\bf 68} (2003)
  064004.

\bibitem{LoboOliveiro2010-067501}
F.~S.~N. Lobo and M.~A. Oliveiro, {\it {General class of~vacuum Brans-Dicke
  wormholes}},  {\em Phys.~Rev.~D} {\bf 81} (2010) 067501.

\bibitem{EiroaSimeone2005-127501}
E.~F. Eiroa and C.~Simeone, {\it {Thin-shell wormholes in~dilaton gravity}},
  {\em Phys.~Rev.~D} {\bf 71} (2005) 127501.

\bibitem{GonzalesGuzmanMontelongoGarciaZannias2009-064027}
J.~A. Gonz{\'a}les, F.~S. Guzm{\'a}n, N.~Montelongo-Garc{\'i}a, and T.~Zannias,
  {\it {On wormholes supported by~phantom energy}},  {\em Phys.~Rev.~D} {\bf
  79} (2009) 064027.

\bibitem{Sushkov2005-043520}
S.~Sushkov, {\it {Wormholes supported by~phantom energy}},  {\em Phys.~Rev.~D}
  {\bf 71} (2005) 043520.

\bibitem{DasKar2005-3045}
A.~Das and S.~Kar, {\it {The Ellis wormhole with tachyon matter}},  {\em
  Class.~Quant.~Grav.} {\bf 225} (2005) 3045.

\bibitem{BalakinLemosZayats2010-084015}
A.~B. Balakin, J.~P.~S. Lemos, and A.~E. Zayats, {\it {Nonminimal coupling
  for~the~gravitational and electromagnetic fields: Traversable electric
  wormholes}},  {\em Phys.~Rev.~D} {\bf 81} (2010) 084015.

\bibitem{KantiKleihausKunz2011-271101}
P.~Kanti, B.~Kleihaus, and J.~Kunz, {\it {Wormholes in~dilatonic
  Einstein-Gauss-Bonnet theory}},  {\em Phys.~Rev.~Lett.} {\bf 107} (2011)
  271101.

\bibitem{KantiKleihausKunz2012-044007}
P.~Kanti, B.~Kleihaus, and J.~Kunz, {\it {Stable Lorentzian wormholes
  in~dilatonic Einstein-Gauss-Bonnet theory}},  {\em Phys.~Rev.~D} {\bf 85}
  (2012) 044007.

\bibitem{HarkoLoboMakSushkov2013-067504}
T.~Harko, F.~S.~N. Lobo, M.~K. Mak, and S.~V. Sushkov, {\it {Modified-gravity
  wormholes without exotic matter}},  {\em Phys.~Rev.~D} {\bf 87} (2013)
  067504.

\bibitem{MehdizadehZangenehLobo2015-084004}
M.~R. Mehdizadeh, M.~K. Zangeneh, and F.~S.~N. Lobo, {\it
  {Einstein-Gauss-Bonnet traversable wormholes satisfying the weak energy
  condition}},  {\em Phys.~Rev.~D} {\bf 91} (2015) 084004.

\bibitem{CataldoMeza2013-064012}
M.~Cataldo and P.~Meza, {\it {Phantom evolving wormholes with big rip
  singularities}},  {\em Phys.~Rev.~D} {\bf 87} (2013) 064012.

\bibitem{ZangenehLoboRiazi2014-024072}
M.~K. Zangeneh, F.~S.~N. Lobo, and N.~Riazi, {\it {Higher-dimensional evolving
  wormholes satisfying the null energy condition}},  {\em Phys.~Rev.~D} {\bf
  90} (2014) 024072.

\bibitem{PanChakraborty2015-1}
S.~Pan and S.~Chakraborty, {\it {Dynamic wormholes with particle creation
  mechanism}},  {\em Eur.~Phys.~J.~C} {\bf 75} (2015) 1.

\bibitem{DamourSolodukhin2007-024016}
T.~Damour and S.~N. Solodukhin, {\it {Wormholes as black hole foils}},  {\em
  Phys.~Rev.~D} {\bf 76} (2007) 024016.

\bibitem{Bambi2013-107501}
C.~Bambi, {\it {Can the~supermassive objects at the~centers of~galaxies be
  traversable wormholes? The first test of~strong gravity for~mm/sub-mm VLBI
  facilities}},  {\em Phys.~Rev.~D} {\bf 87} (2013) 107501.

\bibitem{LiBambi2014-024071}
Z.~Li and C.~Bambi, {\it {Distinguishing black holes and wormholes with
  orbiting hot spots}},  {\em Phys.~Rev.~D} {\bf 90} (2014) 024071.

\bibitem{VirbhadraEllis2000-084003}
K.~S. Virbhadra and G.~F.~R. Ellis, {\it {Schwarzschild black hole lensing}},
  {\em Phys.~Rev.~D} {\bf 62} (2000) 084003.

\bibitem{VirbhadraEllis2002-103004}
K.~S. Virbhadra and G.~F.~R. Ellis, {\it {Gravitational lensing by naked
  singularities}},  {\em Phys.~Rev.~D} {\bf 65} (2002) 103004.

\bibitem{VirbhadraKeeton2008-124014}
K.~S. Virbhadra and C.~R. Keeton, {\it {Time delay and magnification centroid
  due to gravitational lensing by black holes and naked singularities}},  {\em
  Phys.~Rev.~D} {\bf 77} (2008) 124014.

\bibitem{Virbhadra2009-083004}
K.~S. Virbhadra, {\it {Relativistic images of Schwarzschild black hole
  lensing}},  {\em Phys.~Rev.~D} {\bf 79} (2009) 083004.

\bibitem{KovacsHarko2010-124047}
Z.~Kov{\'a}cs and T.~Harko, {\it {Can accretion disk properties observationally
  distinguish black holes from naked singularities?}},  {\em Phys.~Rev.~D} {\bf
  82} (2010) 124047.

\bibitem{JoshiMalafarinaNarayan2014-015002}
P.~S. Joshi, D.~Malafarina, and R.~Narayan, {\it {Distinguishing black holes
  from naked singularities through their accretion disc properties}},  {\em
  Class.~Quant.~Grav.} {\bf 31} (2014) 015002.

\bibitem{SahuPatilNarasimhaJoshi2012-063010}
S.~Sahu, M.~Patil, D.~Narasimha, and P.~S. Joshi, {\it {Can strong
  gravitational lensing distinguish naked singularities from black holes?}},
  {\em Phys.~Rev.~D} {\bf 86} (2012) 063010.

\bibitem{Rakhmanov1994-5155}
M.~Rakhmanov, {\it {Dilaton black holes with electric charge}},  {\em
  Phys.~Rev.~D} {\bf 50} (1994) 5155.

\bibitem{GuptaPrimulandoSaraswat2015-079}
A.~Gupta, R.~Primulando, and P.~Saraswat, {\it {A~new probe of~dark sector
  dynamics at~the~LHC}},  {\em JHEP} {\bf 09} (2015) 079.

\bibitem{SuzukiHorieInoueMinowa2015-042}
J.~Suzuki, T.~Horie, Y.~Inoue, and M.~Minowa, {\it {Experimental search
  for~hidden photon CDM in~the~eV mass range with a~dish antenna}},  {\em JCAP}
  {\bf 09} (2015) 042.

\bibitem{DreinerFortinHanhartUbaldi2014-105015}
H.~K. Dreiner, J.-F. Fortin, C.~Hanhart, and L.~Ubaldi, {\it {Supernova
  constraints on~MeV dark sectors from $e^+e^-$ annihilations}},  {\em
  Phys.~Rev.~D} {\bf 89} (2014) 105015.

\bibitem{Blumlein2011-155}
J.~Bl{\"u}mlein and J.~Brunner, {\it {New exclusion limits for~dark gauge
  forces from beam-dump data}},  {\em Phys.~Lett.~B} {\bf 701} (2011) 155.

\bibitem{Blumlein2014-320}
J.~Bl{\"u}mlein and J.~Brunner, {\it {New exclusion limits on~dark gauge forces
  from~proton Bremsstrahlung in~beam-dump data}},  {\em Phys.~Lett.~B} {\bf
  731} (2014) 320.

\bibitem{AadEtAl2014-088}
G.~Aad et~al., {\it {Search for~long-lived neutral particles decaying into
  lepton jets in~proton-proton collisions at~$\sqrt{s}=8$~TeV with~the~ATLAS
  detector}},  {\em JHEP} {\bf 11} (2014) 088.

\bibitem{Gninenko2012-244}
S.~N. Gninenko, {\it {Constraints on~sub-GeV hidden sector gauge bosons from
  a~search for~heavy neutrino decays}},  {\em Phys.~Lett.~B} {\bf 713} (2012)
  244.

\bibitem{MirizziRedondoSigl2009-026}
A.~Mirizzi, J.~Redondo, and G.~Sigl, {\it {Microwave background constraints
  on~mixing of~photons with hidden photons}},  {\em JCAP} {\bf 03} (2009) 026.

\bibitem{Afanasev2009-317}
A.~Afanasev, O.~Baker, K.~Beard, G.~Biallas, J.~Boyce, M.~Minarni, R.~Ramdon,
  M.~Shinn, and P.~Slocum, {\it {New experimental limit on~photon hidden-sector
  paraphoton mixing}},  {\em Phys.~Lett.~B} {\bf 679} (2009) 317.

\bibitem{Archilli2012-251}
F.~Archilli et~al., {\it {Search for a~vector gauge boson in~$\phi$ meson
  decays with the~KLOE detector}},  {\em Phys.~Lett.~B} {\bf 706} (2012) 251.

\bibitem{Babusci2013-111}
D.~Babusci, {\it {Limit on~the~production of~a~light vector gauge boson
  in~$\phi$ meson decays with the~KLOE detector}},  {\em Phys.~Lett.~B} {\bf
  720} (2013) 111.

\bibitem{Adlarson2013-187}
P.~Adlarson et~al., {\it {Search for a~dark photon in~the~$\pi^0\to
  e^+e^-\gamma$ decay}},  {\em Phys.~Lett.~B} {\bf 726} (2013) 187.

\bibitem{AbrahamyanEtAl2011-191804}
S.~Abrahamyan et~al., {\it {Search for~a~new gauge boson in~electron-nucleus
  fixed-target scattering by~the~APEX experiment}},  {\em Phys.~Rev.~Lett.}
  {\bf 107} (2011) 191804.

\bibitem{Merkel2011-251802}
H.~Merkel et~al., {\it {Search for~light gauge bosons of~the~dark sector
  at~the~Mainz Microtron}},  {\em Phys.~Rev.~Lett.} {\bf 106} (2011) 251802.

\bibitem{Agakishiev2014-265}
G.~Agakishiev et~al., {\it {Searching a~dark photon with HADES}},  {\em
  Phys.~Lett.~B} {\bf 731} (2014) 265.

\bibitem{Ortin}
T.~Ort{\'i}n, {\em {Gravity and Strings}}.
\newblock Cambridge University Press, 2004.

\bibitem{AlvarezConde2002-413}
E.~Alvarez and J.~Conde, {\it {Are the string and Einstein frames
  equivalent?}},  {\em Mod.~Phys.~Lett.~A} {\bf 17} (2002) 413.

\bibitem{Veneziano2002-581}
G.~Veneziano, {\it {String Cosmology: The pre-Big Bang scenario}},  in {\em
  {The primordial universe - L{\rq}univers primordial}} (P.~Bin{\'e}truy,
  R.~Schaeffer, J.~Silk, and F.~David, eds.), vol.~71 of {\em {Les Houches -
  Ecole d{\rq}Ete de Physique Theorique}}, pp.~581--628.
\newblock Springer Berlin Heidelberg, 2002.

\bibitem{Flanagan2004-071101}
{\'E}.~{\'E}. Flanagan, {\it {Palatini form of 1/R gravity}},  {\em
  Phys.~Rev.~Lett.} {\bf 92} (2004) 071101.

\bibitem{Vollick2004-3813}
D.~N. Vollick, {\it {On the viability of the Palatini form of 1/R gravity}},
  {\em Class.~Quant.~Grav.} {\bf 21} (2004) 3813.

\bibitem{DomenechSasaki2015-022}
G.~Dom{\`e}nech and M.~Sasaki, {\it {Conformal frame dependence of inflation}},
   {\em JCAP} {\bf 04} (2015) 022.

\bibitem{WhiteMinamitsujiSasaki2013-015}
J.~White, M.~Minamitsuji, and M.~Sasaki, {\it {Non-linear curvature
  perturbation in multi-field inflation models with non-minimal coupling}},
  {\em JCAP} {\bf 09} (2013) 015.

\bibitem{ChibaYamaguchi2013-040}
T.~Chiba and M.~Yamaguchi, {\it {Conformal-frame (in)dependence of cosmological
  observations in scalar-tensor theory}},  {\em JCAP} {\bf 10} (2013) 040.

\bibitem{MisnerThorneWheeler}
C.~W. Misner, K.~S. Thorne, and J.~A. Wheeler, {\em {Gravitation}}.
\newblock W. H. Freeman and Company, 1973.

\bibitem{InvernoSmallwood1980-1223}
R.~A. d'Inverno and J.~Smallwood, {\it {Covariant 2+2 formulation of~the
  initial-value problem in~general relativity}},  {\em Phys.~Rev.~D} {\bf 225}
  (1980) 1223.

\bibitem{HamadeStewart1996-497}
R.~S. Hamad{\'e} and J.~M. Stewart, {\it {The spherically symmetric collapse
  of~a massless scalar field}},  {\em Class.~Quant.~Grav.} {\bf 13} (1996) 497.

\bibitem{AyalPiran1997-4768}
S.~Ayal and T.~Piran, {\it {Spherical collapse of~a massless scalar field with
  semiclassical corrections}},  {\em Phys.~Rev.~D} {\bf 56} (1997) 4768.

\bibitem{Choptuik1993-9}
M.~W. Choptuik, {\it {Universality and scaling in~gravitational collapse of~a
  massless scalar field}},  {\em Phys.~Rev.~Lett.} {\bf 70} (1993) 9.

\bibitem{FrolovNovikov}
V.~P. Frolov and I.~D. Novikov, {\em {Black hole physics: Basic concepts and
  new developments}}.
\newblock Kluwer Academic Publishers, 1998.

\bibitem{Hawking1968-598}
S.~W. Hawking, {\it {Gravitational radiation in~an expanding universe}},  {\em
  J.~Math.~Phys.} {\bf 9} (1968) 598.

\bibitem{HochbergVisser1998-044021}
D.~Hochberg and M.~Visser, {\it {Dynamic wormholes, antitrapped surfaces
  and~energy conditions}},  {\em Phys.~Rev.~D} {\bf 58} (1998) 044021.

\bibitem{HochbergVisser1998-746}
D.~Hochberg and M.~Visser, {\it {The null energy condition in~dynamic
  wormholes}},  {\em Phys.~Rev.~Lett.} {\bf 81} (1998) 746.

\bibitem{MaedaHaradaCarr2009-044034}
H.~Maeda, T.~Harada, and B.~J. Carr, {\it {Cosmological wormholes}},  {\em
  Phys.~Rev.~D} {\bf 79} (2009) 044034.

\bibitem{Hayward1999-373}
S.~A. Hayward, {\it {Dynamic wormholes}},  {\em Int.~J.~Mod.~Phys.~D} {\bf 8}
  (1999) 373.

\bibitem{Hayward2009-124001}
S.~A. Hayward, {\it {Wormhole dynamics in~spherical symmetry}},  {\em
  Phys.~Rev.~D} {\bf 79} (2009) 124001.

\bibitem{ShinkaiHayward2002-044005}
H.~Shinkai and S.~A. Hayward, {\it {Fate of~the first traversible wormhole:
  black hole collapse or~inflationary expansion}},  {\em Phys.~Rev.~D} {\bf 66}
  (2002) 044005.

\bibitem{KoyamaHaywardKim2003-084008}
H.~Koyama, S.~A. Hayward, and S.-W. Kim, {\it {Construction and enlargement
  of~dilatonic wormholes by~impulsive radiation}},  {\em Phys.~Rev.~D} {\bf 67}
  (2003) 084008.

\bibitem{KoyamaHayward2004-084001}
H.~Koyama and S.~A. Hayward, {\it {Construction and enlargement of~traversable
  wormholes from Schwarzschild black holes}},  {\em Phys.~Rev.~D} {\bf 70}
  (2004) 084001.

\bibitem{OkawaCardosoPani2014-041502}
H.~Okawa, V.~Cardoso, and P.~Pani, {\it {Collapse of~self-interacting fields
  in~asymptotically flat spacetimes: do self-interactions render Minkowski
  spacetime unstable?}},  {\em Phys.~Rev.~D} {\bf 89} (2014) 041502.

\bibitem{HaradaMaedaCarr2006-024024}
T.~Harada, H.~Maeda, and B.~J. Carr, {\it {Nonexistence of~self-similar
  solutions containing a~black hole in~a~universe with a~stiff fluid or~scalar
  field or~quintessence}},  {\em Phys.~Rev.~D} {\bf 74} (2006) 024024.

\bibitem{CarrKohriSendoudaYokoyama2010-104019}
B.~J. Carr, K.~Kohri, Y.~Sendouda, and J.~Yokoyama, {\it {New cosmological
  constraints on~primordial black holes}},  {\em Phys.~Rev.~D} {\bf 81} (2010)
  104019.

\bibitem{EinsteinStraus1945-120}
A.~Einstein and E.~G. Straus, {\it {The~influence of~the~expansion of~space
  on~the~gravitation fields surrounding the~individual stars}},  {\em
  Rev.~Mod.~Phys.} {\bf 17} (1945) 120.

\bibitem{McVittie1933-325}
G.~C. McVittie, {\it {The mass-particle in~an~expanding universe}},  {\em
  Mon.~Not.~Roy.~Astron.~Soc.} {\bf 93} (1933) 325.

\bibitem{KastorTrashen1993-5370}
D.~Kastor and J.~Traschen, {\it {Cosmological multi-black-hole solutions}},
  {\em Phys.~Rev.~D} {\bf 47} (1993) 5370.

\bibitem{GibbonsMaeda2010-131101}
G.~W. Gibbons and K.-i. Maeda, {\it {Black holes in~an~expanding Universe}},
  {\em Phys.~Rev.~Lett.} {\bf 104} (2010) 131101.

\bibitem{Rogatko2001-064014}
M.~Rogatko, {\it {Dilaton black holes on~thick branes}},  {\em Phys.~Rev.~D}
  {\bf 64} (2001) 064014.

\bibitem{Rogatko2004-044022}
M.~Rogatko, {\it {Cosmological black holes on~branes}},  {\em Phys.~Rev.~D}
  {\bf 69} (2004) 044022.

\bibitem{MaedaNozawa2010-044017}
K.-i. Maeda and M.~Nozawa, {\it {Black hole in~the~expanding universe from
  intersecting branes}},  {\em Phys.~Rev.~D} {\bf 81} (2010) 044017.

\bibitem{ChadburnGregory2014-195006}
S.~Chadburn and R.~Gregory, {\it {Time dependent black holes and~scalar hair}},
   {\em Class.~Quant.~Grav.} {\bf 31} (2014) 195006.

\end{thebibliography}\endgroup

\end{document}